\documentclass[11pt]{article}
\usepackage{braket}

\usepackage{amsmath,amsfonts,amssymb,graphicx,epsfig,pdflscape,multirow}
\usepackage{arydshln,cite,enumitem}

\usepackage{amsthm}

\numberwithin{equation}{section}
\graphicspath{{./eps/}}
\usepackage{cite}
\usepackage[usenames]{color}
\usepackage{pstricks}
\usepackage{caption}
\usepackage[backref=false]{hyperref}
\usepackage[capitalise,noabbrev]{cleveref} 
\hypersetup{
colorlinks=true,
citecolor=red,
linkcolor=darkblue,
urlcolor=darkblue
}

\long\def\ignore#1{}
\definecolor{darkblue}{rgb}{0,0,.8}
\definecolor{red}{rgb}{1,0,0}
\definecolor{purple}{rgb}{1,0.4,1}
\definecolor{coloroflink}{rgb}{0.7,0,1}
\definecolor{pink}{rgb}{1,.7,.7}
\definecolor{lightblue}{rgb}{.61,.61,1}
\definecolor{midblue}{rgb}{.7,.7,1}
\definecolor{lightlightblue}{rgb}{.9,.9,1}
\definecolor{lightestblue}{rgb}{.96,.96,1}
\definecolor{lightpurple}{rgb}{1,.65,1}
\definecolor{darkgreen}{rgb}{0.180392, 0.545098, 0.341176}
\definecolor{mybrown}{rgb}{0.69,0.49,0.30}

\newtheoremstyle{smallcaps}{5pt}{5pt}{\itshape}{}{}{}{.5em}
{\scshape\thmname{#1}~\thmnumber{#2}.\thmnote{~\textnormal{(#3)}}}
\theoremstyle{smallcaps}

\newtheorem{Theoreme}{Theorem}

\numberwithin{equation}{section}

\newcommand{\nc}{\newcommand}
\nc{\bib}{\bibitem}
\nc{\be}{\begin{equation}}
\nc{\ee}{\end{equation}}
\nc{\Mod}{\textrm{\,mod\,}}
\nc{\Tt}{\textrm{t}}	

\nc{\ir}{\mathrm{i}}
\nc{\eE}{\mathsf{e}} 
\nc{\dd}{\mathrm{d}}   
\nc{\eps}{\epsilon}

\nc{\elegant}{1.5pt}
\nc{\moyen}{1.0pt}
\nc{\mince}{0.5pt}

\nc{\Db}{\mbox{\boldmath $D$}}
\nc{\Ib}{\mbox{\boldmath $I$}}
\nc{\tl}{\mbox{$\mathsf {TL}$}}
\nc{\chit}{\protect\raisebox{0.25ex}{$\chi$}}

\def\facegrid#1#2{
\psframe[fillstyle=solid,fillcolor=lightlightblue,linewidth=0pt]#1#2
\psgrid[gridlabels=0pt,subgriddiv=1]#1#2}

\def\loopa{
\psframe[linewidth=.25pt](0,0)(1,1)
\psarc[linewidth=1.5pt,linecolor=blue](1,0){.5}{90}{180}
\psarc[linewidth=1.5pt,linecolor=blue](0,1){.5}{-90}{0}
}
\def\loopb{
\psframe[linewidth=.25pt](0,0)(1,1)
\psarc[linewidth=1.5pt,linecolor=blue](0,0){.5}{0}{90}
\psarc[linewidth=1.5pt,linecolor=blue](1,1){.5}{180}{270}
}

\renewcommand{\ge}{\geqslant}
\renewcommand{\le}{\leqslant}

\begin{document}

\topmargin -15mm
\oddsidemargin 05mm

%
%

\title{\mbox{}\vspace{-.2in}
\bf 
\huge Bipartite fidelity of critical dense polymers
}

\date{}
\maketitle

\begin{center}
{\large \textsc{Gilles Parez}, \quad \textsc{Alexi Morin-Duchesne}, \quad \textsc{Philippe Ruelle}
}
\end{center}\vspace{0cm}

\begin{center}
{\em Universit\'e catholique de Louvain \\ Institut de Recherche en Math\'ematique et Physique\\ Chemin du Cyclotron 2, 1348 Louvain-la-Neuve, Belgium}
\end{center}

\begin{center}
{\tt gilles.parez\,@\,uclouvain.be}
\qquad
 {\tt alexi.morin-duchesne\,@\,uclouvain.be}
\qquad
{\tt philippe.ruelle\,@\,uclouvain.be} 
\end{center}
\medskip

%
%
 
\begin{abstract}
We investigate the bipartite fidelity $\mathcal F_d$ for a lattice model described by a logarithmic CFT: the model of critical dense polymers. We define this observable in terms of a partition function on the pants geometry, where $d$ defects enter at the top of the pants lattice and exit in one of the legs. Using the correspondence with the XX spin chain, we obtain an exact closed-form expression for $\mathcal F_d$ and compute the leading terms in its $1/N$ asymptotic expansion as a function of $x = N_A/N$, where $N$ is the lattice width at the top of the pants and $N_A$ is the width of the leg where the defects exit. We find an agreement with the results of St\'ephan and Dubail for rational CFTs, with the central charge and conformal weights specialised to $c=-2$ and $\Delta = \Delta_{1,d+1} = \frac{d(d-2)}8$.

We compute a second instance $\mathcal {\tilde F}_2$ of the bipartite fidelity for $d=2$ by imposing a different rule for the connection of the defects. In the conformal setting, this choice corresponds to inserting two boundary condition changing fields of weight $\Delta = 0$ that are logarithmic instead of primary. We compute the asymptotic expansion in this case as well and find a simple additive correction compared to $\mathcal F_2$, of the form $-2\log((1+x)/(2\sqrt{x}))$. We confirm this lattice result with a CFT derivation and find that this correction term is identical for all logarithmic theories, independently of $c$ and $\Delta$.

\end{abstract}

\vspace{.5cm}
\noindent\textbf{Keywords:} Entanglement, bipartite fidelity, dense loop models, logarithmic conformal field theory.\\

%
%

\newpage

\tableofcontents
\clearpage

\section{Introduction}

A physical system is said to be entangled if performing a measure locally affects the physical behaviour in other parts of the system that are far away. Entanglement is a concept that is purely quantum; it has no classical counterpart. The recent resurgence of interest for entanglement \cite{OAFF02,ON02,VLRK03,CC06,AFOV08,ECP08,ALS09,CCD09,LR10,C11} stems from its importance in quantum information theory, condensed matter physics and high energy physics. 

One way to measure the entanglement of a system is via the {\it entanglement entropy}, or {\it bipartite Von Neumann entropy} \cite{vN55}.
Let us suppose that the physical system is partitioned in two parts $A$ and $B$, and that it is in a pure state $|\phi\rangle$. The entanglement entropy $S_A$ is then defined as
\be
\label{eq:SA}
S_A = - \textrm{tr} (\rho_A \log \rho_A), \qquad \rho_A = \textrm{tr}_B(\rho),\qquad \rho = |\phi\rangle\langle \phi|,
\ee
where $\textrm{tr}_B$ indicates a trace on the degrees of freedom in $B$.
For statistical models that are not critical, the entanglement entropy satisfies an area law \cite{S93,ECP08}:
it is proportional to the area of the boundary between $A$ and $B$. For off-critical quantum models defined on a one-dimensional lattice, this boundary is zero-dimensional and the entanglement entropy saturates to a constant value as the system size $N$ increases to infinity.
In stark contrast, at criticality, the entanglement entropy for these one-dimensional systems diverges logarithmically with $N$. 
The overall constant is predicted by conformal field theory (CFT) \cite{CC06} to be proportional to the central charge $c$,
\be
\label{eq:SA.leading}
S_A =  \frac{a c}6 \log N + \mathcal O(1),
\ee
where $a$ is the number of contact points between $A$ and $B$. In particular, $a$ equals $2$ for periodic boundary conditions, whereas it equals $1$ if one end of $A$ is attached to a boundary. The entanglement entropy is therefore a real-valued observable that allows one to detect quantum phase transitions.

A second observable that shares many features with entanglement entropy is the {\it fidelity} \cite{ZP06,ZB08,S10,G10}. In general, the fidelity is defined as the overlap between the groundstates of two systems that differ by a perturbation. In the case where the system is partitioned into two subsystems $A$ and $B$, one expresses the Hamiltonian $H = H^{AB}$ of the full system as
\be
H^{AB} = H^A + H^B + H^{\textrm{int}},
\ee
where $H^{\textrm{int}}$ is the part of $H^{AB}$ that involves degrees of freedom of both $A$ and $B$, and is taken as the perturbation. The groundstates of $H^{AB}$,  $H^A$ and $H^B$ are denoted by $|\phi^{AB}\rangle$, $|\phi^{A}\rangle$ and $|\phi^{B}\rangle$ respectively. The (logarithmic) {\it bipartite fidelity} $\mathcal F_{A,B}$  is defined as \cite{DS11,SD13}
\be
\label{eq:LBF}
\mathcal F_{A,B} = - \log \bigg|\frac{\langle \phi^A \otimes \phi^B|\phi^{AB}\rangle}{\sqrt{\langle\phi^{AB}|\phi^{AB}\rangle\langle\phi^{A}|\phi^{A}\rangle\langle\phi^{B}|\phi^{B}\rangle}}\bigg|^2,
\ee
where we use the notation $\langle \phi^A \otimes \phi^B | \equiv \langle \phi^A | \otimes \langle \phi^B |$. 
If the system is completely disentangled, then each of these groundstates can be decomposed as a tensor product of states living on the individual sites of the lattice, $| \phi \rangle = \otimes_j |\phi_j\rangle$. If $|\phi_j\rangle$ is independent of $j$ and $N$, then  $\mathcal F_{A,B} = 0$. In the other cases, $\mathcal F_{A,B}$ is a positive real number.

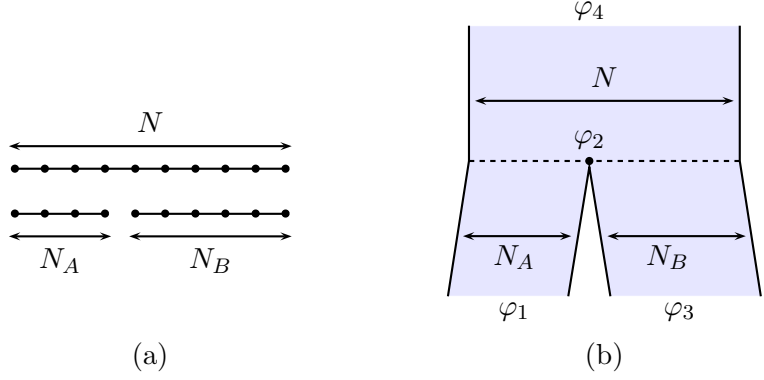
\begin{figure}
\begin{center}
\psset{unit=0.4cm}
\begin{pspicture}[shift=0](0,-6.0)(9,2.5)
\multiput(0,0.75)(1,0){10}{\psarc[fillstyle=solid,fillcolor=black]{-}(0,0){0.1}{0}{360}}
\multiput(0,-0.75)(1,0){10}{\psarc[fillstyle=solid,fillcolor=black]{-}(0,0){0.1}{0}{360}}
\psline{-}(0,0.75)(9,0.75)
\psline{-}(0,-0.75)(3,-0.75)
\psline{-}(4,-0.75)(9,-0.75)
\psline{<->}(-0.2,1.5)(9.2,1.5)\rput(4.5,2.3){$N$}
\psline{<->}(-0.2,-1.5)(3.2,-1.5)\rput(1.5,-2.3){$N_A$}
\psline{<->}(3.8,-1.5)(9.2,-1.5)\rput(6.5,-2.3){$N_B$}
\rput(4.5,-5.5){(a)}
\end{pspicture}
\qquad\qquad\qquad 
\begin{pspicture}(0,-2.5)(9,9)
\pspolygon[fillstyle=solid,fillcolor=lightlightblue,linewidth=0pt,linecolor=white](-0.7,0)(0,4.5)(0,9)(9,9)(9,4.5)(9.7,0)(4.7,0)(4,4.3)(3.3,0)
\psline{-}(-0.7,0)(0,4.5)(0,9)
\psline{-}(9,9)(9,4.5)(9.7,0)
\psline{-}(4.7,0)(4,4.3)(3.3,0)
\psline[linestyle=dashed,dash=2pt 2pt](0,4.5)(9,4.5)
\psline{<->}(0.2,6.5)(8.8,6.5)\rput(4.5,7.3){$N$}
\psline{<->}(-0.2,2.0)(3.4,2.0)\rput(1.5,1.3){$N_A$}
\psline{<->}(4.6,2.0)(9.2,2.0)\rput(6.6,1.3){$N_B$}
\psarc[fillstyle=solid,fillcolor=black]{-}(4,4.5){0.1}{0}{360}
\rput(1.5,-0.5){$\varphi_1$}
\rput(7,-0.5){$\varphi_3$}
\rput(4,5.2){$\varphi_2$}
\rput(4,9.5){$\varphi_4$}
\rput(4.5,-2){(b)}
\end{pspicture}
\caption{(a) A one-dimensional quantum system and its division in two parts $A$ and $B$. (b) The two-dimensional lattice on the pants corresponding to the scalar product $\langle \phi^A \otimes \phi^B|\phi^{AB}\rangle$.}
\label{fig:pants}
\end{center}
\end{figure}

An integrable quantum model in one dimension often underlies a statistical model in two dimensions \cite{Baxterbook}. The transfer matrices $T(u)$ for the latter commute at different values of the spectral parameter $u$, and the Hamiltonian of the former is obtained as a leading term in the expansion of $T(u)$ around $u=0$. From its definition in terms of scalar products of groundstates, the bipartite fidelity then has an interpretation in terms of ratios of partition functions in the two-dimensional model. 
The overlap $\langle \phi^A \otimes \phi^B|\phi^{AB}\rangle$, in particular, is related to the partition function on the so-called {\it pants geometry}, whose lattice is equipped 
with a long vertical slit. This is illustrated in \cref{fig:pants}. The sizes of the subsystems $A$ and $B$ are denoted by $N_A$ and $N_B$ and satisfy $N_A+N_B=N$. Likewise, the scalar products in the denominators in \eqref{eq:LBF} are tied to partition functions on rectangles of width $N$, $N_A$ and~$N_B$. The bipartite fidelity is equal to a ratio of partition functions on these domains, in the limit where the height $M$ of the lattices grows to infinity.

Analogously to the entanglement entropy, the bipartite fidelity can also be used to detect quantum phase transitions. It satisfies an area law away from criticality and has extra logarithmic divergences at the critical point \cite{DS11}. For critical lattice models in $1+1$ dimensions, St\'ephan and Dubail \cite{DS11,SD13} performed a thorough investigation of the universal behaviour of $\mathcal F_{A,B}$ using CFT and obtained expressions for the leading terms in its $1/N$ asymptotic expansion. In the simplest case where all three states correspond to the vacuum of the CFT, as in \eqref{eq:LBF}, their result is an application of the Cardy-Peschel formula \cite{CP88} and the leading $\log N$ term has the coefficient $c/8$, where $c$ is the central charge. More generally, the bipartite fidelity can be defined, as in \eqref{eq:LBF}, in terms of three eigenstates $|\phi^{AB}\rangle$, $|\phi^{A}\rangle$ and $|\phi^{B}\rangle$ that do not correspond to the vacuum. The leading terms then depend on the weights $\Delta_i$ of four fields $\varphi_i$, $i = 1,2,3,4$, that account for changes in the boundary condition, and $\mathcal F_{A,B}$ has an interpretation as a four-point function of these fields. In the two-dimensional model, these fields are inserted at infinity in each extremity of the pants lattice, as well as on the endpoint of the slit, see \cref{fig:pants} (b). The lattice width $N$ is taken to infinity with the aspect ratio $x = N_A/N$ kept fixed in the range $(0,1)$. The resulting $1/N$ expansion reads \cite{SD13}
\be
\label{eq:FAB.DS}
\mathcal F_{A,B} = \Big(\frac c 8 + \Delta_2\Big) \log N + f(x) + g(x) \frac {\log N}N + \mathcal O(N^{-1}),
\ee
where
\begin{alignat}{2}\label{eq:f(x)}
f(x) &= \bigg[\frac c{24}\Big(2x-1+\frac 2 x\Big) + \Big(\frac43 - \frac 2x\Big) \Delta_1 + \frac{\Delta_2}3 - \frac{2 \Delta_3}3 +\Big(\frac 4 3 - 2x\Big)\Delta_4\bigg] \log{(1-x)} - \log \Upsilon(x)\\[0.15cm]
& + \Big\{ x \to 1-x; \Delta_1 \leftrightarrow \Delta_3 \Big\} + C,\nonumber
\end{alignat}
and $C$ is a non-universal constant that is independent of $x$.
Here, $\Upsilon(x)$ is the function that appears in the four point function of the fields $\varphi_1, \dots, \varphi_4$ in the upper-half plane~$\mathbb H$. It depends on the anharmonic ratio and reads
\be\label{eq:4ptfct}
\big\langle\varphi_1(z_1)\varphi_2(z_2)\varphi_3(z_3)\varphi_4(z_4) \big\rangle_{\mathbb H} = \Upsilon(\zeta) \times \prod_{1\le i<j\le 4} z_{ij}^{\hat\Delta/3 - \Delta_i - \Delta_j}
\ee
with
\be
z_{ij} = z_i-z_j, \qquad \zeta = \frac{z_{12}z_{34}}{z_{13}z_{24}}, \qquad \hat\Delta = \Delta_1+\Delta_2+\Delta_3+\Delta_4.
\ee
The notation $\{ x \to 1-x; \Delta_1 \leftrightarrow \Delta_3 \}$ on the second line of \eqref{eq:f(x)} indicates that the content of the first line must be duplicated but with $x$ replaced by $1-x$ and $\Delta_1$ exchanged with $\Delta_3$. The function $g(x)$ in \eqref{eq:FAB.DS} is also given in \cite{SD13} and takes the form of a universal function times a non-universal prefactor known as the {\it extrapolation length} $\Xi$:
\be
\label{eq:g(x)}
g(x) = \Xi \times \bigg[ \Delta_4-\frac c {24} + \Big(\frac c{24}- \Delta_1\Big)\frac 1x + \Big(\frac c{24}- \Delta_3\Big)\frac 1{1-x}\bigg].
\ee

Up until now, there have been only few cases where the bipartite fidelity was computed explicitly via an exact lattice derivation. It was computed for the XX spin chain with free boundary conditions in \cite{SD13} for the special case $x = 1/2$ and $N=4n$ a multiple of $4$. It was also obtained \cite{HL17} for the XXZ chain at the anisotropy $\Delta = -1/2$ with real magnetic fields at the endpoints. This last result is for a single eigenstate, namely the special groundstate that has miraculous combinatorial properties. The bipartite fidelity was also computed by Weston \cite{W11,W12} for the XXZ chain in the infinite chain directly, for generic values of $\Delta$ in the range $-\infty < \Delta < -1$. In this case, the model is not critical and the result is a function of the correlation length $\xi$. A first objective of this paper is to provide a new lattice derivation of the bipartite fidelity for a critical model at finite-size and an investigation of its asymptotics valid for all $x$. We shall achieve this for a family of boundary conditions labeled by a positive integer $d$, with the corresponding observables denoted by~$\mathcal F_d$.

All of the current knowledge regarding the universal behaviour of the bipartite fidelity applies to models for which the underlying CFT is rational. In particular, the conformal prediction \eqref{eq:f(x)} for $f(x)$ was obtained in \cite{SD13} by assuming that the fields $\varphi_1, \dots, \varphi_4$ are primary. It is by now well known that many lattice models in statistical mechanics are described by logarithmic CFTs \cite{GRR13}. Typically, this occurs if the model is defined in terms of non-local observables, or if some of the local Boltzmann weights are not positive real numbers. In these cases, in addition to the primary fields, the CFT can include logarithmic partners for some of the primaries whose correlation functions have logarithmic corrections to the usual power law behaviour \cite{G93}. A second objective of the current paper is thus to provide a first calculation of the bipartite fidelity in the case where the fields inserted on the boundary are logarithmic. The corresponding observable is denoted $\mathcal {\tilde F}_2$.

We pursue these two objectives by defining the bipartite fidelity for the model of critical dense polymers \cite{PR07}. This is a dense loop model in which the loops have a vanishing fugacity: $\beta = 0$. It is the simplest model in the family of logarithmic minimal models \cite{PRZ06}, with the central charge and the conformal weights of the Kac table given by
\be
c = -2, \qquad \Delta_{r,s} = \frac{(2r-s)^2-1}8.
\ee
This model can be mapped to a free fermionic quantum chain: the XX spin chain, with the $U_q(s\ell_2)$-invariant boundary fields of Pasquier and Saleur \cite{PS90} applied to the two endpoints. Our calculation will exploit the integrability of the model and the known diagonalisation of its Hamiltonian and transfer matrix to compute $\mathcal F_d$ and $\mathcal {\tilde F}_2$.

The outline of this paper is as follows. In \cref{sec:TL}, we define the bipartite fidelities $\mathcal F_d$ and $\mathcal {\tilde F}_2$ for the model of critical dense polymers and express these in the framework of the Temperley-Lieb algebra $\tl_N(\beta)$ at $\beta = 0$. We first write these using bilinear forms defined on the standard and projective modules of this algebra, and second as matrix elements in the XX chain. In \cref{sec:ExactResults}, we use the known diagonalisation of the XX spin chain to express $\mathcal F_d$ in terms of overlaps between groundstates of the spin chain. These are expressed as determinants using Wick's theorem and subsequently in closed form using an identity of Cauchy. Our final result for the asymptotic expansion is \cref{thm:Fd}, which is also given in this section. Its technical proof is relegated to \cref{sec:Asymptotics}. \cref{sec:logF} then presents the similar results for $\mathcal {\tilde F}_2$:  it is written in terms of overlaps in the spin chain which are evaluated in closed form, with the final result given in \cref{thm:F2.tilde}. In this section, we also give a derivation of the universal behaviour of $\mathcal {\tilde F}_2$ using CFT arguments. Final comments are given in \cref{sec:Conclusion}.

\section{Bipartite fidelity, the Temperley-Lieb algebra and critical dense polymers}\label{sec:TL}

\subsection{Critical dense polymers on the pants geometry}\label{sec:denseloopmodels}

We study the loop model of critical dense polymers on the pants lattice. In \cref{fig:loop.config}, we depict this lattice as a $4M\times N$ rectangle with a vertical slit. The slit divides the lower segment in two subsegments $A$ and $B$, of respective lengths $N_A$ and $N_B$, and extends halfway across the rectangle. A given configuration of the loop model is a selection, for each face of the lattice, of one of these two tiles: 
$\psset{unit=0.3cm}
\begin{pspicture}[shift=-0.16](0,0)(1,1)
\pspolygon[fillstyle=solid,fillcolor=lightlightblue,linewidth=\mince](0,0)(0,1)(1,1)(1,0)
\psarc[linewidth=\moyen,linecolor=blue](1,0){.5}{90}{180}
\psarc[linewidth=\moyen,linecolor=blue](0,1){.5}{-90}{0}
\end{pspicture}$ \hspace{0.01cm} or \hspace{0.01cm} $\psset{unit=0.3cm}
\begin{pspicture}[shift=-0.16](0,0)(1,1)
\pspolygon[fillstyle=solid,fillcolor=lightlightblue,linewidth=\mince](0,0)(0,1)(1,1)(1,0)
\psarc[linewidth=\moyen,linecolor=blue](0,0){.5}{0}{90}
\psarc[linewidth=\moyen,linecolor=blue](1,1){.5}{180}{-90}
\end{pspicture}$\  .

The boundary condition consists of a collection of arcs connecting neighboring nodes called {\it simple arcs} and vertical loop segments called {\it defects}. The left and right edges of the rectangle are decorated with simple arcs. The same applies to the interior of the slit. The top segment is made of a collection of $d$ defects, which we choose to be adjacent and attached to the leftmost nodes, and $\frac{N-d}2$ arcs. On the bottom segment, the subsegment $A$ is decorated with $d$ defects, also attached to the leftmost nodes, and $\frac{N_A-d}2$ arcs. The subsegment $B$ has only simple arcs.

\begin{figure}
\begin{center}
\psset{unit=0.4cm}
\begin{pspicture}[shift=-9.4](-0.5,-3)(14.5,15.6)
\facegrid{(0,0)}{(14,12)}
\rput(0,11){\loopa}\rput(1,11){\loopb}\rput(2,11){\loopb}\rput(3,11){\loopb}\rput(4,11){\loopb}\rput(5,11){\loopb}\rput(6,11){\loopb}\rput(7,11){\loopb}\rput(8,11){\loopa}\rput(9,11){\loopa}\rput(10,11){\loopa}\rput(11,11){\loopb}\rput(12,11){\loopa}\rput(13,11){\loopa}
\rput(0,10){\loopb}\rput(1,10){\loopb}\rput(2,10){\loopa}\rput(3,10){\loopa}\rput(4,10){\loopa}\rput(5,10){\loopa}\rput(6,10){\loopb}\rput(7,10){\loopa}\rput(8,10){\loopa}\rput(9,10){\loopb}\rput(10,10){\loopa}\rput(11,10){\loopa}\rput(12,10){\loopb}\rput(13,10){\loopa}
\rput(0,9){\loopa}\rput(1,9){\loopa}\rput(2,9){\loopb}\rput(3,9){\loopb}\rput(4,9){\loopa}\rput(5,9){\loopa}\rput(6,9){\loopa}\rput(7,9){\loopb}\rput(8,9){\loopa}\rput(9,9){\loopb}\rput(10,9){\loopa}\rput(11,9){\loopa}\rput(12,9){\loopa}\rput(13,9){\loopb}
\rput(0,8){\loopb}\rput(1,8){\loopa}\rput(2,8){\loopb}\rput(3,8){\loopb}\rput(4,8){\loopa}\rput(5,8){\loopa}\rput(6,8){\loopa}\rput(7,8){\loopb}\rput(8,8){\loopa}\rput(9,8){\loopa}\rput(10,8){\loopa}\rput(11,8){\loopb}\rput(12,8){\loopb}\rput(13,8){\loopb}
\rput(0,7){\loopa}\rput(1,7){\loopa}\rput(2,7){\loopa}\rput(3,7){\loopb}\rput(4,7){\loopb}\rput(5,7){\loopb}\rput(6,7){\loopb}\rput(7,7){\loopb}\rput(8,7){\loopa}\rput(9,7){\loopa}\rput(10,7){\loopb}\rput(11,7){\loopb}\rput(12,7){\loopb}\rput(13,7){\loopb}
\rput(0,6){\loopa}\rput(1,6){\loopb}\rput(2,6){\loopa}\rput(3,6){\loopa}\rput(4,6){\loopa}\rput(5,6){\loopa}\rput(6,6){\loopb}\rput(7,6){\loopb}\rput(8,6){\loopb}\rput(9,6){\loopb}\rput(10,6){\loopb}\rput(11,6){\loopa}\rput(12,6){\loopb}\rput(13,6){\loopb}
\rput(0,5){\loopa}\rput(1,5){\loopb}\rput(2,5){\loopa}\rput(3,5){\loopa}\rput(4,5){\loopa}\rput(5,5){\loopa}\rput(6,5){\loopb}\rput(7,5){\loopb}\rput(8,5){\loopa}\rput(9,5){\loopb}\rput(10,5){\loopa}\rput(11,5){\loopa}\rput(12,5){\loopa}\rput(13,5){\loopa}
\rput(0,4){\loopb}\rput(1,4){\loopb}\rput(2,4){\loopa}\rput(3,4){\loopb}\rput(4,4){\loopb}\rput(5,4){\loopa}\rput(6,4){\loopa}\rput(7,4){\loopb}\rput(8,4){\loopb}\rput(9,4){\loopb}\rput(10,4){\loopa}\rput(11,4){\loopb}\rput(12,4){\loopb}\rput(13,4){\loopa}
\rput(0,3){\loopb}\rput(1,3){\loopb}\rput(2,3){\loopb}\rput(3,3){\loopb}\rput(4,3){\loopa}\rput(5,3){\loopb}\rput(6,3){\loopa}\rput(7,3){\loopa}\rput(8,3){\loopa}\rput(9,3){\loopb}\rput(10,3){\loopa}\rput(11,3){\loopb}\rput(12,3){\loopb}\rput(13,3){\loopa}
\rput(0,2){\loopb}\rput(1,2){\loopa}\rput(2,2){\loopa}\rput(3,2){\loopa}\rput(4,2){\loopa}\rput(5,2){\loopb}\rput(6,2){\loopa}\rput(7,2){\loopa}\rput(8,2){\loopa}\rput(9,2){\loopa}\rput(10,2){\loopb}\rput(11,2){\loopb}\rput(12,2){\loopa}\rput(13,2){\loopa}
\rput(0,1){\loopa}\rput(1,1){\loopb}\rput(2,1){\loopb}\rput(3,1){\loopb}\rput(4,1){\loopa}\rput(5,1){\loopb}\rput(6,1){\loopa}\rput(7,1){\loopb}\rput(8,1){\loopb}\rput(9,1){\loopa}\rput(10,1){\loopa}\rput(11,1){\loopb}\rput(12,1){\loopb}\rput(13,1){\loopb}
\rput(0,0){\loopb}\rput(1,0){\loopb}\rput(2,0){\loopa}\rput(3,0){\loopa}\rput(4,0){\loopb}\rput(5,0){\loopb}\rput(6,0){\loopa}\rput(7,0){\loopa}\rput(8,0){\loopb}\rput(9,0){\loopa}\rput(10,0){\loopa}\rput(11,0){\loopa}\rput(12,0){\loopb}\rput(13,0){\loopa}
\psline[linewidth=3pt]{-}(8,0)(8,6)
\multiput(0,0)(0,2){6}{\psarc[linewidth=\elegant,linecolor=blue](0,1){0.5}{90}{270}\psarc[linewidth=\elegant,linecolor=blue](14,1){0.5}{270}{90}}
\multiput(0,0)(2,0){5}{\psarc[linewidth=\elegant,linecolor=blue](5,0){0.5}{180}{360}\psarc[linewidth=\elegant,linecolor=blue](5,12){0.5}{0}{180}}
\multiput(0,0)(1,0){4}{\psline[linewidth=\elegant,linecolor=blue](0.5,0)(0.5,-1)\psline[linewidth=\elegant,linecolor=blue](0.5,12)(0.5,13)}
\psline{<->}(-1.4,0)(-1.4,12)\rput(-2.3,6){$4M$}
\psline{<->}(0,-2)(8,-2)\rput(4,-2.7){$N_A$}
\psline{<->}(8,-2)(14,-2)\rput(11,-2.7){$N_B$}
\psline{<->}(0,14.8)(14,14.8)\rput(7,15.5){$N$}
\psline{<->}(0,13.5)(4,13.5)\rput(2,14.2){$d$}
\end{pspicture}\qquad\qquad\qquad
\psset{unit=0.5cm}
\begin{pspicture}[shift=-4.4](7,0)(9,8)
\facegrid{(7,0)}{(9,8)}
\psline[linewidth=3pt]{-}(8,0)(8,6)
\rput(7,7){\loopb}\rput(8,7){\loopa}
\rput(7,6){\loopb}\rput(8,6){\loopb}
\rput(7,5){\loopb}\rput(8,5){\loopa}
\rput(7,4){\loopb}\rput(8,4){\loopb}
\rput(7,3){\loopa}\rput(8,3){\loopa}
\rput(7,2){\loopa}\rput(8,2){\loopa}
\rput(7,1){\loopb}\rput(8,1){\loopb}
\rput(7,0){\loopa}\rput(8,0){\loopb}
\psarc[linewidth=\elegant,linecolor=blue](7,0){0.5}{-90}{0}
\psarc[linewidth=\elegant,linecolor=blue](9,0){0.5}{180}{-90}
\rput(6.5,1){...}\rput(6.5,4){...}\rput(6.5,7){...}
\rput(9.5,1){...}\rput(9.5,4){...}\rput(9.5,7){...}
\rput(7.5,8.3){.}\rput(7.5,8.5){.}\rput(7.5,8.7){.}
\rput(8.5,8.3){.}\rput(8.5,8.5){.}\rput(8.5,8.7){.}
\end{pspicture} \qquad$\longleftrightarrow$\qquad
\begin{pspicture}[shift=-4.4](7,0)(11,8)
\facegrid{(7,0)}{(8,8)}\facegrid{(10,0)}{(11,8)}
\rput(7,7){\loopb}\rput(10,7){\loopa}
\rput(7,6){\loopb}\rput(10,6){\loopb}
\rput(7,5){\loopb}\rput(10,5){\loopa}
\rput(7,4){\loopb}\rput(10,4){\loopb}
\rput(7,3){\loopa}\rput(10,3){\loopa}
\rput(7,2){\loopa}\rput(10,2){\loopa}
\rput(7,1){\loopb}\rput(10,1){\loopb}
\rput(7,0){\loopa}\rput(10,0){\loopb}
\psarc[linewidth=\elegant,linecolor=blue](8,1){0.5}{-90}{90}
\psarc[linewidth=\elegant,linecolor=blue](8,3){0.5}{-90}{90}
\psarc[linewidth=\elegant,linecolor=blue](8,5){0.5}{-90}{90}
\psarc[linewidth=\elegant,linecolor=blue](10,1){0.5}{90}{-90}
\psarc[linewidth=\elegant,linecolor=blue](10,3){0.5}{90}{-90}
\psarc[linewidth=\elegant,linecolor=blue](10,5){0.5}{90}{-90}
\psarc[linewidth=\elegant,linecolor=blue](7,0){0.5}{-90}{0}
\psarc[linewidth=\elegant,linecolor=blue](11,0){0.5}{180}{-90}
\psline[linewidth=\elegant,linecolor=blue](8,6.5)(10,6.5)
\psline[linewidth=\elegant,linecolor=blue](8,7.5)(10,7.5)
\rput(6.5,1){...}\rput(6.5,4){...}\rput(6.5,7){...}
\rput(11.5,1){...}\rput(11.5,4){...}\rput(11.5,7){...}
\rput(7.5,8.3){.}\rput(7.5,8.5){.}\rput(7.5,8.7){.}
\rput(10.5,8.3){.}\rput(10.5,8.5){.}\rput(10.5,8.7){.}
\end{pspicture}
\caption{A configuration of the model of critical dense polymers on the pants geometry, with $M=3$, $N=14$, $N_A = 8$ and $N_B=6$.}
\label{fig:loop.config}
\end{center}
\end{figure}
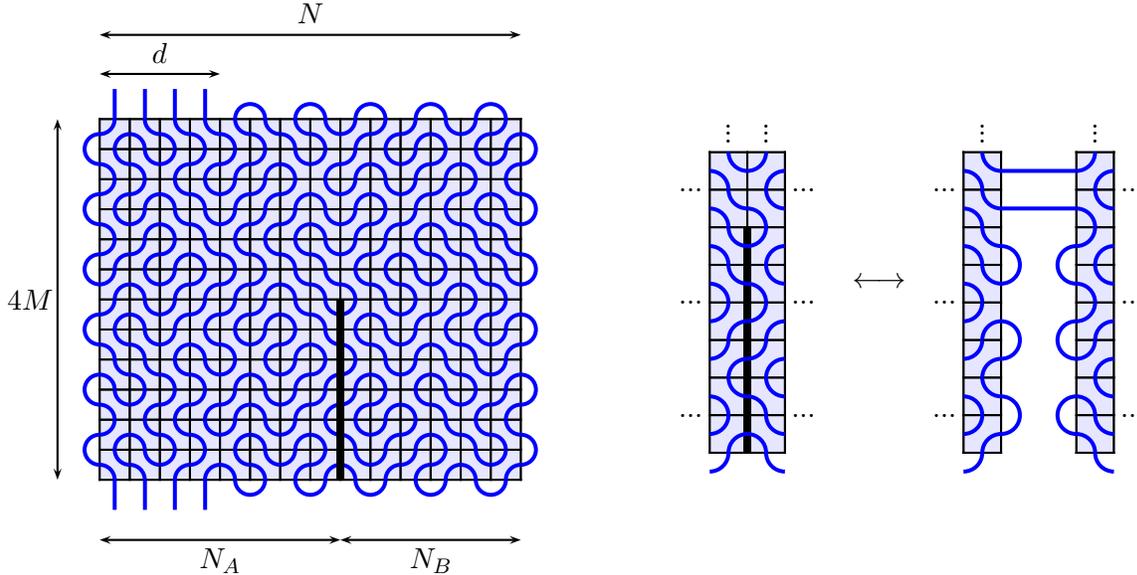

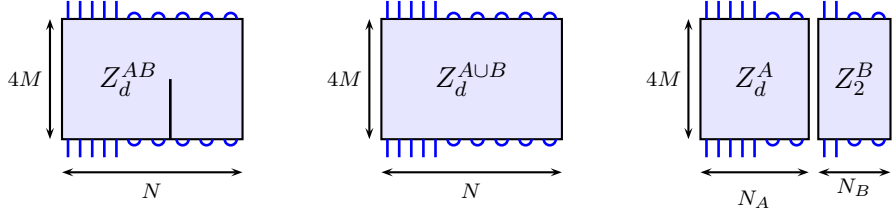
\begin{figure}
\begin{center}
\psset{unit=0.4cm}
\begin{pspicture}[shift=-1.9](-1.4,-2.0)(6,4.6)
\pspolygon[fillstyle=solid,fillcolor=lightlightblue](0,0)(6,0)(6,4)(0,4)
\psline[linewidth=1.0pt]{-}(3.6,0)(3.6,2)
\multiput(0,0)(0.4,0){5}{\psline[linewidth=\moyen,linecolor=blue](0.2,4)(0.2,4.6)}
\multiput(0,0)(0.4,0){5}{\psline[linewidth=\moyen,linecolor=blue](0.2,0)(0.2,-0.6)}
\multiput(0,0)(0.8,0){5}{\psarc[linewidth=\moyen,linecolor=blue](2.4,4){0.2}{0}{180}}
\multiput(0,0)(0.8,0){5}{\psarc[linewidth=\moyen,linecolor=blue](2.4,0){0.2}{180}{0}}
\rput(2.2,2){$Z^{AB}_d$}
\psline{<->}(0,-1.1)(6,-1.1)\rput(3,-1.7){$_N$}
\psline{<->}(-0.4,0)(-0.4,4)\rput(-1.2,2){$_{4M}$}
\end{pspicture}
\qquad\quad
\begin{pspicture}[shift=-1.9](-1.4,-2.0)(6,4.6)
\pspolygon[fillstyle=solid,fillcolor=lightlightblue](0,0)(6,0)(6,4)(0,4)
\multiput(0,0)(0.4,0){5}{\psline[linewidth=\moyen,linecolor=blue](0.2,4)(0.2,4.6)}
\multiput(0,0)(0.4,0){5}{\psline[linewidth=\moyen,linecolor=blue](0.2,0)(0.2,-0.6)}
\multiput(0,0)(0.8,0){5}{\psarc[linewidth=\moyen,linecolor=blue](2.4,4){0.2}{0}{180}}
\multiput(0,0)(0.8,0){5}{\psarc[linewidth=\moyen,linecolor=blue](2.4,0){0.2}{180}{0}}
\rput(3,2){$Z^{A\cup B}_d$}
\psline{<->}(0,-1.1)(6,-1.1)\rput(3,-1.7){$_N$}
\psline{<->}(-0.4,0)(-0.4,4)\rput(-1.2,2){$_{4M}$}
\end{pspicture}
\qquad\quad
\begin{pspicture}[shift=-1.9](-1.4,-2.0)(3.6,4.6)
\pspolygon[fillstyle=solid,fillcolor=lightlightblue](0,0)(3.6,0)(3.6,4)(0,4)
\multiput(0,0)(0.4,0){5}{\psline[linewidth=\moyen,linecolor=blue](0.2,4)(0.2,4.6)}
\multiput(0,0)(0.4,0){5}{\psline[linewidth=\moyen,linecolor=blue](0.2,0)(0.2,-0.6)}
\multiput(0,0)(0.8,0){2}{\psarc[linewidth=\moyen,linecolor=blue](2.4,4){0.2}{0}{180}}
\multiput(0,0)(0.8,0){2}{\psarc[linewidth=\moyen,linecolor=blue](2.4,0){0.2}{180}{0}}
\rput(1.8,2){$Z^{A}_d$}
\psline{<->}(0,-1.1)(3.6,-1.1)\rput(1.8,-2.0){$_{N_A}$}
\psline{<->}(-0.4,0)(-0.4,4)\rput(-1.2,2){$_{4M}$}
\end{pspicture}
\begin{pspicture}[shift=-1.9](0,-2.0)(2.4,4.6)
\pspolygon[fillstyle=solid,fillcolor=lightlightblue](0,0)(2.4,0)(2.4,4)(0,4)
\multiput(0,0)(0.4,0){2}{\psline[linewidth=\moyen,linecolor=blue](0.2,4)(0.2,4.6)}
\multiput(0,0)(0.4,0){2}{\psline[linewidth=\moyen,linecolor=blue](0.2,0)(0.2,-0.6)}
\multiput(0,0)(0.8,0){2}{\psarc[linewidth=\moyen,linecolor=blue](1.2,4){0.2}{0}{180}}
\multiput(0,0)(0.8,0){2}{\psarc[linewidth=\moyen,linecolor=blue](1.2,0){0.2}{180}{0}}
\rput(1.2,2){$Z^{B}_2$}
\psline{<->}(0,-1.1)(2.4,-1.1)\rput(1.2,-1.6){$_{N_B}$}
\end{pspicture}
\caption{The lattices corresponding to the partition functions $Z^{AB}_d$, $Z^{A\cup B}_d$, $Z^{A}_d$ and $Z^{B}_2$}
\label{fig:Z.lattices}
\end{center}
\end{figure}

The loop segments drawn on the tiles and on the boundary form loops. Some of those are closed and are referred to as {\it bulk loops}. In the model of critical dense polymers, bulk loops have the fugacity zero. Loop segments can also connect two defects from the boundary. We refer to this as a {\it boundary loop}. In defining the partition function $Z_d^{AB}$, we assign a fugacity one to a boundary loop if it connects the top and bottom segments, and a fugacity zero if it connects two defects of the same segment. The weight $W_\sigma$ of a configuration $\sigma$ is then the product of the fugacities $w_\ell$ of its loops $\ell$: $W_\sigma = \prod_\ell w_\ell$. In other words, if a configuration $\sigma$ has one or more loops with fugacity zero, its global Boltzmann weight is also zero. It has weight one otherwise. The partition function on the pants geometry, denoted $Z_d^{AB}$, is defined as
\be
Z_d^{AB} = \sum_{\sigma} W_\sigma.
\ee
We define three more partition functions: $Z^{A\cup B}_d$, $Z^{A}_d$ and $Z^{B}_2$ . These are defined with the same choices of fugacities for the bulk and boundary loops, but on lattices without slits, namely on the rectangular lattices given in \cref{fig:Z.lattices}. The logarithmic bipartite fidelity is then defined as
\be
\label{eq:Fd}
\mathcal F_d = -\lim_{M \to \infty} \log \Bigg(\frac{Z_d^{AB}}{\big(Z_d^{A\cup B}Z_d^{A}Z_2^{B}\big)^{1/2}}\Bigg)^2,
\ee
where the defect number $d$ is taken to be greater or equal to one.

The choice to include $Z_2^{B}$ in the denominator is justified as follows. We note that for the other dense loop models for which the fugacity of bulk loops is non-zero, the natural choice is to write $Z_0^{B}$ in the denominator instead of $Z_2^{B}$. For the model of critical dense polymers, this partition function is $Z_0^{B} = 0$. As argued in \cite{MDJ18}, the reference partition function in this case is instead $Z_2^{B}$.

In this same paper, it was also underlined that the boundary condition consisting of two neighboring defects forced to connect together corresponds, from a CFT perspective, to a logarithmic field in a rank-two Jordan cell. Correlation functions involving this field are expected to have logarithmic behaviours. To investigate this in the current context, we study a second instance of the bipartite fidelity for $d=2$, denoted $\tilde {\mathcal F}_2$. We define $\tilde Z^{AB}_2$ to be the partition function on the pants geometry of \cref{fig:loop.config}, but with different fugacities for the boundary loops. In computing the weights of configurations for $\tilde Z^{AB}_2$, a boundary loop is given fugacity zero if it connects a defect from the top segment to one of the bottom segment, and a fugacity one if it connects two defects of the same segment. Bulk loops still have fugacity zero. We similarly define 
$\tilde Z_{2}^{A\cup B}$ and $\tilde Z_{2}^{A}$, with the same fugacities as for $\tilde Z^{AB}_2$, but on the lattices of \cref{fig:Z.lattices}. The second logarithmic bipartite fidelity is then defined as
\be
\label{eq:F2v2}
\tilde{\mathcal F}_2 = -\lim_{M \to \infty} \log \Bigg(\frac{\tilde Z_2^{AB}}{\big(\tilde Z_2^{A\cup B}\tilde Z_2^{A}Z_2^{B}\big)^{1/2}}\Bigg)^2,
\ee 
where we note that the last factor in the denominator is $Z_2^{B}$ and not $\tilde Z_2^{B}$.

\subsection[The Temperley-Lieb algebra at $\beta = 0$]{The Temperley-Lieb algebra at $\boldsymbol{\beta = 0}$}

In this section, we express the various partition functions defined in \cref{sec:denseloopmodels} in the language of the Temperley-Lieb algebra $\tl_N(\beta=0)$. The corresponding expressions involve the standard and projective modules over this algebra, the bilinear forms over these modules and the transfer tangle for the model of critical dense polymers.

\paragraph{The algebra $\boldsymbol{\tl_N(\beta)}$.} The Temperley-Lieb algebra \cite{TL71,J83,M91,GW93,W95,RSA14} is a unital, associative algebra generated by the linear span of connectivities. A connectivity is a diagram drawn inside a rectangular box with $N$ marked nodes on its top segment and $N$ more on its bottom segment. Inside the box, the nodes are connected pairwise by non-intersecting loop segments. A subset of the connectivities, namely the identity connectivity $\boldsymbol I$ and $N-1$ connectivities denoted by $e_j$, with $j = 1, \dots, N-1$, generate this algebra. These are depicted as:
\be
\psset{unit=0.9}
\boldsymbol I =
\ 
\begin{pspicture}[shift=-0.525](0,-0.25)(2.4,0.8)
\pspolygon[fillstyle=solid,fillcolor=lightlightblue,linecolor=black,linewidth=0pt](0,0)(0,0.8)(2.4,0.8)(2.4,0)(0,0)
\psline[linecolor=blue,linewidth=\elegant]{-}(0.2,0)(0.2,0.8)
\psline[linecolor=blue,linewidth=\elegant]{-}(0.6,0)(0.6,0.8)
\psline[linecolor=blue,linewidth=\elegant]{-}(1.0,0)(1.0,0.8)
\rput(1.4,0.4){$...$}
\psline[linecolor=blue,linewidth=\elegant]{-}(1.8,0)(1.8,0.8)
\psline[linecolor=blue,linewidth=\elegant]{-}(2.2,0)(2.2,0.8)
\rput(0.2,-0.25){$_1$}
\rput(0.6,-0.25){$_2$}
\rput(1.0,-0.25){$_3$}
\rput(2.2,-0.25){$_N$}
\end{pspicture}\ \ ,
\qquad
e_j =  \
\begin{pspicture}[shift=-0.525](-0.0,-0.25)(3.2,0.8)
\pspolygon[fillstyle=solid,fillcolor=lightlightblue,linecolor=black,linewidth=0pt](0,0)(0,0.8)(3.2,0.8)(3.2,0)(0,0)
\psline[linecolor=blue,linewidth=\elegant]{-}(0.2,0)(0.2,0.8)
\rput(0.6,0.4){$...$}
\psline[linecolor=blue,linewidth=\elegant]{-}(1.0,0)(1.0,0.8)
\psarc[linecolor=blue,linewidth=\elegant]{-}(1.6,0.8){0.2}{180}{360}
\psarc[linecolor=blue,linewidth=\elegant]{-}(1.6,0){0.2}{0}{180}
\psline[linecolor=blue,linewidth=\elegant]{-}(2.2,0)(2.2,0.8)
\rput(2.6,0.4){$...$}
\psline[linecolor=blue,linewidth=\elegant]{-}(3.0,0)(3.0,0.8)
\rput(0.2,-0.25){$_1$}
\rput(1.4,-0.25){$_{\phantom{+}j\phantom{+}}$}
\rput(3.0,-0.25){$_N$}
\end{pspicture} \ \ .
\ee

The product $a_1 a_2$ of connectivities in $\tl_N(\beta)$ is obtained as follows: one stacks $a_2$ above $a_1$ and straightens the loop segments to obtain a new connectivity. If bulk loops are created, they are erased at the cost of multiplicative factors of $\beta$. Here are some examples for $N=4$:
\begin{subequations}
\be
e_1 e_2 = \ 
\psset{unit=0.9}
\begin{pspicture}[shift=-0.7](-0.0,0)(1.6,1.6)
\multiput(0,0)(0,0.8){2}{\pspolygon[fillstyle=solid,fillcolor=lightlightblue,linecolor=black,linewidth=0pt](0,0)(0,0.8)(1.6,0.8)(1.6,0)(0,0)}
\psarc[linecolor=blue,linewidth=\elegant]{-}(0.4,0.8){0.2}{180}{360}
\psarc[linecolor=blue,linewidth=\elegant]{-}(0.4,0){0.2}{0}{180}
\psline[linecolor=blue,linewidth=\elegant]{-}(1.0,0)(1.0,0.8)
\psline[linecolor=blue,linewidth=\elegant]{-}(1.4,0)(1.4,0.8)
\rput(0,0.8){
\psline[linecolor=blue,linewidth=\elegant]{-}(0.2,0)(0.2,0.8)
\psline[linecolor=blue,linewidth=\elegant]{-}(1.4,0)(1.4,0.8)
\psarc[linecolor=blue,linewidth=\elegant]{-}(0.8,0.8){0.2}{180}{360}
\psarc[linecolor=blue,linewidth=\elegant]{-}(0.8,0){0.2}{0}{180}
}
\end{pspicture} \ = \ 
\begin{pspicture}[shift=-0.525](-0.0,-0.25)(1.6,0.8)
\pspolygon[fillstyle=solid,fillcolor=lightlightblue,linecolor=black,linewidth=0pt](0,0)(0,0.8)(1.6,0.8)(1.6,0)(0,0)
\psarc[linecolor=blue,linewidth=\elegant]{-}(0.8,0.8){0.2}{180}{0}
\psarc[linecolor=blue,linewidth=\elegant]{-}(0.4,0){0.2}{0}{180}
\psline[linecolor=blue,linewidth=\elegant]{-}(1.4,0)(1.4,0.8)
\psbezier[linecolor=blue,linewidth=\elegant]{-}(1.0,0)(1.0,0.4)(0.2,0.4)(0.2,0.8)
\end{pspicture}\ \ ,
\hspace{1.0cm}
e_2 e_1 e_3 = \
\psset{unit=0.9}
\begin{pspicture}[shift=-1.1](-0.0,0)(1.6,2.4)
\multiput(0,0)(0,0.8){3}{\pspolygon[fillstyle=solid,fillcolor=lightlightblue,linecolor=black,linewidth=0pt](0,0)(0,0.8)(1.6,0.8)(1.6,0)(0,0)}
\psline[linecolor=blue,linewidth=\elegant]{-}(0.2,0)(0.2,0.8)
\psline[linecolor=blue,linewidth=\elegant]{-}(1.4,0)(1.4,0.8)
\psarc[linecolor=blue,linewidth=\elegant]{-}(0.8,0.8){0.2}{180}{360}
\psarc[linecolor=blue,linewidth=\elegant]{-}(0.8,0){0.2}{0}{180}
\rput(0,0.8){
\psarc[linecolor=blue,linewidth=\elegant]{-}(0.4,0.8){0.2}{180}{360}
\psarc[linecolor=blue,linewidth=\elegant]{-}(0.4,0){0.2}{0}{180}
\psline[linecolor=blue,linewidth=\elegant]{-}(1.0,0)(1.0,0.8)
\psline[linecolor=blue,linewidth=\elegant]{-}(1.4,0)(1.4,0.8)
}
\rput(0,1.6){
\psarc[linecolor=blue,linewidth=\elegant]{-}(1.2,0.8){0.2}{180}{360}
\psarc[linecolor=blue,linewidth=\elegant]{-}(1.2,0){0.2}{0}{180}
\psline[linecolor=blue,linewidth=\elegant]{-}(0.2,0)(0.2,0.8)
\psline[linecolor=blue,linewidth=\elegant]{-}(0.6,0)(0.6,0.8)
}
\end{pspicture} \ = \ 
\begin{pspicture}[shift=-0.525](-0.0,-0.25)(1.6,0.8)
\pspolygon[fillstyle=solid,fillcolor=lightlightblue,linecolor=black,linewidth=0pt](0,0)(0,0.8)(1.6,0.8)(1.6,0)(0,0)
\psarc[linecolor=blue,linewidth=\elegant]{-}(0.8,0){0.2}{0}{180}
\psarc[linecolor=blue,linewidth=\elegant]{-}(0.4,0.8){0.2}{180}{360}
\psarc[linecolor=blue,linewidth=\elegant]{-}(1.2,0.8){0.2}{180}{360}
\psbezier[linecolor=blue,linewidth=\elegant]{-}(0.2,0)(0.2,0.5)(1.4,0.5)(1.4,0)
\end{pspicture}\ \ ,
\ee
\be
(e_1e_2)(e_2e_1e_3)(e_1e_2) = \ 
\psset{unit=0.9}
\begin{pspicture}[shift=-1.1](-0.0,0)(1.6,2.4)
\multiput(0,0)(0,0.8){3}{\pspolygon[fillstyle=solid,fillcolor=lightlightblue,linecolor=black,linewidth=0pt](0,0)(0,0.8)(1.6,0.8)(1.6,0)(0,0)}
\psarc[linecolor=blue,linewidth=\elegant]{-}(0.8,0.8){0.2}{180}{0}
\psarc[linecolor=blue,linewidth=\elegant]{-}(0.4,0){0.2}{0}{180}
\psline[linecolor=blue,linewidth=\elegant]{-}(1.4,0)(1.4,0.8)
\psbezier[linecolor=blue,linewidth=\elegant]{-}(1.0,0)(1.0,0.4)(0.2,0.4)(0.2,0.8)
\rput(0,0.8){
\psarc[linecolor=blue,linewidth=\elegant]{-}(0.8,0){0.2}{0}{180}
\psarc[linecolor=blue,linewidth=\elegant]{-}(0.4,0.8){0.2}{180}{360}
\psarc[linecolor=blue,linewidth=\elegant]{-}(1.2,0.8){0.2}{180}{360}
\psbezier[linecolor=blue,linewidth=\elegant]{-}(0.2,0)(0.2,0.5)(1.4,0.5)(1.4,0)
}
\rput(0,1.6){
\psarc[linecolor=blue,linewidth=\elegant]{-}(0.8,0.8){0.2}{180}{0}
\psarc[linecolor=blue,linewidth=\elegant]{-}(0.4,0){0.2}{0}{180}
\psline[linecolor=blue,linewidth=\elegant]{-}(1.4,0)(1.4,0.8)
\psbezier[linecolor=blue,linewidth=\elegant]{-}(1.0,0)(1.0,0.4)(0.2,0.4)(0.2,0.8)
}
\end{pspicture}
 \ = \beta^2\
\begin{pspicture}[shift=-0.525](-0.0,-0.25)(1.6,0.8)
\pspolygon[fillstyle=solid,fillcolor=lightlightblue,linecolor=black,linewidth=0pt](0,0)(0,0.8)(1.6,0.8)(1.6,0)(0,0)
\psarc[linecolor=blue,linewidth=\elegant]{-}(0.4,0){0.2}{0}{180}
\psarc[linecolor=blue,linewidth=\elegant]{-}(1.2,0){0.2}{0}{180}
\psarc[linecolor=blue,linewidth=\elegant]{-}(0.8,0.8){0.2}{180}{360}
\psbezier[linecolor=blue,linewidth=\elegant]{-}(0.2,0.8)(0.2,0.3)(1.4,0.3)(1.4,0.8)
\end{pspicture} \ = \beta^2 e_1 e_3 e_2.
\ee
\end{subequations}
With products of the $e_j$ generators, one can produce any connectivity in $\tl_N(\beta)$. The diagrammatic definition for the product of connectivities is then equivalent to a set of relations satisfied by the generators:
\be
\label{eq:TL.def}
(e_j)^2= \beta\, e_j, \qquad e_j e_{j\pm 1} e_j = e_j, \qquad e_j e_k = e_k e_j \quad (|j-k|>1).
\ee
These are the defining relations of $\tl_N(\beta)$. The value of $\beta$ pertaining to the model of critical dense polymers is $\beta = 0$. The algebra $\tl_N(0)$ is non-semisimple. Below, we describe the standard and projective modules over the Temperley-Lieb algebra, which will allow us to compute $\mathcal F_d$ and $\tilde {\mathcal F}_2$ respectively.

\paragraph{The standard modules $\boldsymbol{\mathsf V_{N,d}}$.} The standard modules $\mathsf V_{N,d}$ are built on the vector space generated by link states on $N$ nodes with $d$ defects. These are diagrams drawn above a segment with $N$ marked nodes, wherein nodes are either connected pairwise or occupied by a defect that cannot be overarched. Here are the link states for the standard modules for $N=4,5$:
\begin{subequations}
\begin{alignat}{2}
&\label{eq:V4linkstates}
\psset{unit=0.9}
\mathsf V_{4,0}: \ \begin{pspicture}[shift=-0.08](0.0,0)(1.6,0.5)
\psline[linewidth=\mince](0,0)(1.6,0)
\psarc[linecolor=blue,linewidth=\elegant]{-}(0.4,0){0.2}{0}{180}
\psarc[linecolor=blue,linewidth=\elegant]{-}(1.2,0){0.2}{0}{180}
\end{pspicture} \ \ \
\begin{pspicture}[shift=-0.08](0.0,0)(1.6,0.5)
\psline[linewidth=\mince](0,0)(1.6,0)
\psarc[linecolor=blue,linewidth=\elegant]{-}(0.8,0){0.2}{0}{180}
\psbezier[linecolor=blue,linewidth=\elegant](0.2,0)(0.2,0.6)(1.4,0.6)(1.4,0)
\end{pspicture} \  , \qquad
\mathsf V_{4,2}: \ 
\begin{pspicture}[shift=-0.08](0.0,0)(1.6,0.5)
\psline[linewidth=\mince](0,0)(1.6,0)
\psarc[linecolor=blue,linewidth=\elegant]{-}(0.4,0){0.2}{0}{180}
\psline[linecolor=blue,linewidth=\elegant]{-}(1.0,0)(1.0,0.5)
\psline[linecolor=blue,linewidth=\elegant]{-}(1.4,0)(1.4,0.5)
\end{pspicture} \ \ \ 
\begin{pspicture}[shift=-0.08](0.0,0)(1.6,0.5)
\psline[linewidth=\mince](0,0)(1.6,0)
\psline[linecolor=blue,linewidth=\elegant]{-}(0.2,0)(0.2,0.5)
\psline[linecolor=blue,linewidth=\elegant]{-}(1.4,0)(1.4,0.5)
\psarc[linecolor=blue,linewidth=\elegant]{-}(0.8,0){0.2}{0}{180}
\end{pspicture}\ \ \ 
\begin{pspicture}[shift=-0.08](0.0,0)(1.6,0.5)
\psline[linewidth=\mince](0,0)(1.6,0)
\psline[linecolor=blue,linewidth=\elegant]{-}(0.2,0)(0.2,0.5)
\psline[linecolor=blue,linewidth=\elegant]{-}(0.6,0)(0.6,0.5)
\psarc[linecolor=blue,linewidth=\elegant]{-}(1.2,0){0.2}{0}{180}
\end{pspicture}  \ , \qquad
\mathsf V_{4,4}: \ 
\begin{pspicture}[shift=-0.08](0.0,0)(1.6,0.5)
\psline[linewidth=\mince](0,0)(1.6,0)
\psline[linecolor=blue,linewidth=\elegant]{-}(0.2,0)(0.2,0.5)
\psline[linecolor=blue,linewidth=\elegant]{-}(0.6,0)(0.6,0.5)
\psline[linecolor=blue,linewidth=\elegant]{-}(1.0,0)(1.0,0.5)
\psline[linecolor=blue,linewidth=\elegant]{-}(1.4,0)(1.4,0.5)
\end{pspicture}\  ,\ 
\\[0.35cm]
&
\psset{unit=0.9}
\mathsf V_{5,1}: \  
\begin{pspicture}[shift=-0.08](0.0,0)(2.0,0.5)
\psline[linewidth=\mince](0,0)(2.0,0)
\psarc[linecolor=blue,linewidth=\elegant]{-}(0.4,0){0.2}{0}{180}
\psarc[linecolor=blue,linewidth=\elegant]{-}(1.2,0){0.2}{0}{180}
\psline[linecolor=blue,linewidth=\elegant]{-}(1.8,0)(1.8,0.5)
\end{pspicture}\ \ \
\begin{pspicture}[shift=-0.08](0.0,0)(2.0,0.5)
\psline[linewidth=\mince](0,0)(2.0,0)
\psarc[linecolor=blue,linewidth=\elegant]{-}(0.4,0){0.2}{0}{180}
\psarc[linecolor=blue,linewidth=\elegant]{-}(1.6,0){0.2}{0}{180}
\psline[linecolor=blue,linewidth=\elegant]{-}(1.0,0)(1.0,0.5)
\end{pspicture}\ \ \
\begin{pspicture}[shift=-0.08](0.0,0)(2.0,0.5)
\psline[linewidth=\mince](0,0)(2.0,0)
\psarc[linecolor=blue,linewidth=\elegant]{-}(0.8,0){0.2}{0}{180}
\psarc[linecolor=blue,linewidth=\elegant]{-}(1.6,0){0.2}{0}{180}
\psline[linecolor=blue,linewidth=\elegant]{-}(0.2,0)(0.2,0.5)
\end{pspicture} \ \ \
\begin{pspicture}[shift=-0.08](0.0,0)(2.0,0.5)
\psline[linewidth=\mince](0,0)(2.0,0)
\psarc[linecolor=blue,linewidth=\elegant]{-}(0.8,0){0.2}{0}{180}
\psbezier[linecolor=blue,linewidth=\elegant](0.2,0)(0.2,0.6)(1.4,0.6)(1.4,0)
\psline[linecolor=blue,linewidth=\elegant]{-}(1.8,0)(1.8,0.5)
\end{pspicture} \ \ \
\begin{pspicture}[shift=-0.08](0.0,0)(2.0,0.5)
\psline[linewidth=\mince](0,0)(2.0,0)
\psarc[linecolor=blue,linewidth=\elegant]{-}(1.2,0){0.2}{0}{180}
\psbezier[linecolor=blue,linewidth=\elegant](0.6,0)(0.6,0.6)(1.8,0.6)(1.8,0)
\psline[linecolor=blue,linewidth=\elegant]{-}(0.2,0)(0.2,0.5)
\end{pspicture} \  , \ \ \
\\[0.35cm]
&
\psset{unit=0.9}
\mathsf V_{5,3}: \  
\begin{pspicture}[shift=-0.08](0.0,0)(2.0,0.5)
\psline[linewidth=\mince](0,0)(2.0,0)
\psarc[linecolor=blue,linewidth=\elegant]{-}(0.4,0){0.2}{0}{180}
\psline[linecolor=blue,linewidth=\elegant]{-}(1.0,0)(1.0,0.5)
\psline[linecolor=blue,linewidth=\elegant]{-}(1.4,0)(1.4,0.5)
\psline[linecolor=blue,linewidth=\elegant]{-}(1.8,0)(1.8,0.5)
\end{pspicture}\ \ \
\begin{pspicture}[shift=-0.08](0.0,0)(2.0,0.5)
\psline[linewidth=\mince](0,0)(2.0,0)
\psarc[linecolor=blue,linewidth=\elegant]{-}(0.8,0){0.2}{0}{180}
\psline[linecolor=blue,linewidth=\elegant]{-}(0.2,0)(0.2,0.5)
\psline[linecolor=blue,linewidth=\elegant]{-}(1.4,0)(1.4,0.5)
\psline[linecolor=blue,linewidth=\elegant]{-}(1.8,0)(1.8,0.5)
\end{pspicture} \ \ \
\begin{pspicture}[shift=-0.08](0.0,0)(2.0,0.5)
\psline[linewidth=\mince](0,0)(2.0,0)
\psarc[linecolor=blue,linewidth=\elegant]{-}(1.2,0){0.2}{0}{180}
\psline[linecolor=blue,linewidth=\elegant]{-}(0.2,0)(0.2,0.5)
\psline[linecolor=blue,linewidth=\elegant]{-}(0.6,0)(0.6,0.5)
\psline[linecolor=blue,linewidth=\elegant]{-}(1.8,0)(1.8,0.5)
\end{pspicture} \ \ \
\begin{pspicture}[shift=-0.08](0.0,0)(2.0,0.5)
\psline[linewidth=\mince](0,0)(2.0,0)
\psarc[linecolor=blue,linewidth=\elegant]{-}(1.6,0){0.2}{0}{180}
\psline[linecolor=blue,linewidth=\elegant]{-}(0.2,0)(0.2,0.5)
\psline[linecolor=blue,linewidth=\elegant]{-}(0.6,0)(0.6,0.5)
\psline[linecolor=blue,linewidth=\elegant]{-}(1.0,0)(1.0,0.5)
\end{pspicture} \  ,\qquad
\mathsf V_{5,5}: \ 
\begin{pspicture}[shift=-0.08](0.0,0)(2.0,0.5)
\psline[linewidth=\mince](0,0)(2.0,0)
\psline[linecolor=blue,linewidth=\elegant]{-}(0.2,0)(0.2,0.5)
\psline[linecolor=blue,linewidth=\elegant]{-}(0.6,0)(0.6,0.5)
\psline[linecolor=blue,linewidth=\elegant]{-}(1.0,0)(1.0,0.5)
\psline[linecolor=blue,linewidth=\elegant]{-}(1.4,0)(1.4,0.5)
\psline[linecolor=blue,linewidth=\elegant]{-}(1.8,0)(1.8,0.5)
\end{pspicture} \  .
\end{alignat}
\end{subequations}

The action of a connectivity $a$ in $\tl_N(\beta)$ on a link state $v$ in $\mathsf V_{N,d}$, denoted $a\,v$, uses diagrammatic rules similar to those defined for the product of connectivities. One draws $v$ above $a$, straightens the loop segments and reads the new link state produced at the bottom segment of $a$. If two defects connect, the result is set to zero. Otherwise the result of $a\,v$ is this new link state, with contractible loops replaced by multiplicative factors of $\beta$. For $\beta = 0$, this implies that if one or more contractible loops are formed, then $a\,v = 0$. Here are examples for $N=4$:
\be\label{eq:standard.examples}
\psset{unit=0.9}
\begin{pspicture}[shift=-0.525](-0.0,-0.25)(1.6,1.3)
\pspolygon[fillstyle=solid,fillcolor=lightlightblue,linecolor=black,linewidth=0pt](0,0)(0,0.8)(1.6,0.8)(1.6,0)(0,0)
\psarc[linecolor=blue,linewidth=\elegant]{-}(0.8,0.8){0.2}{180}{0}
\psarc[linecolor=blue,linewidth=\elegant]{-}(0.4,0){0.2}{0}{180}
\psbezier[linecolor=blue,linewidth=\elegant]{-}(0.2,0.8)(0.2,0.4)(1.0,0.4)(1.0,0)
\psline[linecolor=blue,linewidth=\elegant]{-}(1.4,0)(1.4,0.8)
\rput(0,0.8)
{\psline[linecolor=blue,linewidth=\elegant]{-}(0.2,0)(0.2,0.5)
\psarc[linecolor=blue,linewidth=\elegant]{-}(0.8,0){0.2}{0}{180}
\psline[linecolor=blue,linewidth=\elegant]{-}(1.4,0)(1.4,0.5)}
\end{pspicture}
\ = \beta \ 
\begin{pspicture}[shift=-0.08](0.0,0)(1.6,0.5)
\psline[linewidth=\mince](0,0)(1.6,0)
\psarc[linecolor=blue,linewidth=\elegant]{-}(0.4,0){0.2}{0}{180}
\psline[linecolor=blue,linewidth=\elegant]{-}(1.0,0)(1.0,0.5)
\psline[linecolor=blue,linewidth=\elegant]{-}(1.4,0)(1.4,0.5)
\end{pspicture} \ ,
\qquad
\begin{pspicture}[shift=-0.525](-0.0,-0.25)(1.6,1.3)
\pspolygon[fillstyle=solid,fillcolor=lightlightblue,linecolor=black,linewidth=0pt](0,0)(0,0.8)(1.6,0.8)(1.6,0)(0,0)
\psarc[linecolor=blue,linewidth=\elegant]{-}(0.8,0.8){0.2}{180}{0}
\psarc[linecolor=blue,linewidth=\elegant]{-}(0.4,0){0.2}{0}{180}
\psarc[linecolor=blue,linewidth=\elegant]{-}(1.2,0){0.2}{0}{180}
\psbezier[linecolor=blue,linewidth=\elegant]{-}(0.2,0.8)(0.2,0.3)(1.4,0.3)(1.4,0.8)
\rput(0,0.8)
{\psbezier[linecolor=blue,linewidth=\elegant]{-}(0.2,0)(0.2,0.5)(1.4,0.5)(1.4,0)
\psarc[linecolor=blue,linewidth=\elegant]{-}(0.8,0){0.2}{0}{180}
}
\end{pspicture}
\ = \beta^2 \ 
\begin{pspicture}[shift=-0.08](0.0,0)(1.6,0.5)
\psline[linewidth=\mince](0,0)(1.6,0)
\psarc[linecolor=blue,linewidth=\elegant]{-}(0.4,0){0.2}{0}{180}
\psarc[linecolor=blue,linewidth=\elegant]{-}(1.2,0){0.2}{0}{180}
\end{pspicture} \ ,
\qquad
\begin{pspicture}[shift=-0.525](-0.0,-0.25)(1.6,1.3)
\pspolygon[fillstyle=solid,fillcolor=lightlightblue,linecolor=black,linewidth=0pt](0,0)(0,0.8)(1.6,0.8)(1.6,0)(0,0)
\psarc[linecolor=blue,linewidth=\elegant]{-}(0.8,0.8){0.2}{180}{0}
\psarc[linecolor=blue,linewidth=\elegant]{-}(0.4,0){0.2}{0}{180}
\psarc[linecolor=blue,linewidth=\elegant]{-}(1.2,0){0.2}{0}{180}
\psbezier[linecolor=blue,linewidth=\elegant]{-}(0.2,0.8)(0.2,0.3)(1.4,0.3)(1.4,0.8)
\rput(0,0.8)
{
\psline[linecolor=blue,linewidth=\elegant]{-}(0.2,0)(0.2,0.5)
\psline[linecolor=blue,linewidth=\elegant]{-}(0.6,0)(0.6,0.5)
\psarc[linecolor=blue,linewidth=\elegant]{-}(1.2,0){0.2}{0}{180}
}
\end{pspicture}
\ = 0\ .
\ee
\paragraph{The projective modules $\boldsymbol{\mathsf P_{N,2}}$.}
We define the modules $\mathsf P_{N,2}$ for $\tl_N(0)$ with $N$ even. The vector space is spanned by link states with $d=0$ and $d=2$ defects. For example, for $N=4$, $\mathsf P_{N,2}$ has dimension five and its link states are the first five states in \eqref{eq:V4linkstates}.

The action of $a \in \tl_N(0)$ on $v \in \mathsf P_{N,2}$ also uses the diagrammatic construction. One draws $v$ above $a$ and reads the new link state from the diagram. If there are contractible loops in this diagram, the result is set to zero, as it is for the standard action for $\beta = 0$. However, in contrast with the standard action, if two defects connect, the result is {\it not} set to zero. For example, in $\mathsf P_{N,2}$, the calculations of \eqref{eq:standard.examples} become
\be
\psset{unit=0.9}
\begin{pspicture}[shift=-0.525](-0.0,-0.25)(1.6,1.3)
\pspolygon[fillstyle=solid,fillcolor=lightlightblue,linecolor=black,linewidth=0pt](0,0)(0,0.8)(1.6,0.8)(1.6,0)(0,0)
\psarc[linecolor=blue,linewidth=\elegant]{-}(0.8,0.8){0.2}{180}{0}
\psarc[linecolor=blue,linewidth=\elegant]{-}(0.4,0){0.2}{0}{180}
\psbezier[linecolor=blue,linewidth=\elegant]{-}(0.2,0.8)(0.2,0.4)(1.0,0.4)(1.0,0)
\psline[linecolor=blue,linewidth=\elegant]{-}(1.4,0)(1.4,0.8)
\rput(0,0.8)
{\psline[linecolor=blue,linewidth=\elegant]{-}(0.2,0)(0.2,0.5)
\psarc[linecolor=blue,linewidth=\elegant]{-}(0.8,0){0.2}{0}{180}
\psline[linecolor=blue,linewidth=\elegant]{-}(1.4,0)(1.4,0.5)}
\end{pspicture}
\ = 0 \ ,
\qquad
\begin{pspicture}[shift=-0.525](-0.0,-0.25)(1.6,1.3)
\pspolygon[fillstyle=solid,fillcolor=lightlightblue,linecolor=black,linewidth=0pt](0,0)(0,0.8)(1.6,0.8)(1.6,0)(0,0)
\psarc[linecolor=blue,linewidth=\elegant]{-}(0.8,0.8){0.2}{180}{0}
\psarc[linecolor=blue,linewidth=\elegant]{-}(0.4,0){0.2}{0}{180}
\psarc[linecolor=blue,linewidth=\elegant]{-}(1.2,0){0.2}{0}{180}
\psbezier[linecolor=blue,linewidth=\elegant]{-}(0.2,0.8)(0.2,0.3)(1.4,0.3)(1.4,0.8)
\rput(0,0.8)
{\psbezier[linecolor=blue,linewidth=\elegant]{-}(0.2,0)(0.2,0.5)(1.4,0.5)(1.4,0)
\psarc[linecolor=blue,linewidth=\elegant]{-}(0.8,0){0.2}{0}{180}
}
\end{pspicture}
\ = 0 \ ,
\qquad
\begin{pspicture}[shift=-0.525](-0.0,-0.25)(1.6,1.3)
\pspolygon[fillstyle=solid,fillcolor=lightlightblue,linecolor=black,linewidth=0pt](0,0)(0,0.8)(1.6,0.8)(1.6,0)(0,0)
\psarc[linecolor=blue,linewidth=\elegant]{-}(0.8,0.8){0.2}{180}{0}
\psarc[linecolor=blue,linewidth=\elegant]{-}(0.4,0){0.2}{0}{180}
\psarc[linecolor=blue,linewidth=\elegant]{-}(1.2,0){0.2}{0}{180}
\psbezier[linecolor=blue,linewidth=\elegant]{-}(0.2,0.8)(0.2,0.3)(1.4,0.3)(1.4,0.8)
\rput(0,0.8)
{
\psline[linecolor=blue,linewidth=\elegant]{-}(0.2,0)(0.2,0.5)
\psline[linecolor=blue,linewidth=\elegant]{-}(0.6,0)(0.6,0.5)
\psarc[linecolor=blue,linewidth=\elegant]{-}(1.2,0){0.2}{0}{180}
}
\end{pspicture}
\ = \ 
\begin{pspicture}[shift=-0.08](0.0,0)(1.6,0.5)
\psline[linewidth=\mince](0,0)(1.6,0)
\psarc[linecolor=blue,linewidth=\elegant]{-}(0.4,0){0.2}{0}{180}
\psarc[linecolor=blue,linewidth=\elegant]{-}(1.2,0){0.2}{0}{180}
\end{pspicture} \ .
\ 
\ee
The third calculation gives a link state with weight one in $\mathsf P_{N,2}$, whereas this prefactor vanishes in $\mathsf V_{N,2}$. In contrast, the first two are in fact obtained identically in $\mathsf P_{N,2}$ and in $\mathsf V_{N,d}$, but here they vanish because $\beta=0$.

In the modules $\mathsf P_{N,2}$, the number of defects is not a conserved quantity, as it may happen that the initial and final states respectively have $d=2$ and $d=0$ defects. Moreover, we note that $\mathsf P_{N,2}$ has a submodule isomorphic to $\mathsf V_{N,0}$. To distinguish between the actions in $\mathsf V_{N,d}$ and $\mathsf P_{N,2}$, in the following we indicate explicitly which module is involved, and write for instance $a\, v\big|_{\mathsf V_{N,d}}$ or $a\, v\big|_{\mathsf P_{N,2}}$.
\paragraph{Bilinear forms for $\boldsymbol{\tl_N(0)}$.}
For the model of critical dense polymers, the Gram bilinear form is an invariant form on $\mathsf V_{N,d}$ defined as follows. Let $v,v'$ be two link states in $\mathsf V_{N,d}$. Performing a vertical flip of $v$ and connecting its nodes to those of $v'$, we obtain a diagram where the loop segments form closed loops or connect defects together. The Gram product of $v$ and $v'$, denoted $v \cdot v'$, equals $1$ if the number of closed loops is zero and all defects from $v$ are connected to defects of $v'$. Otherwise, $v\cdot v' = 0$.  To illustrate, for
$\psset{unit=0.54}
v_1 = \, 
\begin{pspicture}(0.0,0)(2.4,0.5)
\psline[linewidth=\mince](0,0)(2.4,0)
\psarc[linecolor=blue,linewidth=\elegant]{-}(0.8,0){0.2}{0}{180}
\psarc[linecolor=blue,linewidth=\elegant]{-}(2.0,0){0.2}{0}{180}
\psline[linecolor=blue,linewidth=\elegant]{-}(0.2,0)(0.2,0.5)
\psline[linecolor=blue,linewidth=\elegant]{-}(1.4,0)(1.4,0.5)
\end{pspicture}
$\ ,
$\psset{unit=0.54}
v_2 = \, 
\begin{pspicture}(0.0,0)(2.4,0.5)
\psline[linewidth=\mince](0,0)(2.4,0)
\psarc[linecolor=blue,linewidth=\elegant]{-}(1.2,0){0.2}{0}{180}
\psbezier[linecolor=blue,linewidth=\elegant]{-}(0.6,0)(0.6,0.7)(1.8,0.7)(1.8,0)
\psline[linecolor=blue,linewidth=\elegant]{-}(0.2,0)(0.2,0.5)
\psline[linecolor=blue,linewidth=\elegant]{-}(2.2,0)(2.2,0.5)
\end{pspicture}
$\, and
$\psset{unit=0.54}
v_3 = \, 
\begin{pspicture}(0.0,0)(2.4,0.5)
\psline[linewidth=\mince](0,0)(2.4,0)
\psarc[linecolor=blue,linewidth=\elegant]{-}(0.4,0){0.2}{0}{180}
\psarc[linecolor=blue,linewidth=\elegant]{-}(1.2,0){0.2}{0}{180}
\psline[linecolor=blue,linewidth=\elegant]{-}(1.8,0)(1.8,0.5)
\psline[linecolor=blue,linewidth=\elegant]{-}(2.2,0)(2.2,0.5)
\end{pspicture}
$\ ,
we have 
\be
\label{eq:gram.example}
\psset{unit=0.54}
\begin{pspicture}[shift=-0.6](0.0,-0.7)(2.4,0.7)
\psline[linewidth=\mince](0,0)(2.4,0)
\psarc[linecolor=blue,linewidth=\elegant]{-}(1.2,0){0.2}{0}{180}
\psbezier[linecolor=blue,linewidth=\elegant]{-}(0.6,0)(0.6,0.7)(1.8,0.7)(1.8,0)
\psline[linecolor=blue,linewidth=\elegant]{-}(0.2,0)(0.2,0.5)
\psline[linecolor=blue,linewidth=\elegant]{-}(2.2,0)(2.2,0.5)
\psarc[linecolor=blue,linewidth=\elegant]{-}(0.8,0){-0.2}{0}{180}
\psarc[linecolor=blue,linewidth=\elegant]{-}(2.0,0){-0.2}{0}{180}
\psline[linecolor=blue,linewidth=\elegant]{-}(0.2,0)(0.2,-0.5)
\psline[linecolor=blue,linewidth=\elegant]{-}(1.4,0)(1.4,-0.5)
\end{pspicture}
\quad
\longrightarrow
\quad
v_1 \cdot v_2 = 1,
\qquad
\begin{pspicture}[shift=-0.6](0.0,-0.7)(2.4,0.7)
\psline[linewidth=\mince](0,0)(2.4,0)
\psarc[linecolor=blue,linewidth=\elegant]{-}(0.4,0){0.2}{0}{180}
\psarc[linecolor=blue,linewidth=\elegant]{-}(1.2,0){0.2}{0}{180}
\psline[linecolor=blue,linewidth=\elegant]{-}(1.8,0)(1.8,0.5)
\psline[linecolor=blue,linewidth=\elegant]{-}(2.2,0)(2.2,0.5)
\psarc[linecolor=blue,linewidth=\elegant]{-}(0.8,0){-0.2}{0}{180}
\psarc[linecolor=blue,linewidth=\elegant]{-}(2.0,0){-0.2}{0}{180}
\psline[linecolor=blue,linewidth=\elegant]{-}(0.2,0)(0.2,-0.5)
\psline[linecolor=blue,linewidth=\elegant]{-}(1.4,0)(1.4,-0.5)
\end{pspicture}
\quad
\longrightarrow
\quad
v_1 \cdot v_3 = 0,
\qquad
\begin{pspicture}[shift=-0.6](0.0,-0.7)(2.4,0.7)
\psline[linewidth=\mince](0,0)(2.4,0)
\psarc[linecolor=blue,linewidth=\elegant]{-}(0.4,0){0.2}{0}{180}
\psarc[linecolor=blue,linewidth=\elegant]{-}(1.2,0){0.2}{0}{180}
\psline[linecolor=blue,linewidth=\elegant]{-}(1.8,0)(1.8,0.5)
\psline[linecolor=blue,linewidth=\elegant]{-}(2.2,0)(2.2,0.5)
\psarc[linecolor=blue,linewidth=\elegant]{-}(1.2,0){-0.2}{0}{180}
\psbezier[linecolor=blue,linewidth=\elegant]{-}(0.6,0)(0.6,-0.7)(1.8,-0.7)(1.8,0)
\psline[linecolor=blue,linewidth=\elegant]{-}(0.2,0)(0.2,-0.5)
\psline[linecolor=blue,linewidth=\elegant]{-}(2.2,0)(2.2,-0.5)
\end{pspicture}
\quad
\longrightarrow
\quad
v_2 \cdot v_3 = 0.
\ee
One can also check that $v_1 \cdot v_1 = v_2 \cdot v_2 = v_3 \cdot v_3 = 0$.
We note that $v\cdot w = 0$ if $v,w \in \mathsf V_{N,0}$. The Gram bilinear forms used in the computations below involve $\mathsf V_{N,d}$ with $d \ge 1$.

We define a second bilinear form, defined on the module $\mathsf P_{N,2}$. The link states in this module have either $0$ or $2$ defects. For two such link states $v$ and $v'$, we denote this second product by $v \odot v'$. As for the Gram product, we draw the diagram where $v$ is flipped vertically and its nodes are attached to those of $v'$. If the number of closed loops is zero, the defects of $v$ are connected among themselves, and likewise for those of $v'$, then $v \odot v'=1$. Otherwise, $v \odot v' = 0$. In the examples above, we have
\be
v_1 \odot v_2 = 0, \qquad v_1 \odot v_3 = 1, \qquad v_2 \odot v_3 = 0.
\ee

\paragraph{The transfer tangle.} The double-row transfer tangle for the model of dense polymers is an element of $\tl_N(0)$ defined as \cite{PR07}
\begin{equation} 
\label{eq:Du}
\psset{unit=1.1}
\Db (u) = \frac{1}{\sin 2u}\ \ 
\begin{pspicture}[shift=-1.5](-0.5,-.6)(5.5,2)
\facegrid{(0,0)}{(5,2)}
\psarc[linewidth=0.025]{-}(0,0){0.16}{0}{90}
\psarc[linewidth=0.025]{-}(0,1){0.16}{0}{90}
\psarc[linewidth=0.025]{-}(1,0){0.16}{0}{90}
\psarc[linewidth=0.025]{-}(1,1){0.16}{0}{90}
\psarc[linewidth=0.025]{-}(4,0){0.16}{0}{90}
\psarc[linewidth=0.025]{-}(4,1){0.16}{0}{90}
\rput(2.5,0.5){$\ldots$}
\rput(2.5,1.5){$\ldots$}
\rput(3.5,0.5){$\ldots$}
\rput(3.5,1.5){$\ldots$}
\psarc[linewidth=1.5pt,linecolor=blue]{-}(0,1){0.5}{90}{-90}
\psarc[linewidth=1.5pt,linecolor=blue]{-}(5,1){0.5}{-90}{90}
\rput(0.5,.5){$u$}
\rput(0.5,1.5){$\frac{\pi}{2}-u$}
\rput(1.5,.5){$u$}
\rput(1.5,1.5){$\frac{\pi}{2}-u$}
\rput(4.5,.5){$u$}
\rput(4.5,1.5){$\frac{\pi}{2}-u$}
\rput(2.5,-0.5){$\underbrace{\qquad \hspace{4.1cm} \qquad}_N$}
\end{pspicture}\ \ ,
\qquad 
\psset{unit=0.74}
 \begin{pspicture}[shift=-.40](1,1)
\facegrid{(0,0)}{(1,1)}
\psarc[linewidth=0.025]{-}(0,0){0.16}{0}{90}
\rput(.5,.5){$u$}
\end{pspicture}
\ = \cos u\ \
\begin{pspicture}[shift=-.40](1,1)
\facegrid{(0,0)}{(1,1)}
\rput[bl](0,0){\loopa}
\end{pspicture}
\ + \sin u \ \
\begin{pspicture}[shift=-.40](1,1)
\facegrid{(0,0)}{(1,1)}
\rput[bl](0,0){\loopb}
\end{pspicture}\ \ , 
\end{equation}
where $u$ is the spectral parameter. We refer to the value $u= \frac \pi 4$, for which both tiles have equal weights, as the {\it isotropic point}. Later, we use the shorthand notation $\Db = \Db(\frac \pi4)$. Indeed, the weight $W_\sigma$ of a loop configuration, defined in terms of the loop fugacities only, assumes that the two tiles have the same weight.

The transfer tangle satisfies a number of identities: (i) it satisfies crossing symmetry, namely $\Db(\frac \pi 2 - u) = \Db(u)$, (ii) it satisfies the periodicity property $\Db(u+\pi) = \Db(u)$, (iii) it evaluates to the identity at $u=0$: $\Db(u=0) = \Ib$.
The transfer tangles also commute at different values of the spectral parameter, $[\Db(u),\Db(v)] = 0$, and therefore generate a commuting family of elements of $\tl_N(0)$. The Hamiltonian $\boldsymbol H$ of the model is related to the transfer tangle via the relation
\be
\label{eq:DH}
\Db(u) = \Ib -2 u \boldsymbol{H} + \mathcal \mathcal \mathcal O(u^2), \qquad \boldsymbol{H} = - \sum_{j=1}^{N-1}e_j,
\ee 
and is also in this commuting family. 

Crucially, the transfer tangle satisfies an inversion identity \cite{PR07}:
\be
\Db(u)\Db(u+\tfrac \pi 2) = \Ib\, \bigg( \frac{\cos^{2N}\!u - \sin^{2N}\!u}{\cos^2 u - \sin^2 u}\bigg)^2.
\label{eq:geninv}
\ee
This implies that, in any representation of $\tl_N(0)$, the eigenvalues $\Lambda(u)$ of $\Db(u)$ are of the form
\be
\Lambda(u)= \prod_{j=1}^{\left \lfloor \tfrac {N-1} 2 \right \rfloor} \Big(1 + \epsilon_j \sin2u\, \sin t_j\Big) \Big(1 + \mu_j \sin 2u \,\sin t_j\Big), \ \quad 
t_j = \left\{\begin{array}{c l}
\frac{j \pi}{N} & N\, \textrm{even,} \\[0.2cm]
\frac{(2j-1) \pi}{2N} & N\, \textrm{odd,}
\end{array}\right.
\label{eq:PReigs}
\ee
where $\epsilon_j$ and $\mu_j$ are in $\{+1, -1\}$. Different eigenvalues then correspond to different choices of the signs $\epsilon_j$ and $\mu_j$.

\paragraph{The partition functions.} The various partition functions defined in \cref{sec:denseloopmodels} are written using the tools defined above. We assign the upper labels $AB$, $A$ and $B$ to the transfer tangles and link states to indicate which system or subsystem they pertain to. We find
\begin{subequations}
\label{eq:ZF}
\begin{alignat}{2}
Z^{AB}_d &= 2^{2MN} (v^A_d \otimes v^B_0)\cdot \big(\Db^{A}\otimes \Db^{B}\big)^M \big(\Db^{AB}\big)^M v^{AB}_d\big|_{\mathsf V_{N,d}}\ ,\label{eq:ZFAB}\\[0.2cm]
Z^{A\cup B}_d &=2^{2MN} v^{AB}_d \cdot (\Db^{AB})^{2M} v^{AB}_d\big|_{\mathsf V_{N,d}}\ ,\\[0.2cm] 
Z^{A}_d &=2^{2MN_A} v^A_d \cdot (\Db^{A})^{2M} v^{A}_d\big|_{\mathsf V_{N_A,d}}\ ,\\[0.2cm]
Z^{B}_2 &=2^{2MN_B} v^B_2 \cdot (\Db^{B})^{2M} v^B_2\big|_{\mathsf V_{N_B,2}}\ ,
\end{alignat}
\end{subequations}
and
\begin{subequations}
\label{eq:ZFt}
\begin{alignat}{2}
\tilde Z^{AB}_2 &=2^{2MN} (v^A_2 \otimes v^B_0)\odot \big(\Db^{A}\otimes \Db^{B}\big)^M \big(\Db^{AB}\big)^M v^{AB}_2 \Big|_{\mathsf P_{N,2}}\ ,\\[0.2cm]
\tilde Z^{A\cup B}_2 &= 2^{2MN} v^{AB}_2\odot \big(\Db^{AB}\big)^{2M} v^{AB}_2 \Big|_{\mathsf P_{N,2}}\ ,\\[0.2cm]
\tilde Z^{A}_2 &=2^{2MN_A} v^{A}_2\odot \big(\Db^{A}\big)^{2M} v^{A}_2 \Big|_{\mathsf P_{N_A,2}}\ ,
\end{alignat}
\end{subequations}
where 
\be
\psset{unit=0.9}
v_d = \
\begin{pspicture}[shift=-0.08](0.0,0)(4.8,1.2)
\psline[linewidth=\mince](0,0)(4.8,0)
\psline[linecolor=blue,linewidth=\elegant]{-}(0.2,0)(0.2,0.5)
\psline[linecolor=blue,linewidth=\elegant]{-}(0.6,0)(0.6,0.5)
\psline[linecolor=blue,linewidth=\elegant]{-}(1.0,0)(1.0,0.5)
\psline[linecolor=blue,linewidth=\elegant]{-}(1.4,0)(1.4,0.5)
\psarc[linecolor=blue,linewidth=\elegant]{-}(2.0,0){0.2}{0}{180}
\psarc[linecolor=blue,linewidth=\elegant]{-}(2.8,0){0.2}{0}{180}
\rput(3.6,0.15){...}
\rput(0.8,0.9){$\overbrace{\ \hspace{1.0cm}\ }^d$}
\psarc[linecolor=blue,linewidth=\elegant]{-}(4.4,0){0.2}{0}{180}
\end{pspicture} \ 
\ee
and $v \otimes w$ indicates that $w$ is attached to the right of $v$. The powers of $2$ ensure that each tile has weight $1$ instead of $\frac1{\sqrt2}$ as it does in \eqref{eq:Du} for $u = \frac \pi 4$.

\subsection{Bipartite fidelify from the XX spin chain}\label{sec:spinchain}

The next step of the computation is to rewrite the formulas for the partition functions relevant for the bipartite fidelity in terms of matrix elements in the XX spin chain.
\paragraph{The XX representation.}
The XX representation of $\tl_N(0)$ is defined on the vector space $(\mathbb C^2)^{\otimes N}$. We use the canonical basis 
\be
|{\uparrow}\rangle = \begin{pmatrix}1\\0\end{pmatrix}, \qquad |{\downarrow}\rangle = \begin{pmatrix}0\\1\end{pmatrix}
\ee
for $\mathbb C^2$ and the Pauli matrices $\sigma^x = \left(\begin{smallmatrix}0&1\\1&0\end{smallmatrix}\right)$, $\sigma^y = \left(\begin{smallmatrix}0&-\ir\\\ir&0\end{smallmatrix}\right)$ and $\sigma^z = \left(\begin{smallmatrix}1&0\\0&-1\end{smallmatrix}\right)$.
The generators $e_j$ are represented by the following matrices \cite{PS90}:
\be
\label{eq:X.rep}
\mathsf X_N(e_j) = \underbrace{\mathbb I_2 \otimes \dots \otimes \mathbb I_2}_{j-1} \otimes 
\begin{pmatrix}
0 & 0 & 0 & 0 \\
0 & \ir & 1 & 0 \\
0 & 1 & \ir^{-1} & 0 \\
0 & 0 & 0 & 0
\end{pmatrix}
\otimes \underbrace{\mathbb I_2 \otimes \dots \otimes \mathbb I_2}_{N-j-1}
\ee
where $\mathbb I_2 = \left(\begin{smallmatrix}1&0\\0&1\end{smallmatrix}\right)$. It is indeed not hard to check that these matrices satisfy the defining relations \eqref{eq:TL.def} of $\tl_N(0)$. The generators also commute with the total magnetisation $S^z = \frac12\sum_{i=1}^N \sigma^z_i$. As a consequence, the representation $\mathsf X_N$ splits as a direct sum of smaller representations labelled by the eigenvalues $m$ of $S^z$, given by $-\frac N2$, $-\frac{N-2}2$, \dots, $\frac N2$.

In this representation, the Hamiltonian is the XX Hamiltonian with the $U_q(s\ell_2)$-invariant boundary magnetic fields of Pasquier and Saleur \cite{PS90}:
\be
H = \mathsf X_N(\boldsymbol H) = -\frac{1}{2}\sum_{j=1}^{N-1}\left(\sigma_j^x\sigma_{j+1}^x+\sigma_j^y\sigma_{j+1}^y \right)-\frac{\ir}{2}(\sigma_1^z-\sigma_N^z). 
\ee
Likewise, the representative of $\Db(u)$ in the XX representation, denoted $D(u)$, is the double-row transfer matrix of the six-vertex model: $D(u) = \mathsf X_N(\Db(u))$. We use the notation $D = D(\frac \pi 4)$ for the transfer matrix at the isotropic point.

\paragraph{Embedding $\boldsymbol{\mathsf V_{N,d}}$ and $\boldsymbol{\mathsf P_{N,2}}$ in the spin chain.}
There exists a map from link states to spin states that intertwines the representations $\mathsf V_{N,d}$ and $\mathsf X_N$. For a given link state $v$, we define its image in $(\mathbb C^2)^{\otimes N}$ under this map as $|v\rangle$. Locally, it is defined as
\be
\label{eq:localmaps}
|\,
\psset{unit=0.74}
\begin{pspicture}[shift=-0.08](0.0,0)(0.8,0.5)
\psline[linewidth=\mince](0,0)(0.8,0)
\psarc[linecolor=blue,linewidth=\elegant]{-}(0.4,0){0.2}{0}{180}
\end{pspicture}
\,  \rangle = \omega\, |{\uparrow \downarrow}\rangle + \omega^{-1}\, |{\downarrow \uparrow}\rangle, \qquad
|\,
\begin{pspicture}[shift=-0.08](0.0,0)(0.4,0.5)
\psline[linewidth=\mince](0,0)(0.4,0)
\psline[linecolor=blue,linewidth=\elegant]{-}(0.2,0)(0.2,0.5)
\end{pspicture}
\, \rangle =  |{\downarrow}\rangle, \qquad 
\omega = \eE^{\ir \pi/4}.
\ee
In general, to obtain the spin state corresponding to a given link state, we repeatedly apply the local rules for each arc and each defect. For instance, for $N=4$, we have
\begin{subequations}
\begin{alignat}{2}
&|\,\psset{unit=0.74}\label{eq:spinstate1}
\begin{pspicture}[shift=-0.08](0.0,0)(1.6,0.5)
\psline[linewidth=\mince](0,0)(1.6,0)
\psline[linecolor=blue,linewidth=\elegant]{-}(0.2,0)(0.2,0.5)
\psline[linecolor=blue,linewidth=\elegant]{-}(0.6,0)(0.6,0.5)
\psarc[linecolor=blue,linewidth=\elegant]{-}(1.2,0){0.2}{0}{180}
\end{pspicture}
\, \rangle =  \omega\, |{\downarrow \downarrow \uparrow \downarrow} \rangle + \omega^{-1}\, |{\downarrow \downarrow\downarrow \uparrow} \rangle,\\[0.2cm]
&|\,\psset{unit=0.74}\label{eq:spinstate2}
\begin{pspicture}[shift=-0.08](0.0,0)(1.6,0.5)
\psline[linewidth=\mince](0,0)(1.6,0)
\psarc[linecolor=blue,linewidth=\elegant]{-}(0.4,0){0.2}{0}{180}
\psarc[linecolor=blue,linewidth=\elegant]{-}(1.2,0){0.2}{0}{180}
\end{pspicture}
\, \rangle = \omega^2\, |{\uparrow \downarrow \uparrow \downarrow} \rangle + |{\uparrow \downarrow \downarrow \uparrow } \rangle  + |{\downarrow \uparrow \uparrow \downarrow} \rangle +\omega^{-2}\, |{\downarrow \uparrow\downarrow \uparrow} \rangle,\\[0.2cm]
&|\,\psset{unit=0.74}\label{eq:spinstate3}
\begin{pspicture}[shift=-0.08](0.0,0)(1.6,0.5)
\psline[linewidth=\mince](0,0)(1.6,0)
\psarc[linecolor=blue,linewidth=\elegant]{-}(0.8,0){0.2}{0}{180}
\psbezier[linecolor=blue,linewidth=\elegant]{-}(0.2,0)(0.2,0.6)(1.4,0.6)(1.4,0)
\end{pspicture}
\, \rangle = \omega^2\, |{\uparrow \uparrow \downarrow  \downarrow} \rangle + |{\uparrow \downarrow \uparrow \downarrow} \rangle  + |{\downarrow \uparrow  \downarrow \uparrow} \rangle +\omega^{-2}\, |{\downarrow \downarrow \uparrow \uparrow} \rangle.
\end{alignat}
\end{subequations}
Link states in $\mathsf V_{N,d}$ are therefore mapped to spin states of magnetisation $m = -\frac d2$. The map is indeed a homomorphism, namely one can check that
\be
\mathsf X_N(e_j)|v\rangle = |e_jv\rangle\big|_{\mathsf V_{N,d}} 
\ee
for $j = 1, \dots, N-1$ and $v \in \mathsf V_{N,d}$. This map also has the property that it preserves the bilinear form for the standard modules. Indeed, defining $\langle v| = |v \rangle^{\Tt}$, where the upper label $^{\Tt}$ stands for real transposition, we have $\langle v |v'\rangle = v \cdot v'$, for each $v,v'\in \mathsf V_{N,d}$. For example, 
\be
\psset{unit=0.74}
\langle \, 
\begin{pspicture}[shift=-0.08](0.0,0)(1.6,0.5)
\psline[linewidth=\mince](0,0)(1.6,0)
\psline[linecolor=blue,linewidth=\elegant]{-}(0.2,0)(0.2,0.5)
\psline[linecolor=blue,linewidth=\elegant]{-}(0.6,0)(0.6,0.5)
\psarc[linecolor=blue,linewidth=\elegant]{-}(1.2,0){0.2}{0}{180}
\end{pspicture}
\,|\,
\begin{pspicture}[shift=-0.08](0.0,0)(1.6,0.5)
\psline[linewidth=\mince](0,0)(1.6,0)
\psline[linecolor=blue,linewidth=\elegant]{-}(0.2,0)(0.2,0.5)
\psline[linecolor=blue,linewidth=\elegant]{-}(0.6,0)(0.6,0.5)
\psarc[linecolor=blue,linewidth=\elegant]{-}(1.2,0){0.2}{0}{180}
\end{pspicture}
\, \rangle = 0, \qquad
\langle \, 
\begin{pspicture}[shift=-0.08](0.0,0)(1.6,0.5)
\psline[linewidth=\mince](0,0)(1.6,0)
\psline[linecolor=blue,linewidth=\elegant]{-}(0.2,0)(0.2,0.5)
\psline[linecolor=blue,linewidth=\elegant]{-}(0.6,0)(0.6,0.5)
\psarc[linecolor=blue,linewidth=\elegant]{-}(1.2,0){0.2}{0}{180}
\end{pspicture}
\,|\,
\begin{pspicture}[shift=-0.08](0.0,0)(1.6,0.5)
\psline[linewidth=\mince](0,0)(1.6,0)
\psline[linecolor=blue,linewidth=\elegant]{-}(0.2,0)(0.2,0.5)
\psline[linecolor=blue,linewidth=\elegant]{-}(1.4,0)(1.4,0.5)
\psarc[linecolor=blue,linewidth=\elegant]{-}(0.8,0){0.2}{0}{180}
\end{pspicture}
\, \rangle = 1,
\qquad \langle \, 
\begin{pspicture}[shift=-0.08](0.0,0)(1.6,0.5)
\psline[linewidth=\mince](0,0)(1.6,0)
\psline[linecolor=blue,linewidth=\elegant]{-}(0.2,0)(0.2,0.5)
\psline[linecolor=blue,linewidth=\elegant]{-}(0.6,0)(0.6,0.5)
\psarc[linecolor=blue,linewidth=\elegant]{-}(1.2,0){0.2}{0}{180}
\end{pspicture}
\,|\,
\begin{pspicture}[shift=-0.08](0.0,0)(1.6,0.5)
\psline[linewidth=\mince](0,0)(1.6,0)
\psline[linecolor=blue,linewidth=\elegant]{-}(1.0,0)(1.0,0.5)
\psline[linecolor=blue,linewidth=\elegant]{-}(1.4,0)(1.4,0.5)
\psarc[linecolor=blue,linewidth=\elegant]{-}(0.4,0){0.2}{0}{180}
\end{pspicture}
\, \rangle = 0.
\ee
Clearly, if $v \in \mathsf V_{N,d}$ and $v' \in \mathsf V_{N,d'}$ with $d \neq d'$, then $\langle v|v'\rangle = 0$.

There exists a similar map for $\mathsf P_{N,2}$. For $v \in \mathsf P_{N,2}$, we denote its image under this map as $|\tilde v \rangle$. The local relations are
\be
|\,
\psset{unit=0.74}
\begin{pspicture}[shift=-0.08](0.0,0)(0.8,0.5)
\psline[linewidth=\mince](0,0)(0.8,0)
\psarc[linecolor=blue,linewidth=\elegant]{-}(0.4,0){0.2}{0}{180}
\rput(0.4,0.3){$\widetilde{\phantom{3\,3}}$}
\end{pspicture}
\,  \rangle = \omega\, |{\uparrow \downarrow}\rangle + \omega^{-1}\, |{\downarrow \uparrow}\rangle, \qquad
|\,\begin{pspicture}[shift=-0.08](0.0,0)(0.8,0.5)
\psline[linewidth=\mince](0,0)(0.8,0)
\psline[linecolor=blue,linewidth=\elegant]{-}(0.2,0)(0.2,0.5)
\psline[linecolor=blue,linewidth=\elegant]{-}(0.6,0)(0.6,0.5)
\rput(0.4,0.5){$\widetilde{\phantom{3\,3}}$}
\end{pspicture}
\, \rangle = \frac12 \big(\omega^{-1}|{\uparrow\downarrow}\rangle + \omega\,|{\downarrow\uparrow}\rangle\big). 
\ee
For a given $v \in \mathsf P_{N,2}$, these relations are applied locally to each arc and to the defects, to produce an element of $(\mathbb C^2)^{\otimes N}$. For example, for $N=4$, we have
\begin{subequations}
\begin{alignat}{2}
&|\,\psset{unit=0.74}
\begin{pspicture}[shift=-0.08](0.0,0)(1.6,0.9)
\psline[linewidth=\mince](0,0)(1.6,0)
\psline[linecolor=blue,linewidth=\elegant]{-}(0.2,0)(0.2,0.5)
\psline[linecolor=blue,linewidth=\elegant]{-}(0.6,0)(0.6,0.5)
\psarc[linecolor=blue,linewidth=\elegant]{-}(1.2,0){0.2}{0}{180}
\rput(0.8,0.5){$\widetilde{\phantom{33\,\,\,33}}$}
\end{pspicture}
\, \rangle = \frac12\big(|{\uparrow \downarrow \uparrow \downarrow} \rangle + \omega^{-2}\,|{\uparrow \downarrow \downarrow \uparrow } \rangle  +\omega^{2}\, |{\downarrow \uparrow \uparrow \downarrow} \rangle + |{\downarrow \uparrow\downarrow \uparrow} \rangle\big),\
\end{alignat}
\end{subequations}
which is not identical to \eqref{eq:spinstate1}. The spin states corresponding to 
$
\psset{unit=0.54}
\begin{pspicture}[shift=-0.08](0.0,0)(1.6,0.5)
\psline[linewidth=\mince](0,0)(1.6,0)
\psarc[linecolor=blue,linewidth=\elegant]{-}(0.4,0){0.2}{0}{180}
\psarc[linecolor=blue,linewidth=\elegant]{-}(1.2,0){0.2}{0}{180}
\end{pspicture}
$ 
and
$
\psset{unit=0.54}
\begin{pspicture}[shift=-0.08](0.0,0)(1.6,0.5)
\psline[linewidth=\mince](0,0)(1.6,0)
\psarc[linecolor=blue,linewidth=\elegant]{-}(0.8,0){0.2}{0}{180}
\psbezier[linecolor=blue,linewidth=\elegant]{-}(0.2,0)(0.2,0.6)(1.4,0.6)(1.4,0)\end{pspicture}
$ 
under this map coincide with those in \eqref{eq:spinstate2} and \eqref{eq:spinstate3}. This map is also a homomorphism of representations, namely
\be
\mathsf X_N(e_j)|\tilde v\rangle = |\widetilde{e_jv}\rangle\big|_{\mathsf P_{N,2}}\, .
\ee
It also preserves the bilinear form $\odot$ defined on $\mathsf P_{N,2}$: $\langle \tilde v |\tilde v'\rangle = v \odot v'$ for $v,v' \in \mathsf P_{N,2}$. For instance, we have
\be
\psset{unit=0.74}
\langle \, 
\begin{pspicture}[shift=-0.08](0.0,0)(1.6,0.9)
\psline[linewidth=\mince](0,0)(1.6,0)
\rput(0.8,0.5){$\widetilde{\phantom{33\,\,\,33}}$}
\psline[linecolor=blue,linewidth=\elegant]{-}(0.2,0)(0.2,0.5)
\psline[linecolor=blue,linewidth=\elegant]{-}(0.6,0)(0.6,0.5)
\psarc[linecolor=blue,linewidth=\elegant]{-}(1.2,0){0.2}{0}{180}
\end{pspicture}
\,|\,
\begin{pspicture}[shift=-0.08](0.0,0)(1.6,0.9)
\psline[linewidth=\mince](0,0)(1.6,0)
\rput(0.8,0.5){$\widetilde{\phantom{33\,\,\,33}}$}
\psline[linecolor=blue,linewidth=\elegant]{-}(0.2,0)(0.2,0.5)
\psline[linecolor=blue,linewidth=\elegant]{-}(0.6,0)(0.6,0.5)
\psarc[linecolor=blue,linewidth=\elegant]{-}(1.2,0){0.2}{0}{180}
\end{pspicture}
\, \rangle = 0, \qquad
\langle \, 
\begin{pspicture}[shift=-0.08](0.0,0)(1.6,0.9)
\psline[linewidth=\mince](0,0)(1.6,0)
\rput(0.8,0.5){$\widetilde{\phantom{33\,\,\,33}}$}
\psline[linecolor=blue,linewidth=\elegant]{-}(0.2,0)(0.2,0.5)
\psline[linecolor=blue,linewidth=\elegant]{-}(0.6,0)(0.6,0.5)
\psarc[linecolor=blue,linewidth=\elegant]{-}(1.2,0){0.2}{0}{180}
\end{pspicture}
\,|\,
\begin{pspicture}[shift=-0.08](0.0,0)(1.6,0.9)
\psline[linewidth=\mince](0,0)(1.6,0)
\rput(0.8,0.5){$\widetilde{\phantom{33\,\,\,33}}$}
\psline[linecolor=blue,linewidth=\elegant]{-}(0.2,0)(0.2,0.5)
\psline[linecolor=blue,linewidth=\elegant]{-}(1.4,0)(1.4,0.5)
\psarc[linecolor=blue,linewidth=\elegant]{-}(0.8,0){0.2}{0}{180}
\end{pspicture}
\, \rangle = 0,
\qquad \langle \, 
\begin{pspicture}[shift=-0.08](0.0,0)(1.6,0.9)
\psline[linewidth=\mince](0,0)(1.6,0)
\rput(0.8,0.5){$\widetilde{\phantom{33\,\,\,33}}$}
\psline[linecolor=blue,linewidth=\elegant]{-}(0.2,0)(0.2,0.5)
\psline[linecolor=blue,linewidth=\elegant]{-}(0.6,0)(0.6,0.5)
\psarc[linecolor=blue,linewidth=\elegant]{-}(1.2,0){0.2}{0}{180}
\end{pspicture}
\,|\,
\begin{pspicture}[shift=-0.08](0.0,0)(1.6,0.9)
\psline[linewidth=\mince](0,0)(1.6,0)
\rput(0.8,0.5){$\widetilde{\phantom{33\,\,\,33}}$}
\psline[linecolor=blue,linewidth=\elegant]{-}(1.0,0)(1.0,0.5)
\psline[linecolor=blue,linewidth=\elegant]{-}(1.4,0)(1.4,0.5)
\psarc[linecolor=blue,linewidth=\elegant]{-}(0.4,0){0.2}{0}{180}
\end{pspicture}
\, \rangle = 1.
\ee

\paragraph{The partition functions.} Because the maps defined above are homomorphisms of $\tl_N(0)$ representations and preserve the bilinear forms, we can translate the expressions \eqref{eq:ZF} and \eqref{eq:ZFt} for the partition functions as matrix elements in the XX spin chain. We obtain:
\begin{subequations}\label{eq:ZS}
\begin{alignat}{2}
Z^{AB}_d &= 2^{2MN}\langle v^A_d \otimes v^B_0| \big(D^{A}\otimes D^{B}\big)^M \big(D^{AB}\big)^M |v^{AB}_d\rangle,\label{eq:ZSAB}\\[0.2cm]
Z^{A\cup B}_d &= 2^{2MN}\langle v^{AB}_d| (D^{AB})^{2M} |v^{AB}_d\rangle,\\[0.2cm] 
Z^{A}_d &=2^{2MN_A} \langle v^A_d | (D^{A})^{2M} |v^{A}_d\rangle,\\[0.2cm]
Z^{B}_2 &= 2^{2MN_B}\langle v^B_2| (D^{B})^{2M} |v^B_2\rangle,
\end{alignat}
\end{subequations}
and
\begin{subequations}\label{eq:tZS}
\begin{alignat}{2}
\tilde Z^{AB}_2 &= 2^{2MN}\langle \tilde v^A_2 \otimes \tilde v^B_0| \big(D^{A}\otimes D^{B}\big)^M \big(D^{AB}\big)^M |\tilde v^{AB}_2\rangle,\\[0.2cm]
\tilde Z^{A\cup B}_2 &= 2^{2MN}\langle \tilde v^{AB}_2| \big(D^{AB}\big)^{2M} |\tilde v^{AB}_2\rangle,\\[0.2cm]
\tilde Z^{A}_2 &= 2^{2MN_A}\langle \tilde v^{A}_2| \big(D^{A}\big)^{2M} |\tilde v^{A}_2\rangle,
\end{alignat}
\end{subequations}
where we recall that $\langle v \otimes w | \equiv \langle v | \otimes \langle w |$.

\subsection{Diagonalisation of the Hamiltonian}

The diagonalisation procedure for the XX Hamiltonian is standard and uses the Jordan-Wigner transformation. 
For the $U_q(s\ell_2)$-invariant chain, it was studied in \cite{MD11,GST14}.
The first step is to write $H$ as
\be
H = - \bigg(\sum_{j=1}^{N-1} c^\dagger_{j+1}c_{j} + c^\dagger_j c_{j+1}  \bigg) - \ir \left(c^\dagger_1c_1 -c^\dagger_Nc_N\right)
\ee
where the $c_j$ and $c_j^\dagger$ are the canonical fermionic operators
\be
c_j = (-1)^{j-1}\bigg(\prod_{k=1}^{j-1} \sigma^z_k\bigg) \sigma^-_j, \qquad c_j^\dagger = (-1)^{j-1}\bigg(\prod_{k=1}^{j-1} \sigma^z_k\bigg) \sigma^+_j,
\ee
with $\sigma^{\pm}=(\sigma^x \pm \ir \sigma^y)/2$. Recalling that $\omega = \eE^{\ir \pi/4}$, the second step is to perform a Fourier transform of these operators, by defining
\be
\eta_k = \frac{1}{\kappa_k} \sum_{j=1}^{N-1} \sin(\tfrac {\pi k j}{N})\, a_j, \qquad \eta_k^{\text t} = \frac{1}{\kappa_k} \sum_{j=1}^{N-1} \sin(\tfrac {\pi k j}{N})\, a_j^{\text t}, \qquad \kappa_k = \sqrt{N \cos(\tfrac{\pi k}N)},
\ee
where
\be
a_j = \omega\, c_j + \omega^{-1} c_{j+1}, \qquad a_j^{\text t} = \omega\, c_j^\dagger + \omega^{-1} c_{j+1}^\dagger.
\ee
These operators satisfy the fermionic relations
\be
\{a_j, a_k^\Tt\} = \delta_{j,k-1}+\delta_{j,k+1}, \quad \{\eta_k, \eta_\ell^\Tt\} = \delta_{k,\ell}, \quad  \{a_j, a_k\} = \{a_j^\Tt, a_k^\Tt\} = \{\eta_k, \eta_\ell\} = \{\eta_k^\Tt, \eta_\ell^\Tt\} = 0.
\ee
The Hamiltonian can be expressed in Jordan-normal form using these operators. For $N$ odd, the resulting expression is diagonal and takes the form
\be
\label{eq:Hodd}
H = \sum_{k=1}^{N-1}\lambda_k \eta_k^\Tt \eta_k, \qquad \lambda_k = -2 \cos(\tfrac{\pi k}N) \qquad (N \textrm{ odd}).
\ee
The set of operators $\eta_k$ and $\eta^\Tt_k$, with $k = 1, \dots, N-1$, is complemented with two extra operators,
\be
\label{eq:phi}
\phi = \frac1{\kappa_\phi} \sum_{j=1}^{N}\ir^{-(j-1)} c_j, \qquad \phi^\Tt = \frac1{\kappa_\phi} \sum_{j=1}^{N}\ir^{-(j-1)} c^\dagger_j,
\ee
where 
\be
\kappa_\phi = 1 \qquad (N \textrm{ odd}).
\ee
These operators satisfy the anticommutation relations
\be
\{\phi^\Tt,\phi\} = 1, \qquad \{\phi^\Tt,\eta_k\} = \{\phi,\eta^\Tt_k\} =  \{\phi,\eta_k\} =  \{\phi^\Tt,\eta^\Tt_k\} =  0 \qquad (N \textrm{ odd}).
\ee
The full set thus consists of $N$ creation operators and $N$ annihilation operators. By acting with the $N$ creation operators on the reference state $|0\rangle = |{\downarrow\cdots\downarrow}\rangle$, one obtains a basis for the $2^N$ eigenstates of $H$. In terms of these fermions, the magnetisation operator reads
\be
S^z = - \frac N 2 + \phi^\dagger\phi + \sum_{k=1}^{N-1} \eta_k^\dagger \eta_k \qquad (N \textrm{ odd}).
\ee

For $N$ even, the set of operators $\eta_k$ and $\eta^\Tt_k$, with $k \in \{1, \dots, \frac{N-2}2\} \cup \{\frac{N+2}2, \dots, N-1\}$ is complemented with the operators 
\be
\chit = -\omega\sqrt{\frac 2N} \sum_{j=1}^N \ir^{-(j-1)} \big(\big\lfloor \tfrac j2 \big\rfloor- \tfrac N4\big) \, c_j, \qquad \chit^\Tt =-\omega\sqrt{\frac 2N} \sum_{j=1}^N \ir^{-(j-1)} \big(\big\lfloor \tfrac j2 \big\rfloor- \tfrac N4\big)  \,c_j^\dagger,
\ee
as well as with the operators $\phi$ and $\phi^\Tt$ in \eqref{eq:phi}, with the constant $\kappa_\phi$ set to
\be
\kappa_\phi = \omega\sqrt\frac{N}2 \qquad (N \textrm{ even}).
\ee
The anticommutation relations in this case are
\be
\{\phi, \chit^\Tt\} = \{\phi^\Tt, \chit\} =1, \qquad \{\phi^\Tt, \phi\} = \{\chit^\Tt, \chit\} = \{\phi, \chit\} = \{\phi^\Tt, \chit^\Tt\} =  0 \qquad (N \textrm{ even}).
\ee
All the anticommutators involving the operators $\eta_k$ and $\eta^\Tt_k$ and one of $\phi, \phi^\Tt, \chit$ and $\chit^\Tt$ also vanish.
In terms of these operators, the Hamiltonian takes the form
\be
\label{eq:Heven}
H = \phi^\Tt \phi+ \sum_{\substack{k=1\\ k\neq N/2}}^{N-1}\lambda_k \eta_k^\Tt \eta_k, \qquad \lambda_k = -2 \cos(\tfrac{\pi k}N)  \qquad (N \textrm{ even}).
\ee
The operators $\eta^\Tt_k$, $\phi^\Tt$ and $\chit^\Tt$ form a set of $N$ creation operators. Acting on the reference state $|0\rangle$ with these operators, one obtains a full basis of eigenstates and generalised eigenstates, of dimension $2^N$. The magnetisation operator then reads
\be
S^z = - \frac N 2 + \phi^\dagger\chit + \chit^\dagger\phi + \sum_{\substack{k=1\\k \neq N/2}}^{N-1} \eta_k^\dagger \eta_k  \qquad (N \textrm{ even}).
\ee
\vspace{-0.5cm}

\subsection{Groundstates}\label{sec:g.states}

Except for the case of magnetisation zero, the groundstate eigenspace of $H$ restricted to the sector of magnetisation $m$ has dimension one. Recalling from \cref{sec:spinchain} the relation $d= -2m$, we choose to denote the groundstate of magnetisation $m$ by $|w_{-2m}\rangle = |w_{d}\rangle$ and the corresponding eigenvalue by $\Lambda_d(u)$. In terms of the fermions, this state takes the form
\be
\label{eq:wd}
|w_{d}\rangle = \eta_1^\Tt \eta_2^\Tt \dots \eta^\Tt_{(N-d)/2} |0 \rangle\qquad (d \neq 0).
\ee
For $N$ even and $d=0$, the eigenspace is two-dimensional,
\be
\label{eq:w0}
|w_0 \rangle = \phi^\Tt \eta_1^\Tt \eta_2^\Tt \dots \eta_{N/2-1}^\Tt |0\rangle, \qquad
|\hat w_0 \rangle = \chit^\Tt \eta_1^\Tt \eta_2^\Tt \dots \eta_{N/2-1}^\Tt  |0\rangle, \qquad
\ee
and the states form a rank-two Jordan cell:
\be
\label{eq:HJordan}
H|w_0\rangle = h_0 |w_0\rangle, \qquad H|\hat w_0\rangle = h_0 |\hat w_0\rangle + |w_0 \rangle,
\qquad h_0 = 1- \textrm{cot}(\tfrac\pi{2N}).
\ee
These states are also generalised eigenstates for the transfer matrix $D(u)$. The groundstate eigenvalue corresponds to a specific choice for the unfixed signs in \eqref{eq:PReigs} according to a {\it selection rule} \cite{PR07,MD11}. For the groundstate, these are given by $\epsilon_j = \mu_j = 1$ for $j = 1, \dots, \frac {N-2}2$. The eigenvalue then reads
\be
\label{eq:D0}
\Lambda_0(u) = \prod_{j=1}^{\tfrac {N-2} 2} \big(1 + \sin2u\, \sin t_j\big)^2.
\ee
The transfer tangle $D(u)$ mixes $|w_0\rangle$ and $|\hat w_0\rangle$ in a rank-two Jordan cell:
\be
\label{eq:DJordan}
D(u) |w_0\rangle = \Lambda_0(u) |w_0\rangle, \qquad D(u) |\hat w_0\rangle = \Lambda_0(u) |\hat w_0\rangle + f(u) |w_0\rangle,
\ee
where $f(u)$ is a yet undetermined function of $u$.

To evaluate $f(u)$, we note that $D(u)$, like $\Db(u)$, is a centered Laurent polynomial in the variable $\eE^{\ir u}$, of degree width at most $4N-4$. This also holds true for $\Lambda_0(u)$ and $f(u)$, because $|w_0\rangle$ and $|\hat w_0\rangle$ are independent of $u$. In fact, we see from \eqref{eq:D0} that $\Lambda_0(u)$ has the degree width $4N-8$.
We proceed to compute $D(u)D(u+\frac \pi 2)|\hat w_0\rangle$ in two ways:
\begin{alignat}{2}
D(u)D(u+\tfrac \pi 2)|\hat w_0\rangle &= \bigg( \frac{\cos^{2N}\!u - \sin^{2N}\!u}{\cos^2 u - \sin^2 u}\bigg)^2 |\hat w_0\rangle =  \Lambda_0(u)\Lambda_0(u+\tfrac \pi 2) |\hat w_0\rangle,\\[0.2cm]
D(u)D(u+\tfrac \pi 2)|\hat w_0\rangle &= D(u) \big(\Lambda_0(u+\tfrac \pi 2) |\hat w_0\rangle + f(u+\tfrac \pi 2)|w_0\rangle \big)\nonumber\\[0.2cm]
&= \Lambda_0(u)\Lambda_0(u+\tfrac \pi 2) |\hat w_0\rangle + \big(\Lambda_0(u)f(u+\tfrac \pi 2)+f(u)\Lambda_0(u+\tfrac \pi 2)\big) |w_0\rangle.
\end{alignat}
The two results must coincide, implying that
\be
\label{eq:ffLL}
\frac{f(u)}{f(u+\frac \pi 2)} = - \frac{\Lambda_0(u)}{\Lambda_0(u+\frac \pi 2)}.
\ee
The right-hand side is a ratio of Laurent polynomials in $\eE^{\ir u}$. One computes $\Lambda_0(u+\frac \pi2)$ from \eqref{eq:D0} and finds that there are no factors that cancel out between the numerator and denominator. As a result, we have that $f(u)$, which is itself a Laurent polynomial, must be proportional to $\Lambda_0(u)$:
\be
f(u) = p(u) \Lambda_0(u), \qquad p(u) = \sum_{j=-2}^2 \alpha_j \eE^{\ir j u},
\ee
where the $\alpha_j$ are yet undetermined. The remaining Laurent polynomial $p(u)$ is centered and has degree width at most four, ensuring that $f(u)$ is centered and has degree width at most $4N-4$. To solve for the $\alpha_j$, we note that $f(u)$ satisfies a set of relations due to the symmetry properties of $\Db(u)$ that are listed below \eqref{eq:Du}:
\be
f(u) = f(\tfrac \pi 2 - u), \qquad f(u) = f(u+\pi), \qquad f(u=0) = 0.
\ee
From these, we find that $p(u) = \alpha \sin(2u)$, for some non-zero constant $\alpha$. To determine $\alpha$, we use \eqref{eq:DH}, \eqref{eq:HJordan} and \eqref{eq:DJordan} to obtain a final constraint for $f(u)$,
\be
-\frac12 \frac{\dd f(u)}{\dd u}\Big|_{u=0} = 1,
\ee
from which we obtain $\alpha = -1$. The final result is $f(u) = - \sin (2u) \Lambda_0(u)$ and indeed satisfies \eqref{eq:ffLL}.

\section{Exact results for the bipartite fidelity for primary fields}\label{sec:ExactResults}

In this section, we apply Wick's theorem to derive closed-form expressions for the bipartite fidelity $\mathcal F_d$ for critical dense polymers.

\subsection[Ratios of partition functions in the limit $M \to \infty$]{Ratios of partition functions in the limit $\boldsymbol{M \to \infty}$}\label{sec:ratiosFd}

To compute the bipartite fidelity, we extract the leading behaviours of the partition functions \eqref{eq:ZS} as $M$ tends to infinity. We start with \eqref{eq:ZSAB}:
\be
Z^{AB}_d = 2^{2MN} \Big(\langle v^A_d| (D^A)^M\otimes \langle v^B_0|(D^B)^M\Big) \big(D^{AB}\big)^M |v^{AB}_d\rangle.
\ee
The state $|v_d\rangle$ can be written using the fermions $a_j^\Tt$ as 
\be
\label{eq:vd}
|v_d\rangle = a_{d+1}^\Tt a_{d+3}^\Tt a_{d+5}^\Tt \cdots a_{N-1}^\Tt |0\rangle.
\ee
The identity matrix restricted to the eigenspace of magnetisation $m = -d/2$ is of the form
\be
\mathbb I\big|_{S^z = -d/2} = |w_{d}\rangle\langle w_{d}| + \dots \quad (d \neq 0),\qquad \quad
\mathbb I\big|_{S^z = 0} = |\hat w_{0}\rangle\langle w_{0}| + |w_{0}\rangle\langle \hat w_{0}| + \dots
\ee
where the next terms involve states that are not groundstates. We therefore have 
\begin{subequations}
\begin{alignat}{2}
&(D^{AB})^M |v_d^{AB}\rangle = (\Lambda_d^{AB})^M |w_d^{AB}\rangle \langle w_d^{AB}|v_d^{AB}\rangle + \dots\,,\\[0.2cm]
&\langle v_d^{A}|(D^{A})^M  = (\Lambda_d^{A})^M \langle v_d^{A}|w_d^{A}\rangle \langle w_d^{A}|  + \dots\,,\\[0.2cm]
&\langle v_0^{B}|(D^{B})^M  = (\Lambda_0^{B})^M \langle v_0^{B}|\hat w_0^{B}\rangle \langle w_0^{B}|  + (\Lambda_0^{B})^{M-1} \underbrace{\langle v_0^{B}|w_0^{B}\rangle}_{=0} \Big(\Lambda_0^B \langle \hat w_0^{B}| +  M f^B \langle w_0^{B}|\Big) + \dots\,,
\end{alignat}
\end{subequations}
where we use the notation $\Lambda_d = \Lambda_d(\frac\pi4)$ and $f = f(\frac\pi4)$. The next-order corrections are exponentially small in $M$ compared to the leading terms. The overlap $\langle v_0^{B}|w_0^{B}\rangle$ vanishes because the state $|w_0\rangle$ contains a fermion $\phi^\Tt$ which anticommutes with $a_j$ for each $j$ and satisfies $\langle0|\phi^\Tt = 0$. We thus find
\be
Z^{AB}_d =2^{2MN} (\Lambda_d^{AB}\Lambda_d^{A}\Lambda_0^{B})^M \langle w_d^{AB}|v_d^{AB}\rangle\, \langle v_d^{A}|w_d^{A}\rangle\, \langle v_0^{B}|\hat w_0^{B}\rangle \, \langle w_d^{A} \otimes w_0^{B} |w_d^{AB}\rangle + \dots \ .
\ee
Repeating the same exercise for $Z^{A\cup B}_d$, $Z^{A}_d$ and $Z^{B}_2$, we find
\be
\label{eq:otherZs}
Z^{A\cup B}_d \simeq 2^{2MN} (\Lambda_d^{AB})^{2M}\langle w_d^{AB}|v_d^{AB}\rangle^2, \quad Z^{A}_d \simeq 2^{2MN_A} (\Lambda_d^{A})^{2M}\langle v_d^{A}|w_d^{A}\rangle^2, \quad Z^{B}_2 \simeq 2^{2MN_B} (\Lambda_2^{B})^{2M}\langle v_2^{B}|w_2^{B}\rangle^2,
\ee
where $\simeq$ denotes an equality up to terms that are exponentially small in $M$ compared to the leading term. We also note that the eigenvalues of the sectors $d=0$ and $d=2$ are equal: $\Lambda_2 = \Lambda_0$. From \eqref{eq:Fd}, the bipartite fidelity $\mathcal F_d$ reads
\be
\label{eq:FdOverlaps}
\mathcal F_d = - \log \bigg( \frac{\langle v_0^{B}|\hat w_0^{B}\rangle}{\langle v_2^{B}|w_2^{B}\rangle} \langle w_d^{A} \otimes w_0^{B} |w_d^{AB}\rangle\bigg)^2.
\ee

\subsection{Fermionic operators in the different subsystems}

In this subsection, we introduce the fermionic operators specific to the subsystems $A$ and $B$ and present the anticommutation relations that they satisfy with the fermions of the full system. The computation of \eqref{eq:FdOverlaps} is carried out using the states
\begin{equation}
\ket{w^{AB}_d}=\eta_{1}^{\Tt} \cdots \eta_{\frac{N-d}{2}}^{\Tt} \ket{0}, \qquad
\ket{w^A_d \otimes w^B_0} =  \eta^{A,{\Tt}}_{1}\cdots \eta^{A,{\Tt}}_{\frac{N_A-d}{2}}\phi^{B,{\Tt}} \eta^{B,{\Tt}}_{1}\cdots \eta^{B,{\Tt}}_{\frac{N_B-2}{2}}\ket{0}
\end{equation}
where
\begin{subequations}
\label{eq:OperatorsSubsystems}
\begin{alignat}{3}
\eta_k^{A} &= \frac{1}{\kappa^A_k}\sum_{j=1}^{N_A-1} \sin ( \tfrac{\pi k j}{N_A} )a_j \, , 
\qquad 
&&\phi^{B} = \omega^{-1}\sqrt{\frac{2}{N_B}} \sum_{j=N_A+1}^{N} \ir^{-(j-N_A-1)} \, c_j \, ,
\\[0.15cm]
\eta_k^{B} &= \frac{1}{\kappa^B_k}\sum_{j=N_A+1}^{N-1} \sin ( \tfrac{\pi k (j-N_A)}{N_B})a_j \, ,
\qquad
&&\kappa^A_k = \sqrt{N_A \cos ( \tfrac{\pi k}{N_A})}, 
\qquad 
\kappa^B_k = \sqrt{N_B \cos ( \tfrac{\pi k}{N_B})}.
\end{alignat}
\end{subequations}
The transpose of these fermions are denoted $\eta_k^{A,{\Tt}}$, $\eta_k^{B,{\Tt}}$ and $\phi^{B,{\Tt}}$.

The overlaps are computed using Wick's theorem. Let $\theta_k$ and $\vartheta_\ell^\Tt$ be fermionic annihilation and creation operators. This theorem states that
\be
\langle 0| \theta_L \dots \theta_2\theta_1 \vartheta^\Tt_1 \vartheta^\Tt_2 \dots \vartheta^\Tt_L |0\rangle  = \det_{k,\ell=1}^{L} \{\theta_k,\vartheta^\Tt_\ell\}.
\ee
The ratio $\langle v_0^{B}|\hat w_0^{B}\rangle/\langle v_2^{B}|w_2^{B}\rangle$ is easily computed using this equality. Both the numerator and the denominator are expressed as determinants, and after a simple argument, we find
\begin{equation}
\label{eq:simple.overlap}
\frac{\langle v_0^{B}|\hat w_0^{B}\rangle}{\langle v_2^{B}|w_2^{B}\rangle} =  \omega^{2}\sqrt{\frac{N_B}{2}}.
\end{equation}
The other overlaps involve fermions of both the full system and the subsystems $A$ and $B$. To compute them, we introduce the rescaled operators
\be
\label{eq:twiddle.eta}
\tilde{\eta}_k = \sum_{j=1}^{N-1} \sin ( \tfrac{\pi k j}{N} )\,a_j, \qquad k = 1, \dots, N-1,
\ee
and likewise for $\tilde{\eta}^A_k$ and $\tilde{\eta}^B_k$, by removing the constants $\kappa_k^A$ and $\kappa_k^B$ in \eqref{eq:OperatorsSubsystems}. These operators have the advantage of being well defined for $k=N/2$. All the fermionic operators appearing in $\braket{w^A_d \otimes w^B_0 | w^{AB}_d}$ can in fact be written in terms of the $\tilde{\eta}_k$. Indeed, we have 
$\tilde{\eta}_k=\kappa_k\eta_k$ and, for $N$ even, $\tilde{\eta}_{\frac{N}{2}} =(\omega \kappa_{\phi})\phi$. The same holds for the operators of the subsystems $A$ and $B$.
With this convention, Wick's theorem yields
\begin{equation}\label{eq:overlap1}
\frac{\langle v_0^{B}|\hat w_0^{B}\rangle}{\langle v_2^{B}|w_2^{B}\rangle}\braket{w^A_d \otimes w^B_0 | w^{AB}_d} = \Bigg(\prod_{k=1}^{(N_A-d)/2}\kappa^A_k \prod_{k'=1}^{(N_B-2)/2}\kappa^B_{k'}\prod_{\ell=1}^{(N-d)/2}\kappa_{\ell}\Bigg)^{-1}   \det M_d 
\end{equation}
where 
\begin{equation}
\left[M_d\right]_{k,\ell} = 
\left\{\begin{array}{ll}
\{\tilde{\eta}_k^A, \tilde{\eta}_\ell^t  \} &\quad k=1,\dots,\frac{N_A-d}{2}, \\[0.15cm] 
\{\tilde{\eta}_{k-\frac{N_A-d}{2}}^B, \tilde{\eta}_\ell^t  \} &\quad k=\frac{N_A-d}{2}+1,\dots,\frac{N-d}{2},
\end{array}\right. \qquad \ell = 1, \dots, \tfrac{N-d}2.
\end{equation}
The anticommutators appearing in this matrix are 
\begin{equation}\label{eq:anticomm_eta}
\{\tilde{\eta}_k^A, \tilde{\eta}_\ell^t  \} = (-1)^{k}\frac{\cos(\frac{\pi \ell}{N})\sin\big(\frac{\pi k}{N_A}\big)\sin(\pi \ell x)}{\cos\big(\frac{\pi k}{N_A}\big)-\cos(\frac{\pi \ell}{N})},\qquad
\{\tilde{\eta}_k^B, \tilde{\eta}_\ell^t  \} = -\frac{\cos(\frac{\pi \ell}{N})\sin\big(\frac{\pi k}{N_B}\big)\sin(\pi \ell x)}{\cos\big(\frac{\pi k}{N_B}\big)-\cos(\frac{\pi \ell}{N})},
\end{equation}
where
\be x = \frac{N_A}{N}\ee
is the aspect ratio.

\subsection{Closed form for the overlaps}

In computing the determinant of $M_d$, we find that all the trigonometric functions appearing in the numerators in \eqref{eq:anticomm_eta} factor out. Because $M_d$ is defined in two parts, this is a non-trivial result. It stems from the fact that the $\ell$-dependent factors in the numerators are identical. In contrast, the same factorisation does not occur for the XX spin chain with free boundary conditions. Exact asymptotics in this case were only obtained for $x=1/2$ in \cite{SD13}. The ratio of overlaps in \eqref{eq:overlap1} becomes
\be\label{eq:overlap.and.det}
\frac{\langle v_0^{B}|\hat w_0^{B}\rangle}{\langle v_2^{B}|w_2^{B}\rangle}\braket{w^A_d \otimes w^B_0 | w^{AB}_d} =
\prod_{k=1}^{\frac{N_A-d}{2}}\frac{\sin\big(\frac{\pi k}{N_A}\big)}{\kappa^A_k}  \prod_{k=1}^{\frac{N_B-2}{2}}\frac{\sin\big(\frac{\pi k}{N_B}\big)}{\kappa^B_{k}}
 \prod_{\ell=1}^{\frac{N-d}{2}}\frac{\cos(\frac{\pi \ell}{N})\sin(\pi \ell x)}{\kappa_\ell}
\times  \det C_d 
\ee
where
\begin{equation}
\left[C_d\right]_{k,\ell} = 
\left\{\begin{array}{cll}
\Big[\cos\big(\frac{\pi k}{N_A}\big)-\cos(\frac{\pi \ell}{N})\Big]^{-1}  \quad &k=1,\dots,\frac{N_A-d}{2}, \\[0.15cm]
\bigg[\cos\Big(\frac{\pi \left(k-(N_A-d)/2\right)}{N_B}\Big)-\cos(\frac{\pi \ell}{N})\bigg]^{-1} \quad &k=\frac{N_A-d}{2}+1,\dots,\frac{N-d}{2}, \\
\end{array}\right. \quad  \ell = 1, \dots, \tfrac{N-d}2.
\end{equation}
The determinant is evaluated using Cauchy's identity:
\be
\label{eq:Cauchy}
\det_{k,\ell =a}^b \Big(\frac{1}{w_k-z_\ell}\Big) = \frac{\prod_{a\le k<\ell\le b} (w_\ell-w_k)(z_k-z_\ell)}{\prod_{k,\ell=a}^b (w_k-z_\ell)}.
\ee
The result is
\begin{alignat}{2}
\det C_d &= 
  \prod_{1 \le k<k'\le \frac{N_A-d}{2}}  \left( \cos\big( \tfrac{\pi k'}{N_A}\big) - \cos\big( \tfrac{\pi k}{N_A}\big)  \right)    
  \prod_{1 \le k<k'\le \frac{N_B}2}  \left( \cos\big( \tfrac{\pi k'}{N_B}\big) - \cos\big( \tfrac{\pi k}{N_B}\big)  \right) \nonumber\\ 
& \times \prod_{k=1}^{\frac{N_A-d}{2}}\prod_{k'=1}^{\frac{N_B}2} 
\left( \cos\big( \tfrac{\pi k'}{N_B}\big) - \cos\big( \tfrac{\pi k}{N_A}\big)  \right)  
  \prod_{1 \le \ell<\ell'\le \frac{N-d}{2}}  \left( \cos\big( \tfrac{\pi \ell}{N}\big) - \cos\big( \tfrac{\pi \ell'}{N}\big)  \right) \nonumber\\
& \times     \prod_{k=1}^{\frac{N_A-d}2}\prod_{\ell = 1}^{\frac{N-d}{2}}
\left( \cos\big( \tfrac{\pi k}{N_A}\big) - \cos\big( \tfrac{\pi \ell}{N}\big)  \right)^{-1}\ 
    \prod_{k=1}^{\frac{N_B}2}\prod_{\ell = 1}^{\frac{N-d}{2}}
    \left( \cos\big( \tfrac{\pi k}{N_B}\big) - \cos\big( \tfrac{\pi \ell}{N}\big)  \right)^{-1}.
\label{eq:BFClosedForm}
\end{alignat}

We note that the equalities \eqref{eq:overlap.and.det} and \eqref{eq:BFClosedForm} hold in the case where $\textrm{gcd}(N_A,N) = 1$ for $N$ odd, and $\textrm{gcd}(\frac{N_A}2,\frac N2) = 1$ for $N$ even. If these conditions are not met, it can happen that some factors in the numerator and denominator are zero. When this happens, the number of zeros in the numerator and denominator are always equal and the correct finite result is obtained by taking a limit on $N_A$ in the generic expressions. In fact, our asymptotic analysis in \cref{sec:Asymptotics} presupposes that $N_A$ and $N$ satisfy these coprimality conditions. The resulting functions are continuous in the aspect ratio $x = N_A/N$, and in the scaling limit, any value $x \in (0,1)$ can be approched with these conditions.

\subsection{Asymptotic expansion}

With the closed-form formulas \eqref{eq:overlap.and.det} and \eqref{eq:BFClosedForm}, we derive the asymptotic expansion of $\mathcal F_d$ for large $N$. The aspect ratio $x$ is set to a constant in the interval $(0,1)$.
The result is stated in the next theorem.
\begin{Theoreme}\label{thm:Fd}
The bipartite fidelity $\mathcal F_d$ has the following $1/N$ expansion:
\begin{alignat}{2}
\mathcal{F}_d =&- \frac{2}{8} \log N +  \left[-\frac{1}{12}\left(2 x-1+\frac2x \right) + \left(\frac43-\frac2x \right)\Delta_{1,d+1} +  \left(\frac43-2x \right)\Delta_{1,d+1}\right]\log(1-x) \nonumber\\
& +\left[-\frac{1}{12}\left(2 (1-x)-1+\frac{2}{1-x} \right) -\frac23\Delta_{1,d+1}+\left(\frac43-2(1-x) \right)\Delta_{1,d+1}\right]\log x-2\log\Upsilon_d(x)\nonumber\\
& +3 \log A -\frac14 \left[ 1+\log\left( \frac{\pi}{8}\right)\right]+ \mathcal{O}\left(N^{-1}\right)
\label{eq:AsymptoticsLBF}
\end{alignat}
where 
\begin{equation}
\label{eq:Delta}
\Upsilon_d(x)=(1-x)^{-\frac23 \Delta_{1,d+1}}x^{\frac{1}{3}\Delta_{1,d+1}}, \qquad \Delta_{r,s} =\frac{(2r-s)^2-1}{8},
\end{equation}
and $A \approx 1.282427$ is the Glaisher-Kinkelin constant. 
\end{Theoreme}\medskip

\noindent The proof is given in \cref{sec:Asymptotics}. This result exactly matches the conformal prediction \eqref{eq:FAB.DS} of St\'ephan and Dubail \cite{SD13} for the bipartite fidelity. The coefficient of the leading $\log N$ term is consistent with the known value $c=-2$ of the central charge for critical dense polymers and the conformal weight $\Delta_2 = 0$ of the field inserted at the end of the slit. It is indeed clear that, in the two-dimensional model defined on the pants geometry (see \cref{fig:loop.config}), there is no change of boundary conditions at the end of the slit, and the weight of the identity field indeed vanishes.

The constant term has precisely the predicted form $f(x)$ given in \eqref{eq:f(x)} for a two-point function of primary fields in CFT. The fields $\varphi_1$ and $\varphi_4$ have the dimension $\Delta_1 = \Delta_4 = \Delta_{1,d+1}$, which is the known value for the boundary changing operator that accounts for the insertion of $d$ defects on a boundary \cite{SB89,MDJ18}. The two other fields are identity fields, with $\Delta_2 = \Delta_3 = 0.$ The function $\Upsilon_d(x)$ has the correct form that ensures that the four-point function \eqref{eq:4ptfct} reduces to the usual power-law formula for the two-point function. The non-universal constant in \eqref{eq:f(x)} is given by
\begin{equation}
C= 3 \log A -\frac14 \left( 1+\log\left( \frac{\pi}{8}\right)\right) = -3\zeta'(-1) - \frac 14 \log\left( \frac{\pi}{8}\right),
\end{equation}
where $\zeta(z)$ is the Riemann zeta function.
We also note that $\mathcal F_d$ has no term proportional to $N^{-1}\log N$. Comparing with the predicted form \eqref{eq:g(x)} of the function $g(x)$, we deduce that the extrapolation length $\Xi$ vanishes in the present case.

\section{Exact results for the bipartite fidelity for logarithmic fields}\label{sec:logF}
 
In this section, we calculate the bipartite fidelity $\tilde{\mathcal{F}}_2$ with lattice and field theoretical approaches. In the following, $N$, $N_A$ and $N_B$ are all even integers.
 
\subsection[Ratios of partition functions in the limit $M \to \infty$]{Ratios of partition functions in the limit $\boldsymbol{M \to \infty}$}\label{sec:ratios.tilde}

We start from \eqref{eq:F2v2} and \eqref{eq:tZS} and apply the ideas used in \cref{sec:ratiosFd} to the computation of $\tilde{\mathcal{F}}_2$. For $\tilde Z^{AB}_2$, we have:
\be
\tilde Z^{AB}_2 = 2^{2MN}\Big( \langle \tilde v^A_2| (D^A)^M\otimes \langle \tilde v^B_0|(D^B)^M\Big) \big(D^{AB}\big)^M |\tilde v^{AB}_2\rangle.
\ee
In this case, $\langle \tilde v^A_2|$, $\langle \tilde v^B_0|$ and $|\tilde v^{AB}_2\rangle$ all have zero magnetisation. We find
\begin{subequations}
\begin{alignat}{2}
&(D^{AB})^M |\tilde v_2^{AB}\rangle = (\Lambda_0^{AB})^M |w_0^{AB}\rangle \langle \hat w_0^{AB}|\tilde v_2^{AB}\rangle\nonumber \\ &\hspace{2.42cm}
+ (\Lambda_0^{AB})^{M-1} \Big(\Lambda_0^{AB}| \hat w_0^{AB}\rangle + M f^{AB}| w_0^{AB}\rangle\Big)\langle w_0^{AB}|\tilde v_2^{AB}\rangle+ \dots\,,\\[0.2cm]
&\langle \tilde v_2^{A}|(D^{A})^M  = (\Lambda_0^{A})^M \langle \tilde v_2^{A}|\hat w_0^{A}\rangle \langle w_0^{A}|  + (\Lambda_0^{A})^{M-1} \langle \tilde v_2^{A}|w_0^{A}\rangle \Big(\Lambda_0^A \langle \hat w_0^{A}| +  M f^A \langle w_0^{A}|\Big)+ \dots\,, \\[0.2cm]
&\langle \tilde v_0^{B}|(D^{B})^M  = (\Lambda_0^{B})^M \langle \tilde v_0^{B}|\hat w_0^{B}\rangle \langle w_0^{B}|  + \dots\,,
\end{alignat}
\end{subequations}
where, as compared with the previous calculation, $\langle w_0^{AB}|\tilde v_2^{AB}\rangle$ and $\langle \tilde v_2^{A}|w_0^{A}\rangle$ are non-zero. In computing $\tilde Z^{AB}_2$, the leading-order term in the large-$M$ expansion vanishes because $\langle w_0^A \otimes w_0^B | w_0^{AB}\rangle = 0$. This occurs because $\langle w_0^A \otimes w_0^B| \phi^\Tt = 0$. We therefore compute the next term in the large-$M$ expansion and find
\begin{alignat}{2}
\tilde Z^{AB}_2 &= 2^{2MN}M (\Lambda_0^{AB}\Lambda_0^{A}\Lambda_0^{B})^M \langle w_0^{AB}|\tilde v_2^{AB}\rangle \langle \tilde v_2^{A}|w_0^{A}\rangle \langle \tilde v_0^{B}|\hat w_0^{B}\rangle\nonumber\\[0.2cm] &\times
\Big(\frac{f^{AB}}{\Lambda^{AB}_0} \langle \hat w_0^A \otimes w_0^B | w_0^{AB}\rangle + \frac{f^{A}}{\Lambda^{A}_0} \langle w_0^A \otimes w_0^B |  \hat w_0^{AB}\rangle\Big) + \dots \ .
\end{alignat}
Repeating the derivations for $\tilde Z^{A\cup B}_2$ and $\tilde Z^{A}_2$, we find
\be
\tilde Z^{A\cup B}_2 = 2^{2MN} (2M) (\Lambda_0^{AB})^{2M-1} f_0^{AB} \langle w_0^{AB}|\tilde v_2^{AB}\rangle^2, \qquad \tilde Z^{A}_2 = 2^{2MN_A}(2M) (\Lambda_0^{A})^{2M-1} f_0^{A} \langle \tilde v_2^{A}|w_0^{A}\rangle^2.
\ee
The function $f(u)$ is computed in \cref{sec:g.states}. Setting $u = \frac \pi 4$, we obtain $f(\frac \pi 4)/\Lambda_0(\frac \pi 4) = -1$. Recalling that the denominator in \eqref{eq:F2v2} involves $Z_2^B$ which is computed in \eqref{eq:otherZs}, we find the following expression for $\tilde {\mathcal F}_2$: 
\be
\label{eq:F2TildeOverlaps}
\tilde{\mathcal F}_2 = - \log \bigg( \frac12\frac{\langle v_0^{B}|\hat w_0^{B}\rangle}{\langle v_2^{B}|w_2^{B}\rangle} \big(\langle \hat w_0^{A} \otimes w_0^{B} |w_0^{AB}\rangle + \langle w_0^{A} \otimes w_0^{B} |\hat w_0^{AB}\rangle\big)\bigg)^2.
\ee
We note that, had $\tilde{\mathcal F}_2$ been defined with $\tilde Z_2^B$ in the denominator instead of $Z_2^B$, the argument of the logarithm in \eqref{eq:F2v2} would have vanished trivially in the limit.

\subsection{Closed form for the overlaps}\label{sec:closed.overlaps.tilde}

To obtain a closed-form expression for $\tilde{\mathcal F}_2$, it is easier to compare with \eqref{eq:FdOverlaps} and compute the difference $\tilde{\mathcal F}_2 - \mathcal F_2$:
\be
\tilde{\mathcal F}_2 - \mathcal F_2 = -2 \log \bigg( \frac12 \frac{\langle \hat w_0^{A} \otimes w_0^{B} |w_0^{AB}\rangle + \langle w_0^{A} \otimes w_0^{B} |\hat w_0^{AB}\rangle}{\langle w_2^{A} \otimes w_0^{B} |w_2^{AB}\rangle}\bigg).
\ee
The three overlaps appearing in the right side are expressed in terms of the fermionic operators as
\begin{subequations}
\label{eq:3.overlaps}
\begin{alignat}{2}
\langle \hat w_0^{A} \otimes w_0^{B} |w_0^{AB}\rangle &= \langle 0| \eta_{\frac{N_B}{2}-1}^B \cdots \eta_1^B \phi^B \eta_{\frac{N_A}{2}-1}^A \cdots \eta_1^A \chi^A \  \phi^{\Tt} \eta_1^{\Tt}\cdots \eta_{\frac N2-1}^{\Tt}|0\rangle,\\[0.1cm]
\langle  w_0^{A} \otimes w_0^{B} |\hat w_0^{AB}\rangle &= \langle 0| \eta_{\frac{N_B}{2}-1}^B \cdots \eta_1^B \phi^B \eta_{\frac{N_A}{2}-1}^A \cdots \eta_1^A \phi^A \  \chi^{\Tt} \eta_1^{\Tt}\cdots \eta_{\frac N2-1}^{\Tt}|0\rangle,\\[0.1cm]
\langle  w_2^{A} \otimes w_0^{B} |w_2^{AB}\rangle &= \langle 0| \eta_{\frac{N_B}{2}-1}^B \cdots \eta_1^B \phi^B \eta_{\frac{N_A}{2}-1}^A \cdots \eta_1^A \eta_1^{\Tt}\cdots \eta_{\frac N2-1}^{\Tt}|0\rangle,
\end{alignat}
\end{subequations}
with
\be
\phi^A =  \omega^{-1}\sqrt{\frac{2}{N_A}} \sum_{j=1}^{N_A} \ir^{-(j-1)} \, c_j  \, ,\qquad
\chi^A =  -\omega\sqrt{\frac{2}{N_A}} \sum_{j=1}^{N_A} \ir^{-(j-1)} \big(\big\lfloor \tfrac j2 \big\rfloor- \tfrac{N_A}{4}\big) \, c_j.
\ee
The other fermionic operators for the subsystems $A$ and $B$ are defined in \eqref{eq:OperatorsSubsystems}. We have the anticommutation relations
\begin{subequations}
\begin{alignat}{4}
\{\eta_{j}^A, \phi^{\Tt} \}= 0, \qquad && \{\chi^A, \phi^{\Tt} \}&= \sqrt{x},\qquad &&\{\eta_k^B, \phi^{\Tt} \}= 0, \qquad && \{\phi^B, \phi^{\Tt} \}= 0,
\\[0.15cm]
\{\eta_j^A, \chi^{\Tt}\}=0, \qquad && \{\phi^A, \chi^{\Tt}\} &= \sqrt{x}, \qquad  &&\{\eta_k^B, \chi^{\Tt}\}=0,\qquad &&\{\phi^B, \chi^{\Tt}\} =(-1)^{\frac{N_A}{2}} \sqrt{1-x}.
\end{alignat}
\end{subequations}
Recalling that $N_A$ is even, we use these anticommutation relations to rewrite the first two overlaps in \eqref{eq:3.overlaps} as
\begin{subequations}
\begin{equation}
\langle \hat w_0^{A} \otimes w_0^{B} |w_0^{AB}\rangle =\sqrt{x} \ \langle 0| \eta_{\frac{N_B}{2}-1}^B \cdots \eta_1^B \phi^B \eta_{\frac{N_A}{2}-1}^A \cdots \eta_1^A \ \eta_1^{\Tt}\cdots \eta_{\frac N2-1}^{\Tt}|0\rangle,
\end{equation}
\begin{equation}
\begin{split}
\langle  w_0^{A} \otimes w_0^{B} |\hat w_0^{AB}\rangle = \sqrt{x} \ &\langle 0| \eta_{\frac{N_B}{2}-1}^B \cdots \eta_1^B \phi^B \eta_{\frac{N_A}{2}-1}^A \cdots \eta_1^A \ \eta_1^{\Tt}\cdots \eta_{\frac N2-1}^{\Tt}|0\rangle \\
+ \sqrt{1-x} \ &\langle 0| \eta_{\frac{N_B}{2}-1}^B \cdots \eta_1^B  \eta_{\frac{N_A}{2}-1}^A \cdots \eta_1^A \phi^A \ \eta_1^{\Tt}\cdots \eta_{\frac N2-1}^{\Tt}|0\rangle,
\end{split}
\end{equation}
\end{subequations}
and therefore
\begin{subequations}
\begin{alignat}{2}
\frac{\langle \hat w_0^{A} \otimes w_0^{B} | w_0^{AB}\rangle}{\langle  w_2^{A} \otimes w_0^{B} | w_2^{AB}\rangle} &= \sqrt x, \\[0.15cm]
\frac{\langle  w_0^{A} \otimes w_0^{B} |\hat w_0^{AB}\rangle}{\langle  w_2^{A} \otimes w_0^{B} | w_2^{AB}\rangle} &= \sqrt x + (-1)^{\frac N2}\frac{(1-x)}{\sqrt x} \frac{\langle 0| \tilde{\eta}_{\frac{N_B}{2}-1}^B \cdots \tilde{\eta}_1^B \tilde{\eta}_{\frac{N_A}{2}}^A \cdots \tilde{\eta}_1^A \ \tilde{\eta}_1^{\Tt}\cdots \tilde{\eta}_{\frac N2-1}^{\Tt}|0\rangle}{\langle 0| \tilde{\eta}_{\frac{N_B}{2}}^B \cdots \tilde{\eta}_1^B \tilde{\eta}_{\frac{N_A}{2}-1}^A \cdots \tilde{\eta}_1^A \ \tilde{\eta}_1^{\Tt}\cdots \tilde{\eta}_{\frac N2-1}^{\Tt}|0\rangle}.\label{eq:last.overlap}
\end{alignat}
\end{subequations}
The right side of this last equality is written in terms of the rescaled operators $\tilde{\eta}_k$ defined in \eqref{eq:twiddle.eta}. To evaluate the remaining ratio, we use the identity 
\be
\tilde \eta^A_{\frac {N_A}2}  = \tilde \eta_{\frac N 2} - (-1)^{\frac {N_A} 2} \tilde \eta^B_{\frac {N_B}2}
\ee
in the numerator. The first term with $\tilde \eta_{N/2}$ vanishes because this operator anticommutes with all the other operators. For the second term, by anticommuting $\tilde \eta^B_{N_B/2}$ towards the left across the operators $\tilde \eta_k^{B}$, we find that the numerator is proportional to the denominator, with the overall factor $(-1)^{ N/2}$. The result is
\be
\frac{\langle  w_0^{A} \otimes w_0^{B} |\hat w_0^{AB}\rangle}{\langle  w_2^{A} \otimes w_0^{B} | w_2^{AB}\rangle} = \sqrt x + \frac{(1-x)}{\sqrt x} = \frac1{\sqrt x}.
\ee
Putting the results together, we obtain the following theorem.
\begin{Theoreme}
\label{thm:F2.tilde}
The bipartite fidelity $\tilde{\mathcal F}_2$ satisfies the relation
\begin{equation}
\label{eq:FTildeLattice}
\tilde{\mathcal{F}}_2-\mathcal{F}_2=-2 \log \left( \frac{1+x}{2 \sqrt{x}}\right).
\end{equation}
\end{Theoreme}
\noindent Interestingly, this result holds for finite values of $N$.

\subsection{CFT derivation of the universal behaviour}

In this subsection, we derive the conformal prediction for the bipartite fidelity for logarithmic fields. We consider a logarithmic conformal field theory at an arbitrary value $c$ of the central charge that includes a pair $(\varphi, \omega)$ of boundary fields of conformal weight $\Delta$ in a rank-two Jordan cell, where $\varphi$ is primary and $\omega$ is its Jordan partner. 
On the upper-half plane $\mathbb H$, the correlation functions for these fields are
\begin{equation}
\label{eq:CorrelationsUHP}
\langle \omega(z_0)\omega(z_1)\rangle_{\mathbb{H}} = \frac{\lambda_1\lambda_2-4 \lambda_0 \lambda_1 \log (z_0-z_1)}{(z_0-z_1)^{2 \Delta}},\qquad
\langle \omega(z_0)\varphi(z_1)\rangle_{\mathbb{H}} =\frac{ \lambda_1}{(z_0-z_1)^{2 \Delta}},\qquad
\langle \varphi(z_0)\varphi(z_1)\rangle_{\mathbb{H}}=0.
\end{equation}
Under a conformal transformation $y = y(z)$, the fields $\varphi$ and $\omega$ satisfy the transformation laws \cite{F03}
\begin{equation}
\label{eq:ConfTransFields}
\varphi(y) \rightarrow \Big(\frac{\dd y}{\dd z} \Big)^{-\Delta} \varphi(z), \qquad
\omega(y) \rightarrow \Big(\frac{\dd y}{\dd z} \Big)^{-\Delta} \left( \omega(z) - 2 \lambda_0 \varphi(z)\log \Big ( \frac{\dd y}{\dd z}\Big)\right).
\end{equation}

For the model of critical dense polymers, the central charge is $c=-2$. In \cite{MDJ18}, it is argued that the boundary field that accounts for the insertion of two adjacent defects forced to connect together is a logarithmic field $\omega(z)$ of conformal dimension $\Delta = 0$. In contrast, the field that inserts a simple arc on the boundary is identified with the identity field $\varphi(z)$. It also has weight $\Delta = 0$.

\begin{figure}[h!]
\begin{center}
\begin{pspicture}[shift=-2.0](-3,-0.5)(3,3.5)
\psframe[fillstyle=solid,fillcolor=lightlightblue,linewidth=0pt,linecolor=white](-3,0)(3,3)
\psline[linewidth=1pt]{-}(-3,0)(3,0)
\psarc[fillstyle=solid,fillcolor=black](0.0,0){0.06}{0}{360}\rput(0.0,-0.3){$_{0}$}
\psarc[fillstyle=solid,fillcolor=black](0.6,0){0.06}{0}{360}\rput(0.6,-0.3){$_{x}$}
\psarc[fillstyle=solid,fillcolor=black](1.6,0){0.06}{0}{360}\rput(1.6,-0.3){$_{1}$}
\psarc[fillstyle=solid,fillcolor=black](-1.6,0){0.06}{0}{360}\rput(-1.6,-0.3){$_{-1}$}
\rput(-3,-0.3){$_{-\infty}$}\rput(3,-0.3){$_{\infty}$}
\rput(-3.2,3.3){$z$}
\rput(0,3.3){$\mathbb H$}
\end{pspicture}
\hspace{2.5cm}
\begin{pspicture}[shift=-2.0](-3,-0.5)(3,3.5)
\psframe[fillstyle=solid,fillcolor=lightlightblue,linewidth=0pt,linecolor=white](-3,0)(3,3)
\psline[linewidth=1pt]{-}(-3,0)(3,0)
\psline[linewidth=1pt]{-}(-3,3)(3,3)
\psline[linewidth=1pt]{-}(-3,1.2)(0,1.2)
\psarc[fillstyle=solid,fillcolor=black](0.0,0){0.06}{0}{360}\rput(0.0,-0.3){$_{0}$}
\psarc[fillstyle=solid,fillcolor=black](0.0,1.2){0.06}{0}{360}\rput(0.0,1.5){$_{f(x)}$}
\rput(-3,-0.3){$_{f(1^+)}$}\rput(3,-0.3){$_{f(\infty)}$}
\rput(-3,0.9){$_{f(1^-)}$}
\rput(-3,1.5){$_{f(0^+)}$}
\rput(-3,2.7){$_{f(0^-)}$}
\rput(3,2.7){$_{f(-\infty)}$}
\psline[linewidth=0.5pt]{<->}(1.5,0.1)(1.5,2.9)\rput(1.8,1.5){$N$}
\psline[linewidth=0.5pt]{<->}(-1.5,1.3)(-1.5,2.9)\rput(-1.1,2.1){$Nx$}
\rput(-3.2,3.3){$y$}
\rput(0,3.3){$\mathbb D_{AB}$}
\end{pspicture}\vspace{1.0cm}

\begin{pspicture}[shift=-2.0](-3,-0.5)(3,3.5)
\psframe[fillstyle=solid,fillcolor=lightlightblue,linewidth=0pt,linecolor=white](-3,0)(3,3)
\psline[linewidth=1pt]{-}(-3,0)(3,0)
\psline[linewidth=1pt]{-}(-3,3)(3,3)
\psarc[fillstyle=solid,fillcolor=black](0.0,0){0.06}{0}{360}\rput(0.0,-0.3){$_{0}$}
\rput(3,-0.3){$_{g_N(\infty)}$}
\rput(-3,-0.3){$_{g_N(0^+)}$}
\rput(-3,2.7){$_{g_N(0^-)}$}
\rput(3,2.7){$_{g_N(-\infty)}$}
\psline[linewidth=0.5pt]{<->}(1.5,0.1)(1.5,2.9)\rput(1.8,1.5){$N$}
\rput(-3.2,3.3){$y$}
\rput(0,3.3){$\mathbb D_{A\cup B}$}
\end{pspicture}
\hspace{2.5cm}
\begin{pspicture}[shift=-1.4](-3,-0.5)(3,2.1)
\psframe[fillstyle=solid,fillcolor=lightlightblue,linewidth=0pt,linecolor=white](-3,0)(3,1.8)
\psline[linewidth=1pt]{-}(-3,0)(3,0)
\psline[linewidth=1pt]{-}(-3,1.8)(3,1.8)
\psarc[fillstyle=solid,fillcolor=black](0.0,0){0.06}{0}{360}\rput(0.0,-0.3){$_{0}$}
\rput(3,-0.3){$_{g_{Nx}(\infty)}$}
\rput(-3,-0.3){$_{g_{Nx}(0^+)}$}
\rput(-3,1.5){$_{g_{Nx}(0^-)}$}
\rput(3,1.5){$_{g_{Nx}(-\infty)}$}
\psline[linewidth=0.5pt]{<->}(1.5,0.1)(1.5,1.7)\rput(1.9,0.9){$Nx$}
\rput(-3.2,2.1){$y$}
\rput(0,2.1){$\mathbb D_{A}$}
\end{pspicture}
\end{center}
\medskip

  \caption{The upper-half plane $\mathbb H$ is mapped to the three domains $\mathbb D_{AB}$, $\mathbb D_{A \cup B}$ and $\mathbb D_A$ via the transformations $y=f(z)$, $y=g_N(z)$ and $y=g_{Nx}(z)$.}
  \label{fig:domains}
  \end{figure}
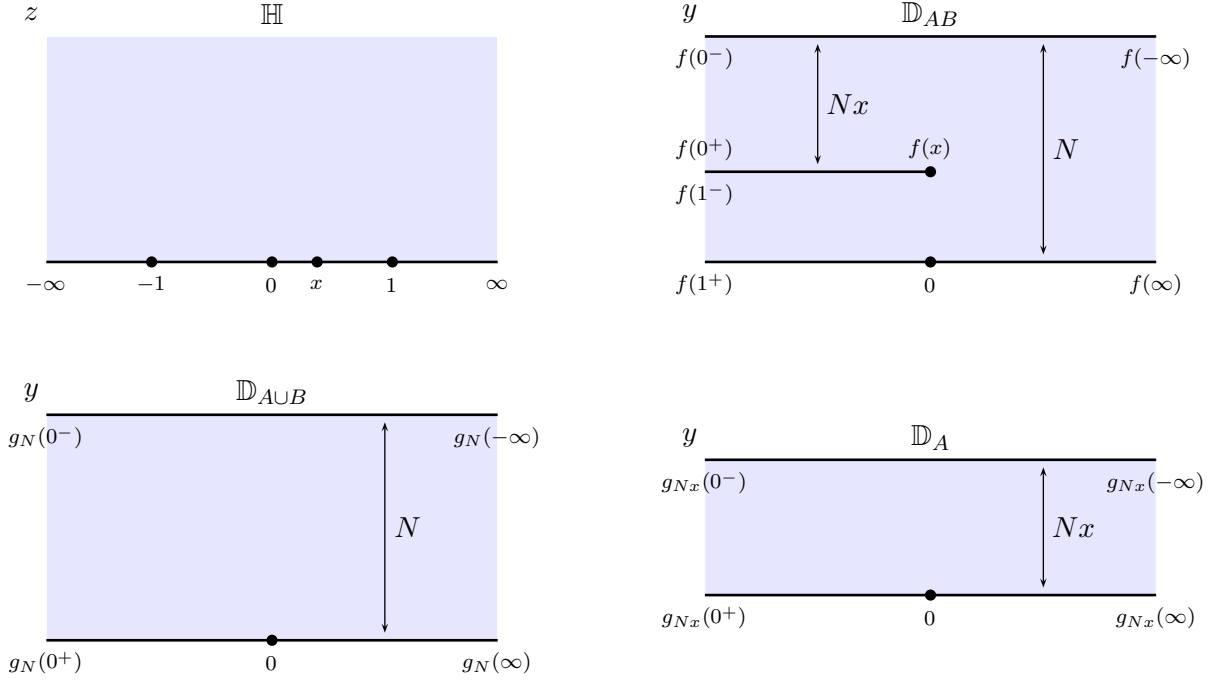

The lattice calculation considered in \cref{sec:ratios.tilde,sec:closed.overlaps.tilde} then corresponds to a ratio of two-point correlation functions of these fields. Indeed, in the context of logarithmic CFT, the bipartite fidelities $\mathcal F_\Delta$ and $\mathcal {\tilde F}_\Delta$ are defined similarly to their lattice analogs \eqref{eq:Fd} and \eqref{eq:F2v2}. They are expressed in terms of partition functions defined on four domains $\mathbb{D}_{AB}$, $\mathbb{D}_{A \cup B}$, $\mathbb{D}_{A}$ and $\mathbb{D}_{B}$, each with specific fields inserted at two ends of the domains. The domains $\mathbb{D}_{A \cup B}$, $\mathbb{D}_A$ and $\mathbb{D}_B$ are infinite horizontal strips where the lower edge is the real axis and the upper edge corresponds to the line with imaginary parts $N$, $N x$ and $N (1-x)$ respectively. The domain $\mathbb{D}_{AB}$ is the pants domain. It has the same upper and lower edges as $\mathbb{D}_{A \cup B}$ but has an added separation between the subsystems $A$ and $B$ in the form of a half-line from $\ir N(1-x)$ to $-\infty + \ir N(1-x)$. The domains $\mathbb{D}_{AB}$, $\mathbb{D}_{A \cup B}$, $\mathbb{D}_{A}$ are depicted in \cref{fig:domains}, along with the upper-half plane that serves as the reference domain for the calculation. The bipartite fidelities $\mathcal F_\Delta$ and $\mathcal {\tilde F}_\Delta$ are then defined as
\be
\mathcal F_\Delta = -2 \log \bigg|\frac{Z_\Delta^{AB}}{(Z_\Delta^{A\cup B}Z_\Delta^{A}Z_\Delta^{B})^{1/2}}\bigg|, \qquad
\mathcal {\tilde F}_\Delta= -2 \log  \bigg|\frac{\tilde Z_\Delta^{AB}}{(\tilde Z_\Delta^{A\cup B}\tilde Z_\Delta^{A}Z_\Delta^{B})^{1/2}}\bigg|.
\ee
In these expressions, $Z_\Delta$ corresponds to a two-point function of the primary field with its logarithmic partner, $\langle\varphi(y_0)\omega(y_1) \rangle$, whereas $\tilde Z_\Delta$ is the correlation function of the logarithmic field with itself, $\langle\omega(y_0)\omega(y_1) \rangle$. In each case, the points $y_0$ and $y_1$ are assigned to the top boundary of the strip, namely 
\be
\textrm{Im}(y_0) = \textrm{Im}(y_1) = \left\{\begin{array}{cl}
N & \textrm{on } \mathbb D_{AB} \textrm{ and } \mathbb D_{A\cup B},\\[0.15cm]
Nx & \textrm{on } \mathbb D_A,\\[0.15cm]
N(1-x) & \textrm{on } \mathbb D_B.\\
\end{array}\right.
\ee
The real parts of $y_0$ and $y_1$ are set to $-m$ and $+m$ and are sent to $- \infty$ and $+\infty$ in the limit. The corresponding expressions are expected to be well-behaved as $m \to \infty$ and to reproduce the results obtained from the lattice.

To reproduce the results of \cref{sec:closed.overlaps.tilde}, we investigate the difference $\mathcal {\tilde F}_\Delta - \mathcal F_\Delta$. Its explicit expression in terms of two-point correlation functions is
\begin{subequations}
\label{eq:DiffF2asLim}
\begin{alignat}{2}
\mathcal {\tilde F}_\Delta - \mathcal F_\Delta &= -2 \log |\Gamma_\Delta|,  \\
\Gamma_\Delta& = \lim_{m\rightarrow\infty}  \frac{\langle \omega(-m+\ir N)\omega(m+\ir N)\rangle_{\mathbb{D}_{AB}} }{\langle \omega(-m+\ir N)\varphi(m+\ir N)\rangle_{\mathbb{D}_{AB}}} \nonumber \\ 
& \times \left(  \frac{\langle \omega(-m+\ir Nx)\varphi(m+\ir Nx)\rangle_{\mathbb{D}_{A}} }{\langle \omega(-m+\ir Nx)\omega(m+\ir Nx)\rangle_{\mathbb{D}_{A}}}\frac{\langle \omega(-m+\ir N)\varphi(m+\ir N)\rangle_{\mathbb{D}_{A \cup B}} }{\langle \omega(-m+\ir N)\omega(m+\ir N)\rangle_{\mathbb{D}_{A \cup B}}}\right)^{1/2}.
\end{alignat}
\end{subequations}
It is independent of the choice of reference partition function on the domain $\mathbb D_B$. 

The map that sends $\mathbb{H}$ to a horizontal strip of height $h$ is
\begin{equation}
\label{eq:gh}
g_h(z)= \frac{h}{\pi}\log z.
\end{equation}
Likewise, the map that sends $\mathbb{H}$ to $\mathbb{D}_{AB}$ is
\begin{equation}
\label{eq:f}
f(z)=\frac{N}{\pi}\big( x \log z +(1-x)\log(z-1)\big)- \frac{N}{\pi}\big(x \log x + (1-x)\log (1-x)\big).
\end{equation}
Along with the known correlation functions \eqref{eq:CorrelationsUHP} on $\mathbb H$ and the transformation laws \eqref{eq:ConfTransFields}, these maps allow us to compute the two-point functions on $\mathbb D_{AB}$, $\mathbb D_{A\cup B}$ and $\mathbb D_{A}$.

\paragraph{Correlation functions on the horizontal strips.} We start by computing the correlation functions on $\mathbb{D}_{A}$. With $y= g_{Nx}(z)$ and $y' = \frac{N x}{\pi z}$, we find
\begin{equation}
\langle \omega(y_0) \varphi(y_1)\rangle_{\mathbb{D}_{A}} =\big(g'_{Nx}(z_0)\big)^{-\Delta}\big(g'_{Nx}(z_1)\big)^{-\Delta} \langle \omega(z_0) \varphi(z_1)\rangle_{\mathbb{H}}\\
 =\left(\frac{z_0z_1 \pi^2}{N^2 x^2 
}\right)^{\Delta}\frac{ \lambda_1}{(z_0-z_1)^{2 \Delta}}
\end{equation}
and
\begin{alignat}{2}
\langle \omega(y_0) \omega(y_1)\rangle_{\mathbb{D}_{A}} &= \big(g'_{Nx}(z_0)\big)^{-\Delta}\big(g'_{Nx}(z_1)\big)^{-\Delta}\nonumber\\[0.15cm]
&\times
\bigg[\langle \omega(z_0) \omega(z_1)\rangle_{\mathbb{H}} - 2\lambda_0 \left( \log\bigg( \frac{N x}{\pi z_0}\bigg) \langle \varphi(z_0) \omega(z_1)\rangle_{\mathbb{H}} +\log\bigg( \frac{N x}{\pi z_1}\bigg) \langle \omega(z_0) \varphi(z_1)\rangle_{\mathbb{H}}\right)\bigg]\nonumber\\
&\hspace{-2.0cm}=\left(\frac{z_0z_1 \pi^2}{N^2x^2
}\right)^{\Delta}\frac{ \lambda_1}{(z_0-z_1)^{2 \Delta}}\left( -4 \lambda_0  \log ( z_0-z_1)+\lambda_2 -4 \lambda_0 \log \left(\frac{N x}{\pi} \right) +2 \lambda_0  \big(\log (z_0) + \log (z_1)\big)\right) \nonumber\\
&\hspace{-2.0cm}=\langle \omega(y_0) \varphi(y_1)\rangle_{\mathbb{D}_{A}}\left( -4 \lambda_0 \log \big( \eE^{\frac{\pi y_0}{N x}}-\eE^{\frac{\pi y_1}{N x}}\big)+\lambda_2 -4 \lambda_0  \log \left(\frac{N x}{\pi} \right) + 2\lambda_0 \frac{\pi}{N x}
(y_0+y_1)
\right).
\end{alignat}
Hence, the ratio is
\begin{equation}
\label{eq:CorrFctDA}
 \frac{\langle \omega(-m+\ir Nx)\omega(m+\ir Nx)\rangle_{\mathbb{D}_{A}}}{\langle \omega(-m+\ir Nx)\varphi(m+\ir Nx)\rangle_{\mathbb{D}_{A}} } = -4 \lambda_0  \log \left(  2 \sinh\left( \frac{\pi m}{Nx}\right)\right)+\lambda_2 -4 \lambda_0  \log \left(\frac{N x}{\pi} \right).
\end{equation}
The same ratio of correlation functions on $\mathbb{D}_{A \cup B}$ is obtained by setting $x=1$ in \eqref{eq:CorrFctDA}:
\begin{equation}
\label{eq:CorrFctDAUB}
 \frac{\langle \omega(-m+\ir N)\omega(m+\ir N)\rangle_{\mathbb{D}_{A\cup B}}}{\langle \omega(-m+\ir N)\varphi(m+\ir N)\rangle_{\mathbb{D}_{A \cup B}} } = -4 \lambda_0  \log  \left(2 \sinh\left( \frac{\pi m}{N}\right)\right)+\lambda_2 -4 \lambda_0  \log \left(\frac{N }{\pi} \right).
\end{equation}
The leading-order terms in the large-$m$ expansion are linear:
\begin{subequations}
\label{eq:CorrFct.leading}
\begin{alignat}{2}
 \frac{\langle \omega(-m+\ir Nx)\omega(m+\ir Nx)\rangle_{\mathbb{D}_{A}}}{\langle \omega(-m+\ir Nx)\varphi(m+\ir Nx)\rangle_{\mathbb{D}_{A}} } &= - 4 \lambda_0 \frac{\pi m}{N x} + \mathcal O(1),
\\[0.15cm]
  \frac{\langle \omega(-m+\ir Nx)\omega(m+\ir Nx)\rangle_{\mathbb{D}_{A\cup B}}}{\langle \omega(-m+\ir Nx)\varphi(m+\ir Nx)\rangle_{\mathbb{D}_{A \cup B}} } &= - 4 \lambda_0 \frac{\pi m}{N} + \mathcal O(1).
\end{alignat}
\end{subequations}

\paragraph{Correlation functions on the pants domain.} We perform the same calculations on the domain $\mathbb D_{AB}$. With $y =f(z)$ and $y'=\frac{N}{\pi}\big(\frac{z-x}{z(z-1)}\big)$, we find
\begin{alignat}{2}
\langle \omega(y_0) \varphi(y_1)\rangle_{\mathbb{D}_{AB}} &=f'(z_0)^{-\Delta}f'(z_1)^{-\Delta} \langle \omega(z_0) \varphi(z_1)\rangle_{\mathbb{H}}\nonumber\\[0.15cm]
&=\left( \frac{\pi^2}{N^2} \frac{z_0(z_0-1)}{z_0-x} \frac{z_1(z_1-1)}{z_1-x}\right)^{\Delta} \frac{ \lambda_1}{(z_0-z_1)^{2 \Delta}}.
\end{alignat}
A calculation similar to the one for the strip domains gives 
\begin{equation}
\label{eq:CorelationwwPant}
\begin{split}
\frac{\langle \omega(y_0) \omega(y_1)\rangle_{\mathbb{D}_{AB}}} {\langle \omega(y_0) \varphi(y_1)\rangle_{\mathbb{D}_{AB}}} =&  -4 \lambda_0 \log ( z_0-z_1)+\lambda_2  - 4 \lambda_0\log \left( \frac{N}{\pi}\right) - 2 \lambda_0 \log \left( \frac{z_0-x}{z_0(z_0-1)}  \frac{z_1-x}{z_1(z_1-1)} \right).
\end{split}
\end{equation}
The next step is to express the right side in terms of $y_0 = -m +\ir N$ and $y_1 = m + \ir N$. Obtaining a closed-form expression for $z(y)$ is not possible, and we instead resort to series expansions for large~$m$. In this regime, the points $z_0$ and $z_1$ approach the values $0^-$ and $-\infty$ respectively. We find
\begin{subequations}
\label{eq:z.expansions}
\begin{alignat}{2}
z_0 &= -\eE^{-\frac{\pi (m+a)}{Nx}} + \Big(\frac {1-x} x\Big) \eE^{-\frac{2\pi (m+a)}{Nx}} + \mathcal O \Big(\eE^{-\frac{3\pi (m+a)}{Nx}}\Big),\\[0.15cm]
z_1 &=-\eE^{\frac{\pi (m-a)}{N}} + (x-1) + \mathcal O\Big(\eE^{-\frac{\pi (m-a)}{N}}\Big),
\end{alignat}
\end{subequations}
where 
\be
a =- \frac N \pi \big(x \log x + (1-x) \log (1-x)\big)
\ee
is the additive constant appearing in \eqref{eq:f}. It turns out that computing $\mathcal {\tilde F}_2 - \mathcal F_2$ only requires the leading-order terms in \eqref{eq:z.expansions}. Inserting the series expansions for $z_0$ and $z_1$ in \eqref{eq:CorelationwwPant}, we again find a leading term that is linear in $m$:
\begin{equation}
\label{eq:CorrFctDAB}
\frac{\langle \omega(-m+\ir N)\omega(m+\ir N)\rangle_{\mathbb{D}_{AB}} }{\langle \omega(-m+\ir N)\varphi(m+\ir N)\rangle_{\mathbb{D}_{AB}}} = - 2 \lambda_0 \frac{\pi m} N \bigg( \frac{1+x}x \bigg) + \mathcal O(1).
\end{equation}

\paragraph{Final result.} We combine \eqref{eq:DiffF2asLim}, \eqref{eq:CorrFct.leading} and \eqref{eq:CorrFctDAB} and find
\be
\mathcal {\tilde F}_\Delta - \mathcal F_\Delta = -2 \log \bigg(\frac{1+x}{2 \sqrt{x}}\bigg).
\ee
This is precisely identical to the lattice result \eqref{eq:FTildeLattice}! Remarkably, the final expression for this difference is independent of the value of the central charge $c$ and of the conformal weight $\Delta$ of the two fields. It also does not depend on the constants $\lambda_0$, $\lambda_1$ and $\lambda_2$ that appear in the two-point function.

\section{Conclusion}\label{sec:Conclusion}

In this paper, we defined and investigated two instances of the bipartite fidelity for the model of critical dense polymers: $\mathcal F_d$ and $\mathcal {\tilde F}_2$. We obtained closed-form expressions at finite lattice size $N$ and computed the leading terms of their asymptotic $1/N$ expansions as a function of the aspect ratio $x = N_A/N$. In both cases, the leading term is $\frac{(-2)}{8} \log N$, which is consistent with the known value $c=-2$ of the central charge for the model of critical dense polymers. The next leading term is constant. For $\mathcal F_d$, we find that it has precisely the form \eqref{eq:f(x)} predicted by St\'ephan and Dubail, with the conformal weights of the field that accounts for the insertion of $d$ defects on the boundary equal to $\Delta_{1,d+1} = \frac{d(d-2)}8$. This is consistent with earlier known results for the weights of these fields \cite{SB89,MDJ18}. The constant term for $\mathcal {\tilde F}_2$ was also computed; it differs from the value of the same constant for $\mathcal F_2$ by the simple factor $-2\log((1+x)/(2\sqrt{x}))$. This lattice result was confirmed by a CFT argument, with the simple factor understood to be universal. Finally, the next-leading term predicted by St\'ephan and Dubail, proportional to $N^{-1}\log N$, is zero for both $\mathcal F_d$ and $\mathcal {\tilde F}_2$, indicating that the (non-universal) extrapolation length $\Xi$ in \eqref{eq:g(x)} vanishes. A similar vanishing of the $N^{-1}\log N$ term was previously observed in \cite{HL17} for the XXZ spin chain at the combinatorial point. In this case, the authors argue that the central charge and the conformal weights are such that the expression inside the bracket in \eqref{eq:g(x)} vanishes, thus preventing a measure of the extrapolation length. 

The extrapolation length plays a prominent role in the study of surface critical phenomena \cite{Diehl97}. In the context of CFT, it is a non-universal constant appearing in a perturbation of the stress-energy tensor that displaces the position of the boundary \cite{SD13}. It also impacts the behaviour of one-point correlation functions near a corner. For the XX chain with no boundary magnetic fields, Dubail and St\'ephan computed the $N^{-1}\log N$ contribution to the bipartite fidelity and found $\Xi = 1$. In this case, the momenta of the fermions are of the form $\frac{\pi j}{N+1}$. In contrast, the fermionic momenta for the chain investigated in the current paper are of the form $\frac{\pi j}{N}$, leading to a vanishing extrapolation length. It is then tempting to conjecture that, for momenta of the form $\frac{\pi j}{N+\delta}$, the quantity $N+\delta$ is an effective length for the system from which the extrapolation length can be read directly. To test this, it would be interesting to investigate the more general case where a field $h$ is applied to the endpoints of the spin chain.

Few observables allow one to measure directly the central charge $c$ of a conformally invariant model. The finite-size corrections to the eigenvalues of the transfer matrix \cite{BCN86,A86}, in particular, only allow a measurement of the effective central charge, $c_{\textrm{eff}}=c-24 \Delta_{\textrm{min}}$, where $\Delta_{\textrm{min}}$ is the smallest conformal weight in the subsector of the theory corresponding to the boundary conditions that are considered. For the model of critical dense polymers, the effective central charge is $c_{\textrm{eff}}=-2$ for $N$ even and $c_{\textrm{eff}}=1$ for $N$ odd. The results of this paper thus suggest that the bipartite fidelity indeed allows one to measure $c$ and not just $c_{\textrm{eff}}$. In parallel, for the entanglement entropy, it was initially argued that for such theories, $S_A$ only allows one to measure the effective central charge \cite{BCDLR15}. However it was later understood \cite{CJS17} that in these cases, because the Hamiltonian $H$ is not hermitian, $S_A$ can be defined in two ways, depending on whether the state $\langle \phi |$ in \eqref{eq:SA} is taken to be the left groundstate of $H$ or the hermitian transpose of the right groundstate. The leading coefficient in \eqref{eq:SA.leading} is $c$ in the former case and $c_{\textrm{eff}}$ in the latter case. In the present paper, the state $\langle \phi^A \otimes \phi^B |$ appearing in \eqref{eq:LBF}
is the left eigenstate of $D^A(u) \otimes D^B(u)$. This is automatically built-in from our choice to define $\mathcal F_d$ in \eqref{eq:Fd} in terms of partition functions of the loop model. Presumably, by repeating our derivation, but with $\langle \phi^A \otimes \phi^B |$ equal to $| \phi^A \otimes \phi^B \rangle^{\dagger}$, one would find a $1/N$ expansion with the first term proportional to the effective central charge.

In terms of the representation theory of the Temperley-Lieb algebra, our definition of $\mathcal F_d$ corresponds to attaching the standard representation $\mathsf V_{N,d}$ at two ends of the pants lattice, and the standard representation $\mathsf V_{N,0}$ at the third end. Likewise, $\mathcal {\tilde F}_2$ corresponds to attaching the projective representation $\mathsf P_{N,2}$ on two ends, and $\mathsf V_{N,0}$ on the third. The zoo of representations of the Temperley-Lieb algebra at $\beta = 0$ includes many more projective representations, as well as zig-zag modules \cite{W95}. One may of course wonder how the asymptotic behaviour of the bipartite fidelity is modified if these other representations are attached to the three ends of the pants geometry. In particular, by attaching non-vacuum representations to all three ends, it may be possible to use the bipartite fidelity to compute the structure constants of the conformal three-point functions. It remains to be seen whether such a computation is again expressible as a matrix entry in the spin-chain representation of $\tl_N(\beta)$.

We conclude by noting that many generalisations of our derivations are worthy of investigation. One could for instance repeat the calculation for the model of polymers in the framework of the one-boundary Temperley-Lieb algebra, wherein the loops touching the boundary are given a weight $\beta'$. The case of the same model on a lattice with periodic boundary conditions is also worthwhile: St\'ephan and Dubail gave a conjecture for the leading universal behaviour for this case \cite{DS11}, and it would allow us to elucidate the question of whether the central charge of the logarithmic minimal models on periodic geometries depends on the fugacity $\alpha$ of the non-contractible loops. One could also try to repeat our calculation away from $\beta = 0$, and we expect that the calculation of the overlaps in this case will be feasible using algebraic Bethe ansatz techniques.

\subsection*{Acknowledgments}

GP, AMD and PR are all Research Fellows of the Fonds de la Recherche Scientifique -- FNRS: GP is supported by the Aspirant fellowship FC23367, AMD is a Postdoctoral Researcher under the project CR28075116 and PR is a Senior Research Associate. They also acknowledge support from the EOS contract O013018F. The authors thank J.~Li\'enardy for carefully reading the manuscript and C.~Hagendorf for useful discussions.

%
%

\bigskip
\bigskip
\appendix
%

\section{Asymptotics}\label{sec:Asymptotics}

In this appendix, we give the computational details needed to obain the asymptotic expansion \eqref{eq:AsymptoticsLBF} of the logarithmic bipartite fidelity from the closed formulas \eqref{eq:FdOverlaps}, \eqref{eq:overlap.and.det} and \eqref{eq:BFClosedForm}. The main ingredients used are: 
\begin{itemize}
\item[(i)] the Euler-Maclaurin formula:
\be
\label{eq:EmL}
\sum_{i=a}^b f(i) = \int_a^b f(x) dx + \frac{f(a) + f(b)}{2} + \sum_{k=1}^\infty \frac{B_{2k}}{(2k)!}(f^{2k-1}(b)-f^{2k-1}(a)),
\ee
where the $B_{k}$ are the Bernoulli numbers,
\item[(ii)] the integral formula for the $\log \Gamma(z)$ function:
\be
\label{eq:LogGamma}
\log \Gamma(z) = \int_0^\infty \frac{dt}t \Big\{(z-1)\eE^{-t} - \frac{\eE^{-t} - \eE^{-zt}}{1 - \eE^{-t}}\Big\}, \qquad {\rm Re}\, z >0,
\ee
\item[(iii)] the exact product formula:
\begin{equation}\label{eq:identity1}
\begin{split}
\prod_{\ell=1}^{\frac{n-2}{2}}\sin\left(\tfrac{\pi \ell}{n}\right) = \prod_{\ell=1}^{\frac{n-2}{2}}\cos\left(\tfrac{\pi \ell}{n}\right)=\frac{ n^{\frac{1}{2}}}{2^{\frac{n-1}{2}}} \qquad (n \textrm{ even}).
\end{split}
\end{equation}
\end{itemize}
The lattice dimensions $N_A$ and $N$ are assumed to satisfy $\textrm{gcd}(N_A,N) =1$ for $N$ odd and $\textrm{gcd}(\frac{N_A}2,\frac N2) =1$ for $N$ even.

\subsection{Closed forms}

With \eqref{eq:overlap.and.det} and \eqref{eq:BFClosedForm}, we write \eqref{eq:FdOverlaps} as
\begin{equation}
\label{eq:FdTP}
\begin{split} 
\mathcal{F}_d &= -2 \Big( \log \big|P_0(N,x,d)\big| - \log \big|P_1(N,x,d)\big| - \log \big|P_2(N,1-x,d)\big| \\[0.15cm]
&+ \log \big|P_2\big(N x,\tfrac{1-x}{x},d\big)\big|+ \log \big|P_3(N,x,d)\big|+ \log \big|P_3(N,1-x,0)\big|+ \log \big|P_3(N,1,d)\big| \Big)
\end{split}
\end{equation}
where
\begin{equation}
\label{eq:ClosedProducts}
\begin{split}
&P_0(N,x,d) = \prod_{k=1}^{\frac{Nx-d}{2}}\frac{\sin\big(\frac{\pi k}{Nx}\big)}{\sqrt{Nx \cos\big( \tfrac{\pi k}{N x}\big)}}  \prod_{k=1}^{\frac{N(1-x)-2}{2}}\frac{\sin\big( \tfrac{\pi k}{N(1-x)}\big)}{\sqrt{N(1-x) \cos\big( \tfrac{\pi k}{N (1-x)} \big)}}
 \prod_{\ell=1}^{\frac{N-d}{2}}\frac{\cos(\frac{\pi \ell}{N})\sin(\pi \ell x)}{\sqrt{N \cos\big( \tfrac{\pi \ell}{N }\big)}}, \\
& P_1(N,x,d) =  \prod_{k=1}^{\frac{Nx-d}2}\prod_{\ell = 1}^{\frac{N-d}{2}}
\left( \cos\big( \tfrac{\pi k}{Nx}\big) - \cos\big( \tfrac{\pi \ell}{N}\big)  \right) ,\\
&P_2 (N,x,d)=\prod_{k=1}^{\frac{Nx}{2}}\prod_{\ell = 1}^{\frac{N-d}{2}}
    \left( \cos\big( \tfrac{\pi k}{Nx}\big) - \cos\big( \tfrac{\pi \ell}{N}\big)  \right) ,\\
& P_3(N,x,d) = \prod_{1 \le k<k'\le \frac{Nx-d}{2}}  \left( \cos\big( \tfrac{\pi k'}{Nx}\big) - \cos\big( \tfrac{\pi k}{Nx}\big)  \right).
\end{split}
\end{equation} 
We recall that $x=N_A/N$, $N_B=N(1-x)$ is even and $N$, $N x$ and $d$ have the same parity so that $N-d$ and $Nx-d$ are even numbers too. 

In the following, we derive the $1/N$ expansion of the logarithm of each product in \eqref{eq:ClosedProducts}. For $P_0(N,x,d)$, we use \eqref{eq:identity1} with $n=N_B$. We express the other products in terms of the cardinal sine function $s[x]=\sin(x)/x$ by transforming the differences of cosines into products of sines. We find
\begingroup
\allowdisplaybreaks
\begin{subequations} 
\label{eq:Pjs}
\begin{alignat}{2}
\label{eq:P0(N,x,d)}
P_0(N,x,d) &= N^{-\frac{2N-2d-3}{4}}x^{-\frac{Nx-d}{4}}(1-x)^{-\frac{N(1-x)-3}{4}}2^{-\frac{N(1-x)-1}{4}}\nonumber \\
&\times \prod_{k=1}^{\frac{Nx-d}{2}}\frac{\sin\left(\tfrac{\pi k}{Nx}\right)}{\sqrt{ \cos\left(\tfrac{\pi k}{Nx}\right)}}  \prod_{\ell=1}^{\frac{N-d}{2}} \sqrt{\cos\left(\tfrac{\pi \ell}{N}\right)} \sin(\pi \ell x) ,
\\[0.35cm]
\label{eq:P1(N,x,d)}
\log |P_1(N,x,d)| &= \left( \frac{Nx-d}{2}\right)\left(\frac{N-d}{2}\right) (-2\log N+2 \log \pi -\log 2)\nonumber\\
&+g(N,x,d,d)+h(N,x,d,d)+Y_+\left( N, x,d\right)+Y_-\left( N, x,d\right) \\
&-2X\left(\frac{1}{4}, N, x,d\right)- 2X\left(\frac{1}{4},N, 1,d\right),\nonumber
\\[0.35cm]
\label{eq:P2(N,x,d)}
\log |P_2(N,x,d)| &= \frac{Nx}{2}\left(\frac{N-d}{2}\right) (-2\log N+2 \log \pi -\log 2)\nonumber\\
&+g(N,x,d,2)+h(N,x,d,2)  + \log \Gamma\left(N-\frac{d}{2}+1\right)-\log N \\
&+ \log 2-\log \Gamma \left( \frac{d}{2}\right)+\tilde{Y}_+\left( N, x,d\right)+\tilde{Y}_-\left( N, x,d\right) \nonumber\\
&- 2X\left(\frac{1}{4},N, x,0\right)- 2X\left(\frac{1}{4}, N, 1,d\right),\nonumber
\\[0.35cm]
\label{eq:P3(N,x,d)}
\log |P_3(N,x,d)| &=\left( \frac{Nx-d}{2}\right)\left( \frac{Nx-d-2}{2}\right)\left(-\log N- \log x+ \log \pi -\frac{1}{2}\log 2\right)\nonumber\\
 &+ i\left( \frac{Nx-d}{2}+1\right) + \frac{1}{2}Y_+\left( Nx, 1, d \right)+\frac{1}{2}Y_-\left( Nx, 1, d \right)\\
&-2X\left(\frac{1}{4}, N, x, d \right)-\frac{1}{2}X\left(\frac{1}{2},N, x, d \right),\nonumber
\end{alignat}
\end{subequations}
where 
\begin{subequations}
\label{eq:XYghi}
  \begin{alignat}{3}
 &X(a,N,x,m) = \sum_{k=0}^{\frac{N x-m}{2}}\log s\left[\frac{2 \pi a}{N}\left(\frac{k}{x}\right) \right], && \qquad g(N,x,d,m) = \sum _{k=1}^{\frac{Nx-m}{2}} \sum _{\ell=1}^{\frac{N-d}{2}} \log \left(\frac{k}{x}+\ell\right), \\
 &Y_{\pm}(N,x,m) =  \sum_{k=0}^{\frac{N x-m}{2}}\sum_{\ell=0}^{\frac{N -m}{2}}\log s\left[\frac{ \pi }{2N}\left(\frac{k}{x} \pm \ell\right) \right], && \qquad h(N,x,d,m) = \sum _{k=1}^{\frac{Nx-m}{2}} \sum _{\ell=1}^{\frac{N-d}{2}} \log \left|\frac{k}{x}-\ell\right|,\\
 &\tilde{Y}_{\pm}(N,x,m) =  \sum_{k=0}^{\frac{N x}{2}}\sum_{\ell=0}^{\frac{N -m}{2}}\log s\left[\frac{ \pi }{2N}\left(\frac{k}{x} \pm \ell\right) \right], && \qquad i(N) = \sum_{k=1}^{N-2}\sum_{\ell=k+1}^{N-1}\log(\ell+k)+\log(\ell-k).
 \end{alignat}
\end{subequations}
\endgroup

\subsection[Asymptotic behaviour of $P_0(N,x,d)$]{Asymptotic behaviour of $\boldsymbol{P_0(N,x,d)}$}

In order to access the $1/N$ expansion of $P_0(N,x,d)$, we need to simplify the trigonometric products in \eqref{eq:P0(N,x,d)}. For that matter we use \eqref{eq:identity1} and similar identities,
\begin{subequations}
\label{eq:identityMixed}
\begin{alignat}{2}
\label{eq:identityMixed_a}
&\prod_{\ell=1}^{\frac{n-2}{2}}\big |\sin(\tfrac{\pi \ell m}{n})\big | = \frac{n}{2^{\frac n2}} \qquad &&(m,  n \textrm{ even}, \ \textrm{gcd}(\tfrac m2, \tfrac n2)=1),
\\[0.15cm]
\label{eq:identityMixed_b}
&\prod_{\ell=1}^{\frac{n-1}{2}}\big |\sin(\tfrac{\pi \ell m}{n})\big | =  \frac{n^{\frac12}} {2^{\frac{n-1}2}} \qquad &&(n \textrm{ odd},
\ \textrm{gcd}(m, n)=1),
\\[0.15cm]
\label{eq:identityMixed_c}
&\prod_{\ell=1}^{\frac{n-1}{2}}\cos(\tfrac{\pi \ell}{n}) = \frac1{2^{\frac{n-1}{2}}} \qquad &&(n \textrm{ odd}).
\end{alignat}
\end{subequations}
We use \eqref{eq:identity1} together with \eqref{eq:identityMixed_c} to find the intermediate results
\begin{equation}
\label{eq:identityIntermediate}
\begin{split}
&\log  \prod_{\ell=1}^{\frac{N-d}{2}} \cos\left( \tfrac{\pi \ell}{N}\right)  = \frac{d-1}{2}\log N - \frac{N-1}{2} \log 2 - \frac{d-2}{2}\log \pi - \log\Gamma\left(\frac{d}{2}\right)+ \mathcal{O}\left( N^{-2}\right),\\
&\log \prod_{\ell=1}^{\frac{N-d}{2}} \sin\left( \tfrac{\pi \ell}{N}\right)  = \frac{1}{2}\log N - \frac{N-1}{2} \log 2 + \mathcal{O}\left( N^{-2}\right),
\end{split}
\end{equation}
for both parities of $N$. We use \eqref{eq:identityMixed_a} and \eqref{eq:identityMixed_b} to derive an exact expression for the term $ \prod_{\ell=1}^{(N-d)/2}\sin(\pi \ell x)$ in \eqref{eq:P0(N,x,d)}. We find
\begin{equation}
\label{eq:partialSumSines}
 \log \prod_{\ell=1}^{\frac{n-d}{2}} \left|\sin\left( \tfrac{\pi \ell m}{n}\right) \right|=  
  \left\{\begin{array}{ll}
\displaystyle \log n - \tfrac{n}{2} \log 2 - \sum _{\ell=1}^{\frac{d}{2}-1} \log \left|\sin (\tfrac{\pi \ell m}{n})\right| \quad &\Big(\!\begin{array}{c}m,  n \textrm{ even} \\ \textrm{gcd}(\tfrac m2,\tfrac n2)=1\end{array}\!\Big),\\[0.55cm]
\displaystyle \tfrac{1}{2} \log n - \tfrac{n-1}{2} \log 2-\sum _{\ell=1}^{\frac{d-1}{2}} \log \left| \cos \Big(\tfrac{\pi m}{n} (\ell-\tfrac{d}{2})\Big)\right|   \quad &\Big(\!\begin{array}{c} m, n \textrm{ odd}\\ \textrm{gcd}(m,n)=1\end{array}\!\Big),
\\[0.55cm]
\displaystyle \tfrac{1}{2} \log n - \tfrac{n-1}{2} \log 2-\sum _{\ell=1}^{\frac{d-1}{2}} \log \left| \sin \Big(\tfrac{\pi m}{n} (\ell-\tfrac{d}{2})\Big)\right|   \quad &\Big(\!\begin{array}{c}m \ \textrm{even}, \, n \textrm{ odd}\\ \textrm{gcd}(m,n)=1  \end{array} \!\Big).
\end{array}\right.
\end{equation}
The identities \eqref{eq:identityIntermediate} and \eqref{eq:partialSumSines} applied in \eqref{eq:P0(N,x,d)} imply
\begin{equation}
\label{eq:P0Asymptotics}
\begin{split}
\log P_0(N,x,d) &=\frac{N}{4} (-2 \log N-(1-x) \log (1-x)-x \log x-4 \log 2)\\[0.15cm]
&+\frac{1}{4} \big((2 d+9) \log N+3 \log (1-x)+3 \log x+5 \log 2\big)\\[0.15cm]
&-\begin{cases}
\displaystyle \;\;\ \frac12 \log 2 + \sum _{j=1}^{\frac{d}{2}-1} \log \left|\sin (\pi  j x)\right| + \mathcal{O}\left( N^{-2}\right) & \qquad \textrm{for } N, d \textrm{ even},\\
\displaystyle \;\;\ \frac12 \log N +  \sum _{j=1}^{\frac{d-1}{2}} \log \left| \cos \Big(\pi  x(j-\tfrac{d}{2})\Big)\right| + \mathcal{O}\left( N^{-2}\right) & \qquad \textrm{for } N, d \textrm{ odd}.
\end{cases}
\end{split}
\end{equation}

\subsection[Asymptotic behaviour of $P_1(N,x,d)$, $P_2(N,x,d)$ and $P_3(N,x,d)$]{Asymptotic behaviour of $\boldsymbol{P_1(N,x,d)}$, $\boldsymbol{P_2(N,x,d)}$ and $\boldsymbol{P_3(N,x,d)}$}

In order to understand the asymptotic behaviour of $P_1(N,x,d)$, $P_2(N,x,d)$ and $P_3(N,x,d)$ in \eqref{eq:Pjs}, we need to understand the functions $X$, $Y_{\pm}$,$\tilde{Y}_{\pm}$, $g$, $h$ and $i$ in \eqref{eq:XYghi}.

\paragraph{The functions $\boldsymbol{X}$, $\boldsymbol{Y_{\pm}}$ and  $\boldsymbol{\tilde{Y}_{\pm}}$.} We first define the integrals
 \begin{subequations}
  \label{eq:I0Jpm}
 \begin{alignat}{2}
 &\mathcal{I}_0(a) = \int_0^1 \log s\left[\pi a z\right] \dd z, \\
  &\mathcal{I}_{\pm}(x) = \int_0^x \log s\left[ \frac{\pi (z \pm x)}{4x}\right] \dd z, \\
  &\mathcal{J}_{\pm}(x) = \int_0^x  \int_0^x \log s\left[ \frac{\pi  (y\pm z)}{4x}\right] \dd y \ \dd z,
 \end{alignat}
 \end{subequations}
 where we recall that $s[x] = \frac{\sin(x)}{x}$. We make repeated use of Euler-MacLaurin formula \eqref{eq:EmL} and obtain the $1/N$ expansions
 \begingroup
 \allowdisplaybreaks
 \begin{subequations}
 \label{eq:XYAsymptotics}
 \begin{alignat}{2}
 X(a,N,x,m) &= \frac{Nx}{2} \mathcal{I}_0(a) + \frac{1-m}{2} \log s[\pi a]+ \mathcal{O}\left(N^{-1}\right), 
\\[0.4cm]%
 Y_+(N,x,m)&= \frac{N^2}{4 x} \mathcal{J}_+(x)  +\frac{Nx}{2} \left(\frac{1}{2 x}+\frac{1}{2}\right) \mathcal{I}_0\Big(\frac 14\Big) 
 + \frac{N}{2} \left(\frac{1}{2 x}+\frac{1}{2}-\frac{m}{2}\left( 1+\frac{1}{x}\right) \right) \mathcal{I}_+(x) \nonumber\\
 &-\frac{4-6m(x-1)^2+4x(-3+x)+3m^2(1+x^2)}{24 x}\log s\Big[\frac{\pi}{4}\Big] \nonumber \\
 &+ \frac{2-6m(1+x)^2+3m^2(1+x)^2+2x(3+x)}{24 x}\log s\Big[\frac{\pi}{2}\Big] 
 + \mathcal{O}\left(N^{-1}\right),
\\[0.4cm]%
  Y_-(N,x,m) &=  \frac{N^2}{4 x} \mathcal{J}_-(x) +\frac{Nx}{2} \left(\frac{1}{2 x}+\frac{1}{2}\right) \mathcal{I}_0\Big(\frac 14\Big) 
  + \frac{N}{2} \left(\frac{1}{2 x}+\frac{1}{2}-\frac{m}{2}\left( 1+\frac{1}{x}\right) \right) \mathcal{I}_-(x)\nonumber\\
  &+\frac{4-6m(x+1)^2+4x(3+x)+3m^2(1+x^2)}{24 x}\log s\Big[\frac{\pi}{4}\Big] + \mathcal{O}\left(N^{-1}\right),
\\[0.4cm]%
 \tilde{Y}_+(N,x,m) &= \frac{N^2}{4 x} \mathcal{J}_+(x)  +\frac{Nx}{2} \left(\frac{1}{2 x}+\frac{1}{2}\right) \mathcal{I}_0\Big(\frac 14\Big) 
+ \frac{N}{2} \left(\frac{1}{2 x}+\frac{1}{2}-\frac{m}{2} \right) \mathcal{I}_+(x) \nonumber\\
 &-\frac{4+4x(-3+x)+3mx(2+(-2+m)x)}{24 x}\log s\Big[\frac{\pi}{4}\Big] \nonumber\\
 &+ \frac{2+2x(3+x)+3mx(-2+(-2+m)x)}{24 x}\log s\Big[\frac{\pi}{2}\Big] + \mathcal{O}\left(N^{-1}\right),
 \\[0.4cm]%
  \tilde{Y}_-(N,x,m) &=  \frac{N^2}{4 x} \mathcal{J}_-(x) +\frac{Nx}{2} \left(\frac{1}{2 x}+\frac{1}{2}\right) \mathcal{I}_0\Big(\frac 14\Big)
  + \frac{N}{2} \left(\frac{1}{2 x}+\frac{1}{2}-\frac{m}{2}\right) \mathcal{I}_-(x)\nonumber\\
  &+\frac{4+4x(3+x)+3mx(-2+(-2+m)x)}{24 x}\log s\Big[\frac{\pi}{4}\Big] + \mathcal{O}\left(N^{-1}\right).
 \end{alignat}
 \end{subequations}
 \endgroup
Remarkably, all the terms in \eqref{eq:XYAsymptotics} that contain the integrals defined in \eqref{eq:I0Jpm} identically cancel in the combination \eqref{eq:FdTP}.

\paragraph{The function $\boldsymbol{i}$.} 
We express $i(N)$ in \eqref{eq:XYghi} in terms of the Gamma function and Barnes' G-function:
 \begin{equation}
 \label{eq:Gamma}
 \Gamma(z+1)=z\, \Gamma(z), \qquad G(z+1) = \Gamma(z) G(z).
 \end{equation}
 Indeed, using the duplication formula for the Gamma function, 
 \be
 \Gamma(2k) = \frac{2^{2k-1}}{\sqrt{\pi}}\Gamma(k)\Gamma\left(k+\frac{1}{2}\right),
 \ee
 we obtain
  \begin{equation}
 \begin{split}
i(N)= &-(N-2)(N-1)\log 2 + \frac{N-2}{2}\log \pi + \log G(2N-1)-\log G(N+1)\\
&-\log G\left(N-\frac{1}{2}\right)+\log G\left(\frac{3}{2}\right).
\end{split}
 \end{equation}
The asymptotic behaviour of the Barnes' G-function is known, 
 \begin{equation}\label{eq:AsymptoticsGammaG}
 \begin{split}
 &\log G(z) = \left(\frac{(z-1)^2}{2} -\frac{1}{12}\right) \log (z-1)-\frac{3(z-1)^2}{4} +\frac{z-1}{2} \log (2 \pi )+\frac{1}{12}-\log A+ \mathcal{O}\left( z^{-1}\right),
 \end{split}
 \end{equation}
 where $A \simeq 1.282427$ is the Glaisher-Kinkelin constant. With \eqref{eq:AsymptoticsGammaG}, we find
 \begin{equation}\label{eq:iAsymptotics}
 \begin{split}
 i(N)=&N^2 \left(\log N-\frac{3}{2}+\log 2 \right)+\frac{N}{2}  \left(-5 \log N+5+\log \left(\frac{\pi }{4}\right)\right)-\log \pi  \\
  +& \frac{1}{24} (-12 \log A+23 \log N+1-7 \log 2)+\mathcal{O}\left( N^{-1}\right).
 \end{split}
 \end{equation}

\paragraph{The functions $\boldsymbol{g}$ and $\boldsymbol{h}$.} 
An inspection of \eqref{eq:FdTP}, \eqref{eq:P1(N,x,d)}, \eqref{eq:P2(N,x,d)} and \eqref{eq:P3(N,x,d)} shows that the two functions $g$ and $h$ only appear through their sum, which we denote by $gh(N,x,d,m)$. Moreover, and this will be crucial, we only need to compute the following specific combination:
\begin{equation}
\label{eq:CombinationGH}
C_{gh}(N,x,d) \equiv  2\Big( gh(N,x,d,d) + gh(N,1-x,d,2) - gh(N x,\tfrac{1-x}{x},d,2)\Big).
\end{equation}

The first step is to rewrite the functions $g$ and $h$ as integrals. The way this can be done differs for the two functions, and for $h$, also depends on the value of $m$, equal to $d$ or 2.
In each case, the passage to an integral expression is achieved using the integral representation of $\log\Gamma(z)$ given in \eqref{eq:LogGamma}. In doing so, we obtain sums of integrals which are separately divergent (the integrands have non-integrable singularities at $t=0$). Because the functions $g$ and $h$ are finite, these divergencies must cancel when summed up. As we intend to compute the integrals separately, we regularise them by restricting the domain of integration to $[\epsilon,+\infty)$. For each such integral, we will compute the first terms in an expansion in $\epsilon$ at order $\mathcal O(\epsilon)$. All terms that are divergent as $\epsilon \to 0$ should eventually cancel out. We denote the regularised functions by $g_\epsilon$ and $h_\epsilon$.

By using the identity $z = \Gamma(z+1)/\Gamma(z)$, the summation over $\ell$ in $g$ is telescopic and yields
\be
g(N,x,d,m) = \sum_{k=1}^{\frac{Nx-m}2} \: \big[\log{\Gamma(\tfrac kx + 1 + \tfrac{N-d}2)} - \log\Gamma(\tfrac kx + 1)\big].
\ee
With \eqref{eq:LogGamma}, the summation over $k$ can be explicitly carried out, leading to
\be
g_\epsilon(N,x,d,m) = \tfrac{(Nx-m)(N-d)}4 \int_\epsilon^\infty \frac{dt}t \eE^{-t} - \int_\epsilon^\infty \frac{dt}t \frac{1-\eE^{-(N-d)t/2}}{\eE^{t}-1} \cdot\frac{1-\eE^{-(Nx-m)t/2x}}{\eE^{t/x}-1} .\label{eq:g.int}
\ee

For $m=d$, the summation over $\ell$ in the function $h$ yields
\be
h(N,x,d,d) = \sum_{k=1}^{\frac{Nx-d}2} \big[\log{\Gamma(\tfrac kx)} - \log\big|\Gamma(\tfrac kx - \tfrac{N-d}2)\big|\big] = \sum_{k=1}^{\frac{Nx-d}2} \big[\log{\Gamma(\tfrac kx)} - \log\big|\Gamma(\tfrac d2 + \tfrac{2-d}{2x} - \tfrac kx)\big|\big],
\ee
where the last equality follows from the change of variable $k \to \frac{Nx-d}2 + 1 - k$. Using the identity $\Gamma(z)\Gamma(1-z) = \frac\pi{\sin{\pi z}}$, and then the integral representation of $\log \Gamma(z)$, we obtain
\be
h(N,x,d,d) = \sum_{k=1}^{\frac{Nx-d}2} \log\frac{\big|\sin\big(\pi(\tfrac kx + \tfrac{(2-d)(x-1)}{2x})\big)\big|}\pi + \sum_{k=1}^{\frac{Nx-d}2} \big[\log{\Gamma(\tfrac kx)} + \log\Gamma(\tfrac kx + \tfrac{(2-d)(x-1)}{2x})\big],
\ee
and subsequently
\begin{alignat}{2}
h_\epsilon(N,x,d,d) &= \sum_{k=1}^{\frac{Nx-d}2} \log\frac{\big|\sin\big(\pi(\tfrac kx + \tfrac{(2-d)(x-1)}{2x})\big)\big|}\pi \, + 
\tfrac{(Nx-d)(N-d-2)}4 \int_\epsilon^\infty \frac{dt}t \,\eE^{-t}  \nonumber\\
& - (Nx-d) \int_\epsilon^\infty \frac{dt}t \frac 1{\eE^t-1} + \int_\epsilon^\infty \frac{dt}t \: \frac{1+\eE^{-(d-2)(1-x)t/2x}}{1-\eE^{-t}} \cdot \frac{1-\eE^{-(Nx-d)t/2x}}{\eE^{t/x}-1}.
\end{alignat}

For $m=2$, the same procedure would lead to using the integral representation of $\log\Gamma(z)$ for Re$\,z < 0$. Instead we carry out the summation over $k$ first. The other steps are similar and yield
\begin{alignat}{2}
h(N,x,d,2) &= -\tfrac{(Nx-2)(N-d)}4 \log{x} + \sum_{\ell=1}^{\frac{N-d}2} \big[\log\Gamma(\tfrac{Nx}2 - \ell x) - \log\big|\Gamma(1 - \ell x)\big| \big]\\
& =  -\tfrac{(Nx-2)(N-d)}4 \log{x} + \sum_{\ell=1}^{\frac{N-d}2} \log \frac{|\sin(\pi\ell x)|}\pi  + \sum_{\ell=1}^{\frac{N-d}2} \big[\log\Gamma(\ell x) + \log\Gamma\big((\ell+\tfrac d2-1)x\big)\big],\nonumber
\end{alignat}
and
\begin{alignat}{2}
h_\epsilon(N,x,d,2) &=  \sum_{\ell=1}^{\frac{N-d}2} \log\frac{|\sin(\pi\ell x)|}\pi -\tfrac{(Nx-2)(N-d)}4 \log{x} + 
\tfrac{(Nx-4)(N-d)}4 \int_\epsilon^\infty \frac{dt}t \,\eE^{-t}  \nonumber\\
& - (N-d) \int_\epsilon^\infty \frac{dt}t \frac 1{\eE^t-1} + \int_\epsilon^\infty \frac{dt}t \: \frac{1+\eE^{-(d-2)xt/2}}{1-\eE^{-t}} \cdot \frac{1-\eE^{-(N-d)xt/2}}{\eE^{xt}-1}.\label{eq:hm2}
\end{alignat}

In the above expressions for $g_\epsilon$ and $h_\epsilon$, one can distinguish two types of integrals: those for which the integrand decays exponentially fast with $N$ (because they contain a factor $\eE^{-aNt}$), and those which do not depend on $N$ at all. We start with the first category, bearing in mind that we are interested in the asymptotic value of the integrals to order $N^{-1}\log N$. All terms of order $N^{-1}$ or smaller will be neglected.

The integrals we have to evaluate are of the form $\int_\epsilon^\infty \eE^{-aNt}F(t)$ where the function $F(t)$ decays exponentially fast at infinity but generically has a pole at the origin of finite order (smaller or equal to 3 in all cases). Expanding $F(t)$ in a Laurent series around the singularity $t=0$, the positive power part of the expansion can be neglected since a power of $t^k$, upon integration, contributes a finite term proportional to $N^{-k-1}$. In particular, no term proportional to $N^{-1}\log N$ can be produced. For the negative powers of $t$, we use the following results, valid for any positive $n$, large or not:
\begin{subequations}
\label{eq:int123}
\begin{eqnarray}
\int_\eps^\infty \; \frac{dt}t \eE^{-nt} &\!\!\!=\!\!\!& -\log{\eps} - \log{n} - \gamma + \mathcal{O}\left(\eps\right),\\
\int_\eps^\infty \; \frac{dt}{t^2} \eE^{-nt} &\!\!\!=\!\!\!& \frac 1\eps + n\log\eps + n\log{n} +n(\gamma-1) + \mathcal{O}\left(\eps\right),\\
\int_\eps^\infty \; \frac{dt}{t^3} \eE^{-nt} &\!\!\!=\!\!\!& \frac 1{2\eps^2} - \frac n\eps - \frac{n^2}2 \log{\eps} -\frac{n^2}2 \Big(\log{n} - \frac 32 + \gamma\Big) + \mathcal{O}\left(\eps\right),
\end{eqnarray}
\end{subequations}
where $\gamma \simeq 0.57722$ is the Euler constant. 

For some of the integrals which do not depend on $N$, we also use the following integral:
\be
\label{eq:int.alpha}
\int_\eps^\infty \; \frac{dt}t \: \frac 1{\eE^{\alpha t}-1} = \frac 1{\alpha\eps} + \frac 12 \log{\eps} + \frac 12 \log{\alpha} + \frac\gamma2 - \frac 12 \log(2\pi) + \mathcal{O}\left(\eps\right), \qquad \alpha > 0.
\ee
The formulas \eqref{eq:int123} and \eqref{eq:int.alpha} are easily derived from the following two convergent integrals, given in \cite[eqs. (3.429) and (3.427.4)]{GR07}:
\begin{alignat}{2}
\int_0^\infty  \frac{dt}t \Big(\eE^{-t}-\frac1{1+t} \Big) = - \gamma,\qquad \int_0^\infty \frac{dt}t \eE^{-t} \Big(\frac1{\eE^t-1} - \frac 1t + \frac 12\Big) = 1 - \frac12 \log(2 \pi).
\end{alignat}

When computing $gh(N,x,d,m)$, we are free to regularise the functions $g_\epsilon$ and $h_{\epsilon'}$ with two parameters $\epsilon$ and $\epsilon'$ that can differ. The convenient choice turns out to be $\epsilon = \epsilon'$ for $m=d$ and $\epsilon = x\epsilon'$ for $m=2$. For instance, we combine an integral from \eqref{eq:g.int} with a second integral from \eqref{eq:hm2} and find
\be
- \int_\epsilon^\infty \frac{dt}t \frac{1}{\eE^{t}-1} \cdot\frac{1}{\eE^{t/x}-1} + \int_{\epsilon'}^\infty \frac{dt}t \: \frac{1}{1-\eE^{-t}} \cdot \frac{1}{\eE^{xt}-1} = \int_\epsilon^\infty \frac{dt}t \frac{1}{\eE^{t}-1} 
\ee
and the resulting integral is of the form \eqref{eq:int.alpha} with $\alpha = 1$.
We then encounter only two integrals that are not expressible as the integrals in \eqref{eq:int123} and \eqref{eq:int.alpha}. Both are independent of $N$. They read, respectively for $m=d$ and $m=2$,
\be
\int_\epsilon^\infty \frac{dt}t \: \frac{\eE^{-(\frac d2-1)(1-x)t/x}}{(1-\eE^{-t})(\eE^{t/x}-1)} \qquad {\rm and} \qquad 
\int_\epsilon^\infty \frac{dt}t \: \frac{\eE^{-(\frac d2-1)xt}}{(1-\eE^{-t})(\eE^{xt}-1)}.
\ee
By subtracting counterterms, their $\eps$-expansion can be expressed in terms of the following two convergent integrals $I_1(x)$ and $I_2(x)$,
\begin{subequations}
\begin{eqnarray}
I_1(x) &\!\!\!=\!\!\!& \int_0^\infty \frac{dt}t \Big\{\frac{\eE^{-(\frac d2-1)(1-x)t/x}}{(1-\eE^{-t})(\eE^{t/x}-1)} - A(t;x) \, \eE^{-t(1+x)/(2x)} \Big\},\\
\noalign{\smallskip}
A(t;x) &\!\!\!=\!\!\!& \Big[\frac x{t^2} + \frac{1-\frac d2(1-x)}{t} + \frac{11-12d+3d^2+6xd(2-d)+x^2(3d^2-1)}{24x}\Big],
\end{eqnarray}
\end{subequations}
and
\begin{subequations}
\begin{eqnarray}
I_2(x) &\!\!\!=\!\!\!& \int_0^\infty \frac{dt}t \Big\{ \frac{\eE^{-(\frac d2-1)xt}}{(1-\eE^{-t})(\eE^{xt}-1)} - B(t;x) \, \eE^{-t(1+x)/2}\Big\}, \\
\noalign{\smallskip}
B(t;x) &\!\!\!=\!\!\!& \Big[\frac 1{xt^2} +\frac{1-\frac d2 + \frac 1x}t + \frac{11+12x(2-d)+x^2(11-12d+3d^2)}{24x}\Big],
\end{eqnarray}
\end{subequations}
to which must be added the divergent integrals of the counterterms, well under control by using \eqref{eq:int123}. 

Putting all together, we obtain the following expression of $gh(N,x,d,m)$ for $m=d$,
\begin{eqnarray}
\label{eq:ghd}
&& \hspace{-1.5cm}gh(N,x,d,d) = \frac{N^2 x}2 \big(\log N - \tfrac 32\big) + \frac N2 \big[d(x+1) + \log 2 + x \log(4\pi) - d (x+1) \log N\big] \nonumber\\[0.15cm]
&& \hspace{-1cm} + \: \frac{(1-3d+\frac 32 d^2)(1+x^2)-3x(3-2d-d^2)}{12x} \bigg(\log N - \log\Big(\frac{1+x}{2x}\Big)\bigg) + \frac{d^2-1}2 \,\log\Big(\frac{1+x}{2x}\Big)\nonumber\\[0.15cm]
&& \hspace{-1cm}  +\: (1-d) \log 2 - \frac{d+1}2 \log(2\pi) - \frac 12 \log x - \frac{(x+1)(5-3x+4d(x-1))}{16x} + I_1(x)  \nonumber\\[0.15cm]
&& \hspace{-1cm}  +\: \sum_{k=1}^{\frac{Nx-d}2} \log\frac{\big|\sin\big(\pi(\tfrac kx + \tfrac{(2-d)(x-1)}{2x})\big)\big|}\pi + \mathcal{O}\left( N^{-1}\right),
\end{eqnarray}
and for $m=2$,
\begin{eqnarray}
\label{eq:gh2}
&& \hspace{-1.5cm}gh(N,x,d,2) = \frac{N^2 x}2 \big(\log N - \tfrac 32\big) + \frac N2 \big[(dx+2) + (x+1)\log 2 + \log(2\pi) - (dx+2) \log N\big] \nonumber\\[0.15cm]
&& \hspace{-.5cm} + \: \frac{1-3x(1-3d)+x^2(1-3d+\frac 32 d^2)}{12x} \bigg(\log N - \log\Big(\frac{1+x}{2x}\Big)\bigg) + (d-\tfrac 12) \,\log\Big(\frac{1+x}{2x}\Big) - \frac d2 \log 2\nonumber\\[0.15cm]
&& \hspace{-.5cm} - \: \frac{d+1}2 \log(2\pi) + \frac{(x+1)(4dx-5(x+1))}{16x} + I_2(x) + \sum_{\ell=1}^{\frac{N-d}2} \log\frac{|\sin(\pi\ell x)|}\pi + \mathcal{O}\left( N^{-1}\right).
\end{eqnarray}

Finally, in order to evaluate the combination $C_{gh}(N,x,d)$, we need to compute the combination $I_1(x) + I_2(1-x) - I_2(\tfrac{1-x}x)$. Since both $I_1$ and $I_2$ are convergent, we can make the change of variable $t \to xt$ in the first term, $I_1(x)$, and in the third one, $I_2(\tfrac{1-x}x)$. Doing this yields
\begin{eqnarray}
I_1(x) + I_2(1-x) - I_2(\tfrac{1-x}x) && \hspace{-6mm} = \int_0^\infty \frac{dt}t \Bigg\{\frac{\eE^{-(\frac d2-1)(1-x)t}}{1-\eE^{-t}} \Big[\frac{\eE^{-t}}{1-\eE^{-xt}} + \frac 1{\eE^{(1-x)t}-1}\Big] - \frac{\eE^{-(\frac d2-1)(1-x)t}}{(1-\eE^{-xt})(\eE^{(1-x)t}-1)} \nonumber\\
&& \hspace{-2cm} -A(xt;x) \, \eE^{-t(1+x)/2} - B(t;1-x) \, \eE^{-t(2-x)/2} + B(xt;\tfrac{1-x}x) \, \eE^{-t/2} \Bigg\}.
\end{eqnarray}
The three terms on the first line miraculously cancel out, leaving the integral of the three terms on the second line. It is easily carried out by using the formulas \eqref{eq:int123}, 
\begin{alignat}{2}
I_1(x) + I_2(1-x) - I_2(\tfrac{1-x}x) & = \frac{17}{16} - \frac x2 + \frac d2(x-1) + \frac{(1-3d+\frac 32 d^2)(1+x^2) - 3x(1-d)^2}{12x} \log{(1+x)} \nonumber\\
\noalign{\medskip}
&  + \: \frac{5-6d+\frac 32 d^2 - x(5-9d+3d^2) + x^2(1-3d+\frac 32 d^2)}{12(1-x)} \log{(2-x)}.
\end{alignat}

The function $C_{gh}(N,x,d)$ may then be computed. We simplify the sum of logarithms of trigonometric functions in \eqref{eq:gh2} with \eqref{eq:partialSumSines}. A similar identity is used for \eqref{eq:ghd}. The result reads
\begin{alignat}{2}
C_{gh}(N,x,d) =\,&N^2 \Big((1-x+x^2)(\log N - \tfrac 32) + x(x-1)x \log x\Big)\nonumber\\
& \hspace{-2cm} +\,N \Big((2-2x+d+dx)(1 - \log N) + (d+2x-dx) \log x + 2\log 2\Big) + d^2 \log N - 3d\log 2 - \log \Big(\frac \pi 2\Big) \nonumber\\
& \hspace{-2cm} + \, \frac{2x-(2-6d+3d^2)(1-x)^2}{12 x} \log(1-x) - 
\frac{10-10x+2x^2 + 6d(2-x-x^2) + 3d^2(1-x)^2}{12(1-x)}\log x\nonumber\\
\noalign{\medskip}
& + \left\{\begin{array}{ll}
\displaystyle \;\;\log N - 2\sum _{j=1}^{\frac{d}{2}-1} \log \left|\sin (\pi  j x)\right| + \mathcal{O}\left( N^{-1}\right) & \qquad \textrm{for } N, d \textrm{ even},\\
\displaystyle \;\;\log 2 - 2\sum _{j=1}^{\frac{d-1}{2}} \log \left| \cos \Big(\pi  x(j-\tfrac{d}{2})\Big)\right| + \mathcal{O}\left( N^{-1}\right) & \qquad \textrm{for } N, d \textrm{ odd}.
\end{array}\right.
\label{eq:CghAsymptotics}
\end{alignat}
In computing \eqref{eq:FdTP}, we find that the sums of logarithms of trigonometric functions from \eqref{eq:CghAsymptotics} simplify with those in \eqref{eq:P0Asymptotics}. Combining \eqref{eq:FdTP}, \eqref{eq:P1(N,x,d)} to  \eqref{eq:P3(N,x,d)}, \eqref{eq:P0Asymptotics}, \eqref{eq:XYAsymptotics}, \eqref{eq:iAsymptotics} and \eqref{eq:CghAsymptotics}, we obtain \eqref{eq:AsymptoticsLBF} for both parities of $N$ and $d$, ending the proof of \cref{thm:Fd}. As noted earlier, the final result contains no $N^{-1}\log N$ term.


\end{document}